\documentclass[]{aa}  

\usepackage{graphicx}
\usepackage{txfonts}
\usepackage{multirow}
\usepackage[final]{hyperref}
\hypersetup{colorlinks=true, linkcolor=red,citecolor=blue,urlcolor=magenta} 

\usepackage{comment}
\usepackage{longtable}
\usepackage{rotating}
\usepackage{placeins}
\usepackage{tikz}
\usepackage{booktabs}
\usepackage{multirow}
\usepackage{rotating}
\usepackage{pdflscape}
\usepackage{xcolor} 
\usepackage[normalem]{ulem} 
\begin{document}

   \title{Living on the edge of the Central Molecular Zone:} 
   \subtitle{G1.3 is the more likely candidate for gas accretion into the CMZ}

   \author{Laura\,A.\,Busch
          \inst{1}\thanks{Member of the International Max\,Planck Research School\,(IMPRS) for Astronomy and Astrophysics at the Universities of Bonn and Cologne.}
          \and
          Denise\,Riquelme\inst{1}
          \and
          Rolf\,Güsten\inst{1}
          \and
          Karl\,M.\,Menten\inst{1}
          \and 
          Thushara\,G.\,S.\,Pillai\inst{2}$^,$\inst{1}
          \and 
          Jens\,Kauffmann\inst{3}$^,$\inst{1}
          }

   \institute{Max-Planck-Institut f\"ur Radioastronomie, Auf dem H\"ugel 69, 53121 Bonn, Germany\\
              \email{labusch@mpifr-bonn.mpg.de}
              \and 
              Institute for Astrophysical Research, 725 Commonwealth Ave, Boston University Boston, MA 02215, USA
              \and
              Haystack Observatory, Massachusetts Institute of Technology, 99 Millstone Road, Westford, MA 01886, USA 
             }

   \date{Received ; accepted }

 
  \abstract
   {The 1$\overset{\circ}{.}$3 (G1.3) and 1$\overset{\circ}{.}$6 (G1.6) cloud complexes in the Central Molecular Zone (CMZ) of our Galaxy have been proposed to possibly reside at the intersection region of the X1 and X2 orbits for several reasons. This includes the detection of co-spatial low- and high-velocity clouds, high velocity dispersion, high fractional molecular abundances of shock-tracing molecules, and kinetic temperatures that are higher than for usual CMZ clouds. } 
   {By investigating the morphology and deriving physical properties as well as chemical composition, we want to find the origin of the turbulent gas and, in particular, whether evidence of interaction between clouds can be identified.} 
   {We mapped both cloud complexes in molecular lines in the frequency range from 85 to 117\,GHz with the IRAM 30\,m telescope. The APEX 12\,m telescope was used to observe higher frequency transitions between 210 and 475\,GHz from selected molecules that are emitted from higher energy levels. We performed  non-local thermodynamic equilibrium (non-LTE) modelling of the emission of an ensemble of CH$_3$CN lines to derive kinetic temperatures and H$_2$ volume densities. These were used as starting points for non-LTE modelling of other molecules, for which column densities and abundances were determined and compared with values found for other sources in the CMZ.}
   {The kinematic structure of G1.3 reveals an `emission bridge' at intermediate velocities ($\sim$150\,km\,s$^{-1}$) connecting low-velocity ($\sim$100\,km\,s$^{-1}$) and high-velocity ($\sim$180\,km\,s$^{-1}$) gas and an overall fluffy shell-like structure. These may represent observational evidence of cloud-cloud interactions. Low- and high-velocity gas components in G1.6 do not show such evidence of interaction, suggesting that they are spatially separated. We selected three positions in each cloud complex for further analysis. Each position reveals several gas components at various peak velocities and of various line widths. We derived kinetic temperatures of 60--100\,K and H$_2$ volume densities of 10$^4$--10$^5$\,cm$^{-3}$ in both complexes. Molecular abundances relative to H$_2$ suggest a similar chemistry of the two clouds, which is moreover similar to that of other GC clouds and, especially, agrees well with that of G$+$0.693 and G$-$0.11.}
   {We conclude that G1.3 may indeed exhibit signs of cloud-cloud interactions. In particular, we propose an interaction of gas that is accreted from the near-side dust lane to the CMZ, with gas pre-existing at this location. Low- and high-velocity components in G1.6 are rather coincidentally observed along the same line of sight. They may be associated with either overshot decelerated gas from the far-side dust line or actual CMZ gas and high-velocity gas moving on a dust lane. These scenarios would be in agreement with numerical simulations. }

    \authorrunning{L.A.\,Busch et al.}
    \titlerunning{ISM in G1.3 and G1.6}
   \maketitle
%

\section{Introduction}
\begin{figure*}
    \begin{minipage}[b]{0.43\textwidth}
       \includegraphics[width=1.1\textwidth]{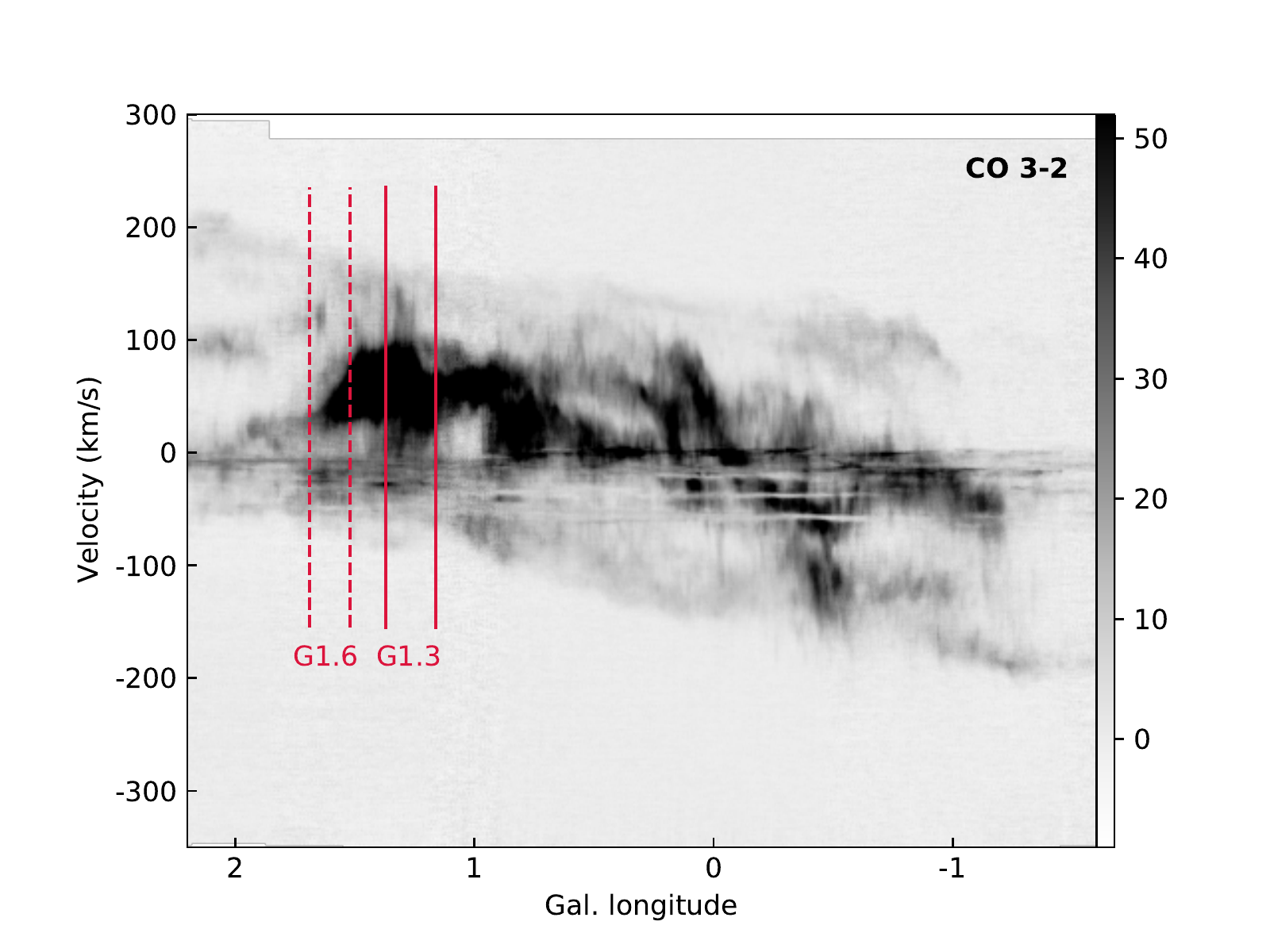}%
       \vspace*{0.5cm}
    \end{minipage}%
    \hfill
    \begin{minipage}[b]{0.57\textwidth}
        \centering
        \includegraphics[width=.95\textwidth]{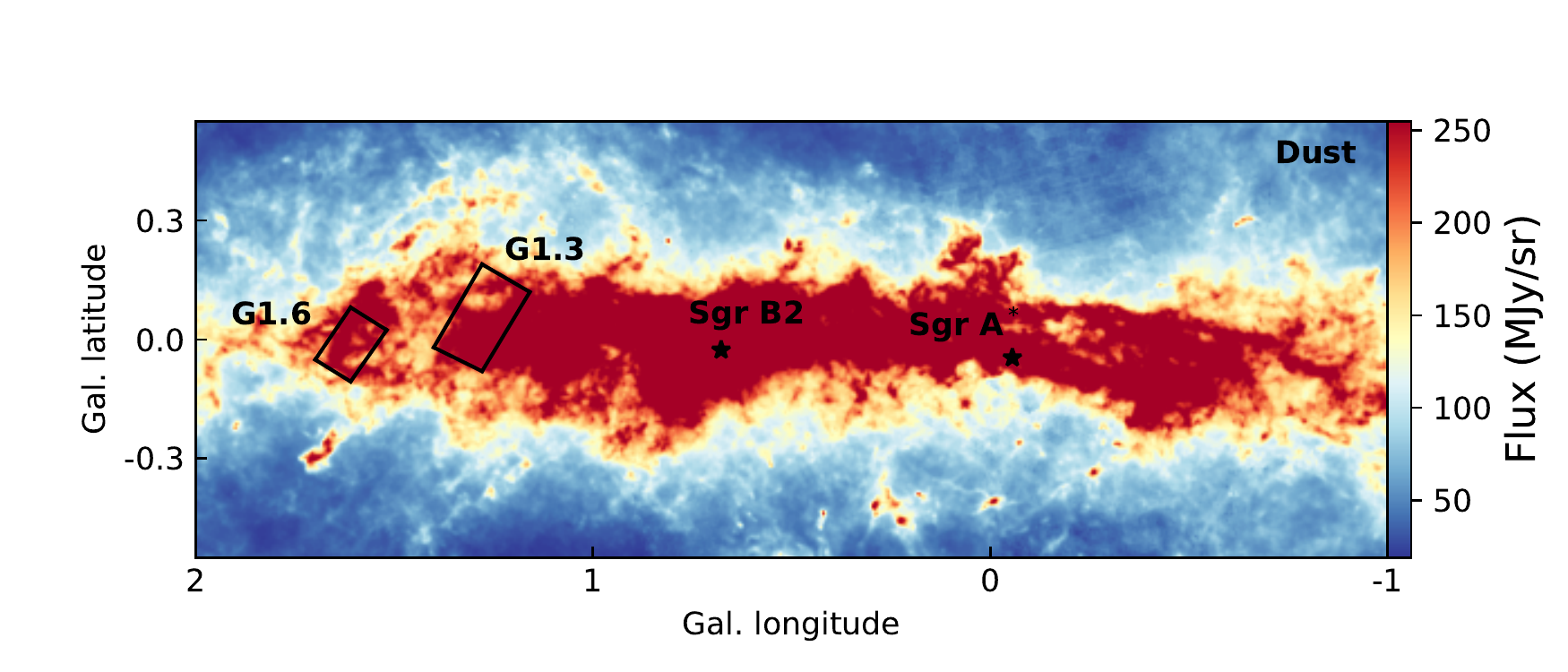}\\
        \includegraphics[width=.9\textwidth]{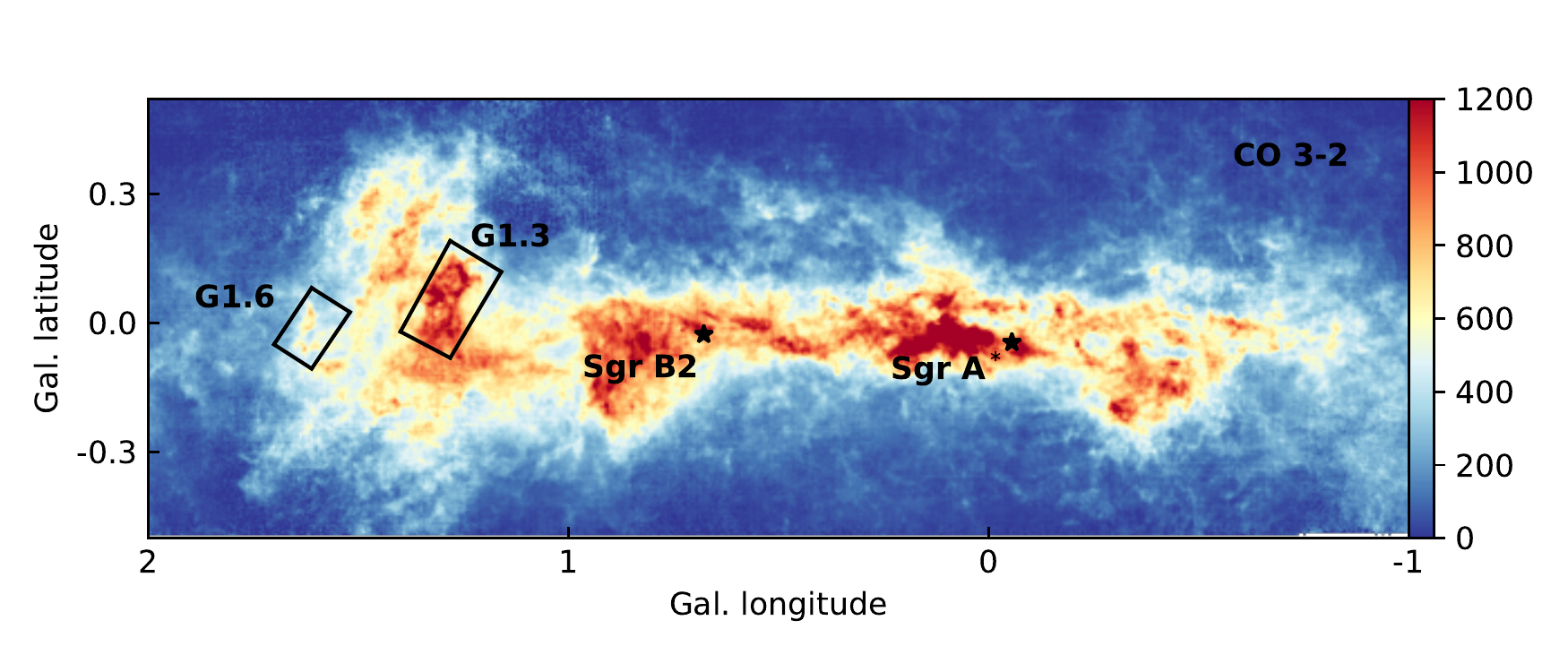}
    \end{minipage}
    \caption{\textit{Left panel:} CHIMPS\,2 longitude–velocity diagram of $^{12}$CO 3\,--\,2 intensity integrated over $|b| < 0.5^\circ$ Galactic latitude \citep[][]{Eden20}. Solid and dashed red lines indicate the observed longitude range for G1.3 and G1.6, respectively. \textit{Right panels:} Longitude-latitudes maps of the CMZ observed in warm dust emission with Hi-Gal/Herschel \citep[\textit{top},][]{Molinari16} and in $^{12}$CO 3\,--\,2 emission as part of the CHIMPS\,2 survey \citep[\textit{bottom},][]{Eden20}. The regions of G1.3 and G1.6 that were covered during our IRAM\,30m observations are indicated in black }
    \label{fig:cmz-overview}
\end{figure*}
In a radius of 200\,\,--\,\,250\,pc,  the Galactic Centre (GC) region  of our galaxy harbours a huge amount of molecular gas. This region is known as the Central Molecular Zone (CMZ, see Fig.\,\ref{fig:cmz-overview}). 
Physical properties of the molecular gas in the CMZ differ significantly from those of gas in the Galactic disk  \citep[e.g.][]{Morris96, Gusten04}. The gas in the CMZ is observed to generally be much denser ($\gtrsim$\,10$^4$\,cm$^{-2}$) and hotter \citep[$\gtrsim$\,50\,\,--\,\,70\,K;][]{Huttemeister93,Ao13,ott14,Ginsburg16,Immer16}.
Given these high densities, GC giant molecular clouds (GMCs) should  present a perfect nursery for stars. However, the overall star formation rate in the CMZ is unexpectedly low \citep[$< 0.1 M_\odot\rm{yr}^{-1}$, e.g.][]{Immer2012,Longmore13,Kaufmann17a,Kauffmann17b,Nguyen2021}. 
Active and recent high-mass star formation is localised in a few spectacular regions,
for example Sagittarius\,B2 \citep{Bally10,Kruijssen14} or the Arches and Quintuplet clusters \citep[][\citeyear{Figer02}]{Figer99} while much of the volume of the CMZ is devoid of massive young stellar objects \citep[early B- and O-type stars,][]{Immer2012, Nguyen2021}.
As a consequence, ultraviolet radiation emanating from young stars does not seem to be the dominating heating mechanism in the global view of the CMZ \citep[e.g.][]{Ao13}. 

Dust temperatures in this region do generally not exceed a value of 20\,K \citep{Lis99,Rodriguez-Fernandez02,Molinari11} suggesting that the dust is decoupled from the gas and hence, does not account for the globally high kinetic gas temperatures, either \citep[see also][]{Clark13}.
Widespread and intense emission of SiO suggests the omnipresence of shocks \citep{Martin-Pintado97,Huettemeister98,Riquelme10} as this molecule is formed from atomic silicon that was freed by the sputtering of (silicate) dust grains in shocks \citep{Schilke1997}. 
Shocks introduce turbulence in the gas, which is evident in broad line widths of observed spectral lines \citep[$\sim$10\,\,--\,\,30\,km\,s$^{-1}$; e.g.][]{Morris96} in the CMZ. 
Energy released by subsequent dissipation of the turbulence is converted to heat, which may be the dominant heating source. Shocks in the CMZ may be produced by, for example, interaction of clouds or by large-scale dynamics induced by the galactic gravitational potential or as a consequence of the central black hole's activity. 

The large velocity dispersion embodies a variety of velocity components in GMCs in the CMZ suggesting complex kinematics. The left panel of Fig.\,\ref{fig:cmz-overview} shows the distribution of gas in the CMZ with longitude over the radial velocity range from $-$200 to $+$200\,km\,s$^{-1}$. Many observational and theoretical studies have been trying to explain the motions of dense and diffuse gas. For example, the gas might move in two spiral arms \citep[][]{Sofue95,Sawada04,Ridley17} within the CMZ, a twisted ring \citep[][]{Molinari11}, on open orbits \citep[][]{Kruijssen15}, or/and on the so-called EMR \citep[Expanding Molecular Ring,][]{Kaifu72,Scoville72,Oka22,Sofue22}. \citet{Binney91} modelled the so-called X1 and X2 orbits, which arise as a consequence of the barred potential \citep[][]{Contopoulos77,Sormani2015}. 
According to this theory, there is a region in which higher-velocity gas from larger Galactic scales (associated with the X1 orbits) enters the CMZ (associated with the X2 orbits) at the point of intersection of the innermost X1 and outermost X2 orbits. The gas in this region is expected to be highly turbulent and may even be experiencing cloud-cloud interactions, which should be reflected in the physical and chemical properties of the interstellar medium.

At the edge of the CMZ, two cloud complexes at longitudes $\sim$1$\overset{\circ}{.}$3 and $\sim$1$\overset{\circ}{.}$6 (in the following G1.3 and G1.6) were proposed to be promising candidates of this intersection region, that is, the point at which gas enters the CMZ  \citep[][]{Huettemeister98,Rodriguez06,Rodriguez08,Riquelme10iso}. The few observations that were conducted towards G1.3 and G1.6 revealed their molecular diversity and large spectral line widths, which led to their assignment to the environment of the GC, for which a distance of 8.2\,kpc was recently accurately determined \citep{gravity19,gravity21}. We highlight their locations at the edge of the CMZ in Fig.\,\ref{fig:cmz-overview}.

G1.3 was observed already early, for example by \citet{Bally88} and \citet[][]{Oka01} in emission of CS and CO and showed a remarkable extension from the Galactic plane to high latitudes.
Significant amounts of gas in this region were found not only at velocities of $\lesssim$100\,km\,s$^{-1}$ but also at higher velocities of $\sim$180\,km\,s$^{-1}$ \citep[][\citeyear{Riquelme13}]{Tanaka07,Riquelme10}. 
Each of these velocity components seems to be comprised of warm ($\sim$30\,\,--\,\,40 K) and dense ($n(\mathrm{H}_2)\gtrsim10^5$\,cm$^{-3}$) gas, which is typical for GC GMCs, and hotter but less dense gas, which is suggested to have a different origin \citep{Huettemeister98,Riquelme13}. \citet{Riquelme13} derived kinetic temperatures of up to 300\,K.
G1.3 shows the largest fractional abundance of SiO in the CMZ \citep[$\sim3\times10^{-9}$,][]{Riquelme18} indicating that shocks play a major role in 
the heating of the gas and the formation and excitation of molecules. The presence of at least two gas components at different velocities suggests the possibility of cloud-cloud collisions \citep[e.g.][]{Habe92,Fukui21}, which might be the result of gas plunging from the innermost X1 onto the outermost X2 orbit \citep[e.g.][]{Mulder86} and encountering gas that already resides on the X2 orbit. \citet{Riquelme10iso} reported on $^{12}$C/$^{13}$C isotopic abundance ratios in G1.3 and found a low value of $\sim$\,20 in the low-velocity component and a higher value of $\sim$\,50 in the high-velocity
component. Low ratios around 20 are typically attributed to the CMZ gas, which is significantly processed
by star formation, while for less-processed gas, originating at larger Galacto-centric distances, this isotopic ratio is higher \citep[][and references therein]{Riquelme10iso}.
Thus, this is consistent with the view that the
high-velocity gas has been accreted into the GC. 

\begin{table*}[]
    \centering
    \caption{Telescopes, receivers, and backends used in this project along with observed frequency ranges.}
    \begin{tabular}{lllcccrrrr}
    \hline\hline\\[-0.3cm]
        Telescope & Receiver & Backend & $\nu\tablefootmark{a}$ & \multicolumn{2}{c}{Map coverage} & \multicolumn{2}{c}{Average rms$\tablefootmark{b}$} \\
        & & & & G1.3 & G1.6 & G1.3 & G1.6 \\
        & & & (GHz) & $(\arcsec)$ & $(\arcsec)$ & $T_\mathrm{mB}$\,(mK) & $T_\mathrm{mB}$\,(mK) \\\hline\\[-0.3cm]
        IRAM 30\,m & EMIR & FFTS & 85\,--\,93 & $860\times 550$ & $550\times 330$ & 73 & 43 \\
        & (E090) & & \,\,\,93\,--\,101 & $850\times 440$ & $550\times 330$ & 88 & 51 \\
        & & & 101\,--\,109 & $860\times 550$ & $550\times 330$ & 95 & 55 \\
        & & & 109\,--\,117 & $850\times 440$ & $550\times 330$  & 110 & 71 \\
        APEX & PI230 & FFTS4G & 
          217\,--\,221 & $780\times 500$ & $600\times 300$ & 254 & 266 \\ 
        & & & 229\,--\,233 & $780\times 500$ & $600\times 300$ & & \\
          & SHeFI (APEX2) & XFFTS & 265\,--\,269 & \,--\, & $650\times 340$ & \,--\, & 182 \\
          & FLASH+(345) & XFFTS & 289.5\,--\,293.5 & $150\times 150$ & $220\times 160$ & 70 & 73 \\
          & & & 301.5\,--\,305.5 & $150\times 150$ & $220\times 160$ & 63 & 65 \\
          & & & 328\,--\,332 & \,--\, & $600\times 300$ & \,--\, & 115 \\
          & & & 340\,--\,344 & \,--\, & $600\times 300$ & \,--\, & 116 \\
          & & & 341.5\,--\,345.5 & $250\times 250$ & $400\times 400$ & 168 & 116  \\
          & & & 344\,--\,348 & \,--\, & $500\times 300$ & \,--\, & 189 \\
          & & & 353.5\,--\,357.5 & -- & $400\times 400$ & -- & 141 \\
          & & & 356\,--\,360 & \,--\, & $500\times 300$ & \,--\, & 152 \\
          & FLASH+(460) & XFFTS & 
          433\,--\,437 & $150\times 150$ & $220\times 160$ & 209 & 511 \\
          & & & 459\,--\,463 & \,--\, & $600\times 300$ & \,--\, & 260 \\
          \hline\hline
    \end{tabular}
    \label{tab:obs_freqs}
    \tablefoot{\tablefoottext{a}{Covered frequency range.} \tablefoottext{b}{Average rms value in the reduced cubes after smoothing to a spectral resolution of 1\,km\,s$^{-1}$ and after spatial regridding to a common pixel size of 14.5\arcsec}\,(see Sect.\,\ref{ss:datared}). The centers of the mapped regions are given in the text.} 
\end{table*}

G1.6 is located at $\sim$1$\overset{\circ}{.}$6 Galactic longitude and shows similar characteristics as G1.3. Observations revealed gas components at velocities of $\sim$60\,km\,s$^{-1}$ and $\sim$160\,km\,s$^{-1}$ \citep{Whiteoak89,Salii02,Menten09}. 
These two components possess different physical properties, which resemble the picture 
found in the G1.3 region: While the low-velocity component is warm ($\sim$40\,--\,60 K) and dense ($n(\mathrm{H}_2)\lesssim10^5$\,cm$^{-3}$), the other is hot ($\sim$200\,K) and less dense ($\sim$10$^4$\,cm$^{-3}$) \citep[][\citeyear{Gardner87}]{Menten09,Gardner85}.
Observations of SiO rotational transitions \citep[e.g.][]{Menten09,Martin-Pintado97,Bally87} showed that this molecule's fractional abundance in the G1.6 region is almost as high as in the G1.3 \citep[$\sim2.5\times10^{-9}$,][]{Riquelme18}, suggesting the presence of shocks. These might as well occur as a consequence of a cloud-cloud collision \citep{Haschick93,Salii02,Menten09}, which makes
G1.6 another candidate for the X1\,--\,X2 intersection region.

To investigate the nature of these two cloud complexes, we conducted observations of a large number of molecules. For several species, we covered multiple lines allowing for excitation studies, for many of them for the first time. We mapped the G1.3 and G1.6 regions in various molecular transitions using the IRAM 30\,m and the APEX 12\,m telescopes. The observed regions are indicated in the right panels in Fig.\,\ref{fig:cmz-overview}.
In this paper we report on the overall morphology of both cloud complexes, their physical properties, and the chemical composition of their gas. Based on the results, we discuss the sources' appearances in the CMZ and possible origins.
In Sect.\,\ref{s:observations} we describe the observations and the data reduction. The results are presented in Sect.\,\ref{s:results} and discussed in Sect.\,\ref{s:discussion}. The conclusions are given in Sect.\,\ref{s:conclusion}.

\section{Observations \& Data reduction}\label{s:observations}

The data used in this work were acquired by a combination of observations carried out with the Atacama Pathfinder Experiment (APEX) 12\,m single dish \citep{Gusten06} and the IRAM 30\,m telescopes \citep{Baars87} in order to cover specific frequency segments, in which molecules of interest (see Sect.\,\ref{ss:lines}) emit. While with the IRAM 30\,m telescope the EMIR multi-band mm-waver receiver \citep{EMIR12,kramer16} was used to cover the complete frequency range between 85 and 117\,GHz, multiple receivers were used in the observations with APEX to be able to cover frequency segments over a total range of $\sim$\,250\,GHz. 
The details on the observations are summarised in the following.

\subsection{Observations with the APEX telescope}

\subsubsection{FLASH+ \& SHeFI}
G1.3 and G1.6 were observed in various frequency ranges between 213 and 475\,GHz with the APEX telescope in order to capture higher-excitation transitions of molecules, whose lower-excitation transitions are detected in a 3\,mm survey subsequently performed with the IRAM 30\,m telescope (see Sect.\,\ref{ss:iram}). 
The observations were performed using the FLASH+ \citep[First Light at APEX Submm Heterodyne,][]{Klein14} and SHeFI receivers \citep[Swedisch Heterodyne Facility Instrument,][]{shefi08}.
Covered frequency ranges used in the subsequent analysis and corresponding receivers are summarised in Table\,\ref{tab:obs_freqs}.
FLASH+ comprises two modules that detect different frequency ranges. FLASH345 covers a total frequency range of 268\,--\,374\,GHz while FLASH460 covers 374\,--\,516\,GHz. The receiver detects a lower and an upper sideband, each with a bandwidth of 4\,GHz. 
We used only one sideband of SHeFI (APEX2), which covers a frequency range of 267\,--\,378\,GHz (Table\,\ref{tab:obs_freqs}).
The backend used during the observations with FLASH+ and SHeFI is a digital Fast-Fourier Transform Spectrometer \citep[XFFTS, which is an evolved version of the original APEX FFTS,][]{Klein12}. It is composed of four single boards, each of which has a bandwidth of 2.5\,GHz. The generic spectral resolutions provided by FLASH345, FLASH460, and APEX2 are 38, 76, and 76\,kHz, which translate to (average) velocity resolutions of 0.04, 0.05, and 0.09\,km\,s$^{-1}$, respectively.
 
The observations were carried out during seven days in 2013 June and August for both cloud complexes and G1.6 was additionally observed during another five days in 2015 June and August. The sources were mapped by observing several sub-maps using the on-the-fly (OTF) observing mode. 
The observations of G1.3 are centered at \hbox{$(l,b)=$}(1$\overset{\circ}{.}$28, 0$\overset{\circ}{.}$070), 
those of G1.6 at $(1\overset{\circ}{.}59, 0\overset{\circ}{.}015)$. 
The off-positions used during the 
OTF observations were placed at $(1\overset{\circ}{.}28, 0\overset{\circ}{.}073)$ and $(1\overset{\circ}{.}75, 0\overset{\circ}{.}600)$ for G1.3 and G1.6, respectively. Because these primary off-positions showed emission features in some strong lines, we observed secondary off-positions farther away at $(1\overset{\circ}{.}27, -0\overset{\circ}{.}248)$ for G1.3 and $(1\overset{\circ}{.}63, 0\overset{\circ}{.}083)$ for G1.6 to correct for this contamination in the primary off-position. For more details see Sect.\,\ref{ss:datared}. The pointing uncertainty was usually less than 5\arcsec.

Table\,\ref{tab:obs_freqs} lists the observed map sizes for each observed frequency range together with average noise levels in the maps. 
The data were automatically calibrated after the observations using the APEX online calibrator tool \citep{apex-cali}. We assume the standard calibration uncertainty of 10\%.

\subsubsection{PI230}
To cover additional transitions that emit in the frequency range of 217--233\,GHz we used data obtained with the PI230 receiver. These data were observed as part of a large survey covering the whole CMZ (Riquelme et al. in prep.). The achieved spectral resolution is 61\,kHz ($\sim$0.08\,km\,s$^{-1}$). 
The covered frequency range and the average noise levels of the maps are listed in Table\,\ref{tab:obs_freqs}. 

 
\subsection{Observations with the IRAM 30\,m telescope}\label{ss:iram}
We mapped G1.3 and G1.6 in the 3\,mm spectral window (85\,--\,117\,GHz) with the IRAM 30\,m telescope.
The observations were carried out during seven days in 2016 February, May, August, and November. G1.3 and G1.6 were mapped using OTF. 
Each source was covered by several sub-maps yielding a total map with a coverage of a few hundred arcseconds per side. 
Table\,\ref{tab:obs_freqs} lists the map sizes together with an average noise level for each frequency range. 
The map centres and primary off-positions are at the same positions as for the APEX observations for both sources. Moreover, the secondary off-position is the same for G1.3, while for G1.6, an observation of
a secondary off-position was not necessary because the primary off-position was verified to be free of emission in all lines (except for the $^{12}$CO and $^{13}$CO transitions). The pointing accuracy during the observations was better than 5\%.  

For the observations, we used the E090 module of the EMIR receiver. It is comprised of two (separated) sidebands, each of which operates in horizontal and vertical polarisation modes. One sideband covers a bandwidth of 8\,GHz. 
An FFTS \citep[][]{Klein12,kramer16} was used as backend. It is composed of 24 single modules working at 200\,kHz spectral resolution, which translates to 0.57\,--\,0.65\,km\,s$^{-1}$ depending on the observed frequency.
The data was calibrated by applying the standard procedure \citep{iram-cali} using the MIRA software\footnote{\href{https://www.iram.fr/IRAMFR/GILDAS/doc/html/mira-html/mira.html}{https://www.iram.fr/IRAMFR/GILDAS/doc/html/mira-html/mira.html}} provided by IRAM. The uncertainty on the calibration is assumed to be about the standard value of 10\%.


\begin{table*}[htpb]
\caption{Molecular transitions selected for analysis.}
\centering
\small
\begin{tabular}{lcrrrrr}
        \hline\hline\\[-0.3cm]
         Molecule &  Transition & $\nu_\mathrm{ul}\tablefootmark{a}$ & $E_\mathrm{u}/k\tablefootmark{b}$ & $A_\mathrm{ul}\tablefootmark{c}$ & $C_\mathrm{ul}\tablefootmark{d}$ & $n_\mathrm{crit}\tablefootmark{e}$ \\
         & & (GHz) & (K) & (s$^{-1}$) & ($10^{-11}$\,cm$^{-3}$\,s$^{-1}$) & (cm$^{-3}$) \\\hline\\[-0.3cm]
         CO & 1\,--\,0 & 115.271 & 5.5 & $7\times10^{-8}$ & 3.5 & $2\times10^3$ \\ 
         & 2\,--\,1 & 230.538 & 16.6 & $7\times10^{-7}$ & 6.3 & $1\times10^4$ \\
         & 3\,--\,2 & 345.796 & 33.2 & $2\times10^{-6}$ & 6.9 & $3\times10^4$ \\
         & 4\,--\,3 & 461.041 & 55.3 & $6\times10^{-6}$ & 7.0 & $9\times10^4$ \\
         $^{13}$CO$^*$ & 1\,--\,0 & 110.201 & 5.3 & $6\times10^{-8}$ & 3.5 & $2\times10^3$ \\
         & 2\,--\,1 & 220.299 & 15.9 & $6\times10^{-7}$ & 6.3 & $1\times10^4$ \\
         & 3\,--\,2 & 330.588 & 31.7 & $2\times10^{-6}$ & 6.9 & $3\times10^4$ \\
         HC$_3$N$^*$ & 10\,--\,9 & 90.979 & 24.0 & $6\times10^{-5}$ & 5.1 & $1\times10^6$ \\
         & 11\,--\,10 & 100.076 & 28.8 & $8\times10^{-5}$ & 5.1 & $2\times10^6$ \\
         & 12\,--\,11 & 109.174 & 34.1 & $1\times10^{-4}$ & 5.1 & $2\times10^6$ \\
         & 24\,--\,23 & 218.325 & 131.0 & $8\times10^{-4}$ & 5.2 & $2\times10^7$ \\
         p-H$_2$CO$^*$ & 3$_{0,3}$\,--\,2$_{0,2}$ & 218.222 & 21.0 & $3\times10^{-4}$ & 8.4 & $3\times10^6$ \\
         & 3$_{2,1}$\,--\,2$_{2,0}$ & 218.760 & 68.1 & $2\times10^{-4}$ & 4.9 & $3\times10^6$ \\
         & 4$_{0,4}$\,--\,3$_{0,3}$ & 290.623 & 34.9 & $7\times10^{-4}$ & 8.9 & $8\times10^6$ \\
         & 4$_{2,2}$\,--\,3$_{2,1}$ & 291.948 & 82.1 & $5\times10^{-4}$ & 6.2 & $8\times10^6$ \\
         CS$^*$ & 2\,--\,1 & 97.981 & 7.1 & $2\times10^{-5}$ & 4.4 & $5\times10^5$ \\
         & 7\,--\,6 & 342.882 & 65.8 & $8\times10^{-4}$ & 5.7 & $1\times10^7$ \\
         $^{13}$CS & 2\,--\,1 & 92.494 & 6.7 & $1\times10^{-5}$ & 4.4 & $2\times10^5$ \\
         N$_2$H$^{+\,*}$ & 1\,--\,0 & 93.173 & 4.5 & $4\times10^{-5}$ & 20.0 & $2\times10^5$ \\
         HCO$^{+\,*}$ & 1\,--\,0 & 89.189 & 4.3 & $4\times10^{-5}$ & 18.0 & $2\times10^5$ \\
         & 3\,--\,2 & 267.558 & 25.7 & $1\times10^{-3}$ & 39.0 & $4\times10^6$ \\
         & 4\,--\,3 & 356.734 & 42.8 & $4\times10^{-3}$ & 40.0 & $9\times10^6$ \\
         H$^{13}$CO$^+$ & 1\,--\,0 & 86.754 & 4.2 & $4\times10^{-5}$ & 18.0 & $1\times10^5$ \\
         HCN$^*$ & 1\,--\,0 & 88.632 & 4.3 & $2\times10^{-5}$ & 0.9 & $2\times10^6$ \\
         & 3\,--\,2 & 265.886 & 25.5 & $8\times10^{-4}$ & 1.2 & $7\times10^7$ \\
         & 4\,--\,3 & 354.505 & 42.5 & $2\times10^{-3}$ & 1.2 & $2\times10^8$ \\
         H$^{13}$CN & 1\,--\,0 & 86.340 & 4.1 & $2\times10^{-5}$ & 0.7 & $3\times10^6$ \\
         HNC$^*$ & 1\,--\,0 & 90.664 & 4.4 & $3\times10^{-5}$ & 6.5 & $4\times10^5$ \\
         SiO$^*$ & 2\,--\,1 & 86.847 & 6.3 & $3\times10^{-5}$ & 9.6 & $3\times10^5$ \\
         & 5\,--\,4 & 217.105 & 31.3 & $5\times10^{-4}$ & 11.0 & $5\times10^8$ \\
         & 7\,--\,6 & 303.927 & 58.3 & $1\times10^{-3}$ & 11.0 & $1\times10^9$ \\
         & 8\,--\,7 & 347.331 & 75.0 & $2\times10^{-3}$ & 11.0 & $2\times10^9$ \\
         & 10\,--\,9 & 434.120 & 114.6 & $4\times10^{-3}$ & 11.0 & $4\times10^9$ \\
         SO$^*$ & 2$_{3}$\,--\,1$_{2}$ & 109.252 & 21.1 & $1\times10^{-5}$ & 5.2 & $2\times10^5$ \\
         & 3$_{2}$\,--\,2$_{1}$ & 99.300 & 9.2 & $1\times10^{-5}$ & 3.8 & $3\times10^5$ \\
         & 3$_{4}$\,--\,4$_{3}$ & 267.198 & 28.7 & $7\times10^{-7}$ & 0.8 & $1\times10^3$ \\
         & 6$_{5}$\,--\,5$_{4}$ & 219.949 & 35.0 & $1\times10^{-4}$ & 5.8 & $2\times10^6$ \\
         HNCO$^*$ & 4$_{0,4}$\,--\,3$_{0,3}$ & 87.925 & 10.5 & $7\times10^{-6}$ & 6.3 & $1\times10^5$ \\
         & 5$_{0,5}$\,--\,4$_{0,4}$ & 109.906 & 15.8 & $2\times10^{-5}$ & 6.7 & $3\times10^5$ \\
         & 10$_{0,10}$\,--\,9$_{0,9}$ & 219.798 & 58.0 & $2\times10^{-4}$ & 7.6 & $2\times10^6$ \\
         OCS$^*$ & 8\,--\,7 & 97.301 & 21.0 & $3\times10^{-6}$ & 8.0 & $4\times10^4$ \\
         & 9\,--\,8 & 109.463 & 26.3 & $4\times10^{-6}$ & 8.1 & $5\times10^4$ \\
         & 18\,--\,17 & 218.903 & 100.0 & $3\times10^{-5}$ & 7.5 & $4\times10^5$ \\
         A-CH$_3$OH$^*$ & 2$_{1,2}$\,--\,1$_{1}$ & 95.914 & 21.4 & $2\times10^{-6}$ & 6.3 & $3\times10^4$ \\
         & 2$_{1,1}$\,--\,1$_{1,0}$ & 97.583 & 21.6 & $3\times10^{-6}$ & 5.5 & $5\times10^4$ \\
         & 6$_{1,5}$\,--\,5$_{1,4}$ & 292.673 & 63.7 & $1\times10^{-4}$ & 9.5 & $1\times10^6$ \\
         CH$_3$CN & 5$_{4}$\,--\,4$_{4}$ & 91.959 & 85.6 & $1\times10^{-5}$ & 18.0 & $6\times10^4$ \\ 
         & 5$_{3}$\,--\,4$_{3}$ & 91.971 & 50.8 & $4\times10^{-5}$ & 18.0 & $2\times10^5$ \\
         & 5$_{2}$\,--\,4$_{2}$ & 91.980 & 26.0 & $5\times10^{-5}$ & 19.0 & $3\times10^5$ \\
         & 5$_{1}$\,--\,4$_{1}$ & 91.985 & 11.1 & $5\times10^{-5}$ & 19.0 & $3\times10^5$ \\
         & 5$_{0}$\,--\,4$_{0}$ & 91.987 & 6.1 & $6\times10^{-5}$ & 20.0 & $3\times10^5$ \\
         & 6$_{5}$\,--\,5$_{5}$ & 110.330 & 133.3 & $2\times10^{-5}$ & 18.0 & $1\times10^5$  \\
         & 6$_{4}$\,--\,5$_{4}$ & 110.349 & 88.6 & $6\times10^{-5}$ & 18.0 & $3\times10^5$ \\
         & 6$_{3}$\,--\,5$_{3}$ & 110.364 & 53.9 & $7\times10^{-5}$ & 18.0 & $4\times10^5$ \\
         & 6$_{2}$\,--\,5$_{2}$ & 110.375 & 29.1 & $9\times10^{-5}$ & 19.0 & $5\times10^5$ \\
         & 6$_{1}$\,--\,5$_{1}$ & 110.381 & 14.2 & $9\times10^{-5}$ & 20.0 & $5\times10^5$ \\
         & 6$_{0}$\,--\,5$_{0}$ & 110.384 & 9.2 & $1\times10^{-4}$ & 20.0 & $6\times10^5$ \\
         \hline\hline
\end{tabular}
\tablefoot{\tablefoottext{a}{Rest frequency from CDMS.} \tablefoottext{b}{Upper level energy from CDMS.} \tablefoottext{c}{Einstein A coefficient from CDMS.} \tablefoottext{d}{Collisional rate coefficient at 100\,K from LAMDA database \citep[][]{Schoier05}.} \tablefoottext{e}{Critical density $n_\mathrm{crit}=A_\mathrm{ul}/C_\mathrm{ul}$ at 100\,K.}\tablefoottext{*}{Molecules for which LVG modelling was performed (see Sect.\,\ref{ss:lvg}).} }\\
\label{tab:lines}
\end{table*}

\subsection{Data reduction of the APEX and IRAM data}\label{ss:datared}
The data reduction was done with the Continuum and Line Analysis Single-dish Software (CLASS), which is part of the 
GILDAS software created and developed by IRAM\footnote{\url{https://www.iram.fr/IRAMFR/GILDAS/}}.
We resampled all spectra to a common resolution of 1\,km\,s$^{-1}$ and subtracted baselines generally of degree $n=1$. In rare cases we used higher order polynomials of $n=2$ or 3 to ensure flat baselines across the map.
The post-calibration data were converted from antenna temperature $T_A^*$ to main-beam temperature $T_\mathrm{MB}$ using $T_\mathrm{mB}=\frac{F_\mathrm{eff}}{B_\mathrm{eff}}T_A^*$,
where $F_\mathrm{eff}$ and $B_\mathrm{eff}$ are forward and beam efficiencies of the telescopes, respectively.
The forward efficiency for the IRAM 30\,m and the APEX telescopes is 0.95\footnote{\url{http://www.apex-telescope.org/telescope/efficiency/}}$^,$\footnote{\url{http://www.iram.es/IRAMES/mainWiki/Iram30mEfficiencies}}. 
The beam efficiency correction $B_\mathrm{eff}$ is computed using the Ruze formula $B_\mathrm{eff}(\nu) = B_0\cdot \mathrm{exp}\left(-\frac{4\pi c \sigma}{\nu}\right)^2$ \citep[][]{Ruze66}, where $c$ is the speed of light, $B_0=B(\nu=0)$, and $\sigma$ is the antenna surface rms value accounting for the deviations from an ideal telescope surface. 
For the IRAM 30\,m telescope, the Ruze formula was applied using values of $B_0=0.86$ and $\sigma = 66\,\mu$m, which are provided on the IRAM webpage\footnotemark[4]$^,$\footnote{The values for $B_0$ and $\sigma$ were computed for compact sources. Therefore, given that the CMZ sources are extended, the values may be higher as emission from the error beam may be picked up.}. For the APEX telescope, we used $B_0 = 0.7$ and $\sigma = 19\,\mu$m for the observations with FLASH+ and SHeFI\footnotemark[5] and a beam efficiency value of $B_\mathrm{eff} = 0.73$ for the observations performed with PI230 (internal communication at MPIfR). 

In order to correct for subtraction of off-source emission during the calibration process, we used a spline interpolation method, with which we modelled the line features of the off-position while keeping the intensity of channels without any features at zero. The method is explained in more detail in Appendix\,\ref{s:spline}.
The lowest achieved angular resolution is $\sim$30$''$ for both telescopes, which translates to a spatial resolution of $\sim1$\,pc at a distance of 8\,kpc. The highest resolution of $\sim$13$''$ ($\sim$0.5\,pc at 8\,kpc) is achieved at highest frequencies observed with the APEX telescope. To analyse and compare various spectra taken at different frequencies and observed with different telescopes, the angular resolution of all maps was smoothed with a 2D Gaussian kernel to the lowest resolution of 30$''$.

In addition, we regridded the maps such that the pixel size is the same for all maps. The default size of one grid cell is half of the original beam size. Therefore, all pixel sizes were degraded to the largest beam size of our observations, which is $\sim$14.5$''$. 

\section{Results}\label{s:results}

\begin{figure*}[htpb]
    \centering
    \includegraphics[width=\textwidth]{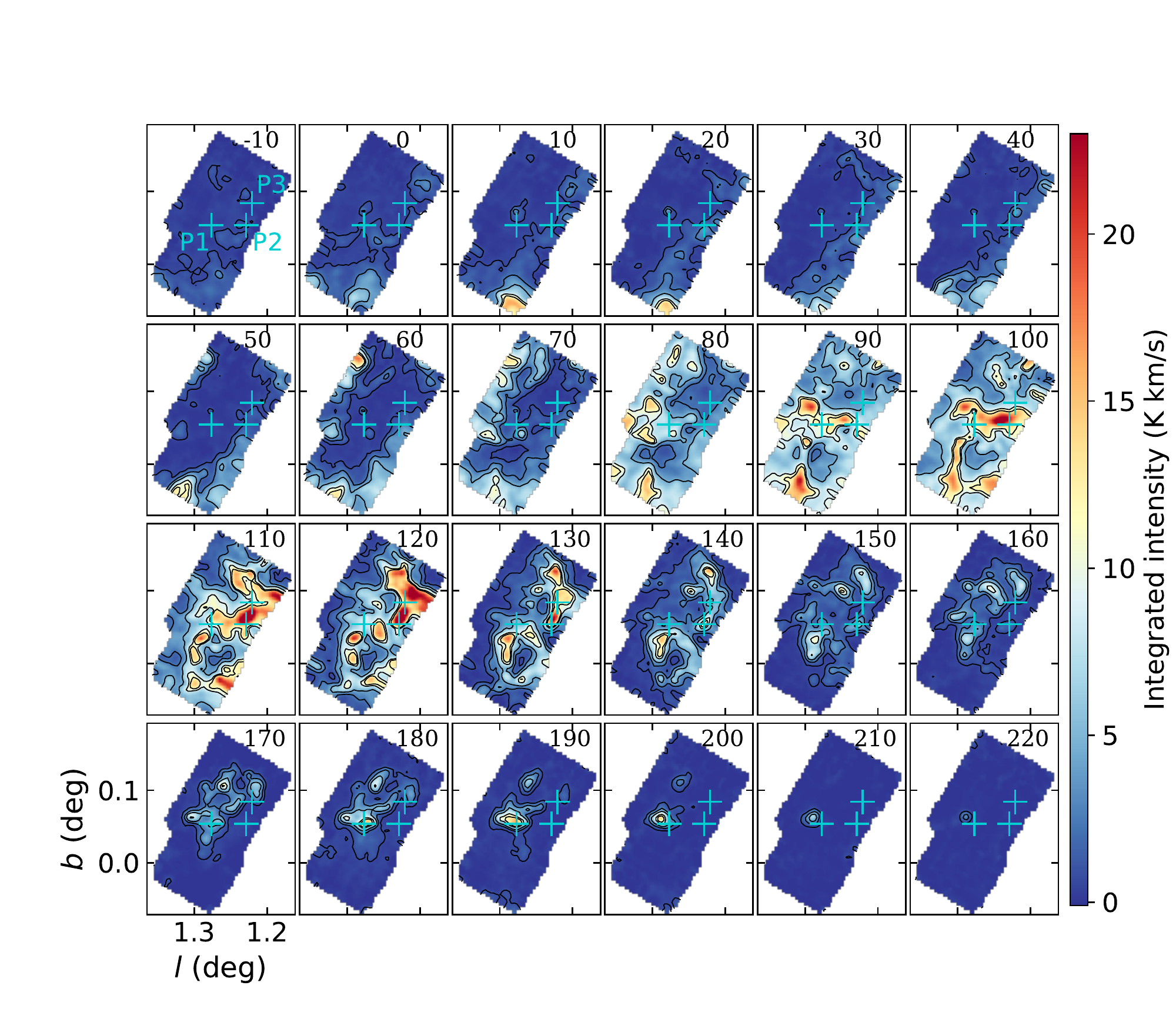}
    \caption{Velocity-channel maps of CS 2\,--\,1 emission towards G1.3. Intensities are integrated in velocity steps of 10\,km\,s$^{-1}$. The central velocity is displayed in km\,s$^{-1}$ in the upper right corner of each map. 
    The contour steps are $3\sigma$, $9\sigma$, $18\sigma$, $36\sigma$, and $45\sigma$ with $\sigma=0.25$\,K\,km\,s$^{-1}$. Blue crosses labelled P1\,--\,P3 indicate positions selected for further analysis.
    }
    \label{fig:cmaps13}
\end{figure*}
\begin{figure}
    \hspace{-1.8cm}
    \includegraphics[width=0.66\textwidth]{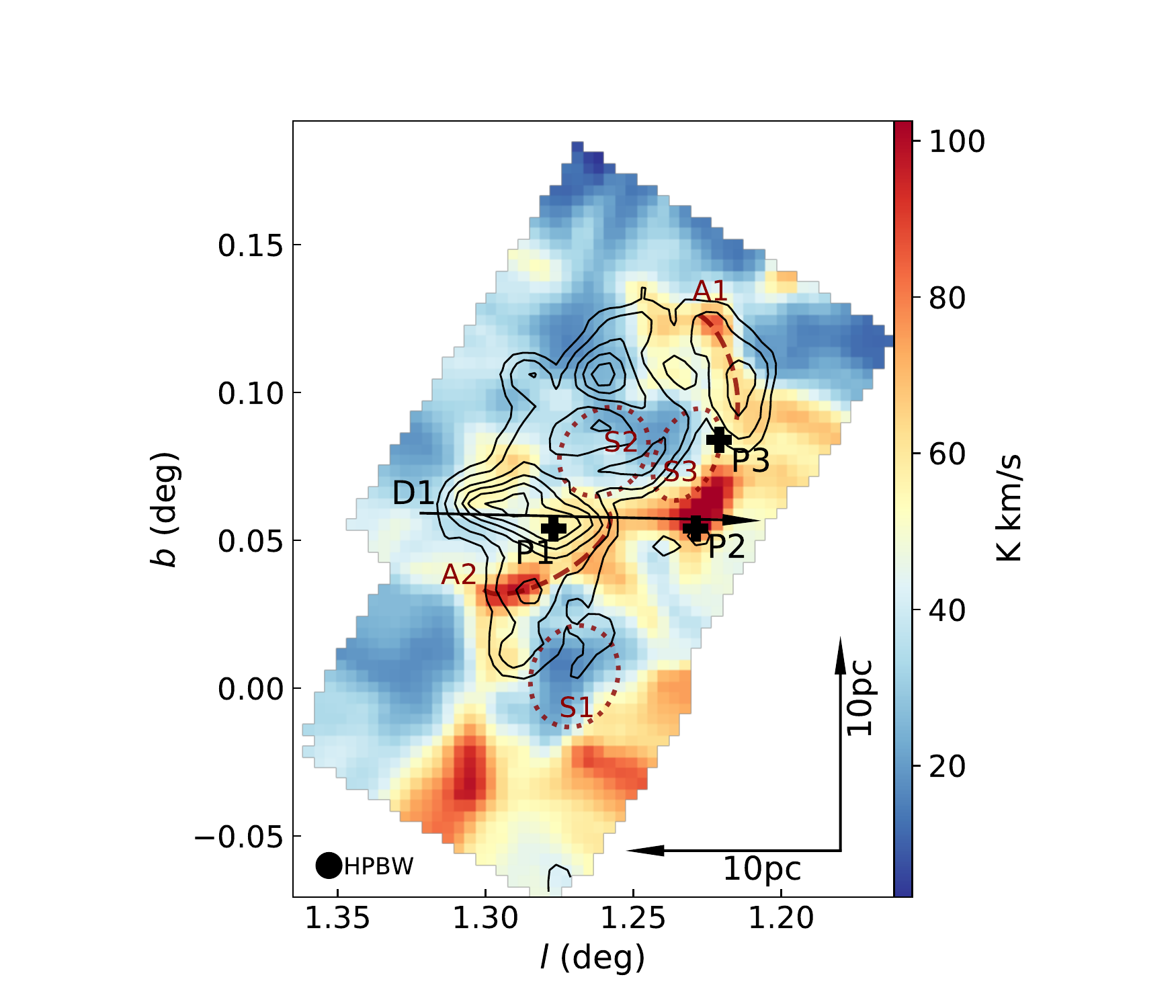}
    \caption{CS 2\,--\,1 intensities towards G1.3 integrated from 35 to 150\,km\,s$^{-1}$ (low- and intermediate-velocity components) are shown in colour-scale and from 150 to 220\,km\,s$^{-1}$ (high-velocity component) are shown with black contours. Contour levels are $6\sigma$, $12\sigma$, and then increase by $12\sigma$ with $\sigma=1.0$\,K\,km\,s$^{-1}$. The HPBW is shown in the bottom left corner. The black arrow labelled D1 indicates the position axis along which the PV diagram in Fig.\,\ref{fig:13slice}a was taken. Black crosses labelled P1\,--\,P3 indicate positions selected for further analysis. Dark red ellipses labelled S1--S3 correspond to shells A, C, and C1, respectively, identified by \citet{Tanaka07}. Arcs A1 and A2 indicate regions of seemingly spatial coincidences of the low- and high-velocity gas (see text).}
    \label{fig:13x1x2}
\end{figure}
\begin{figure}
    \includegraphics[width=0.5\textwidth]{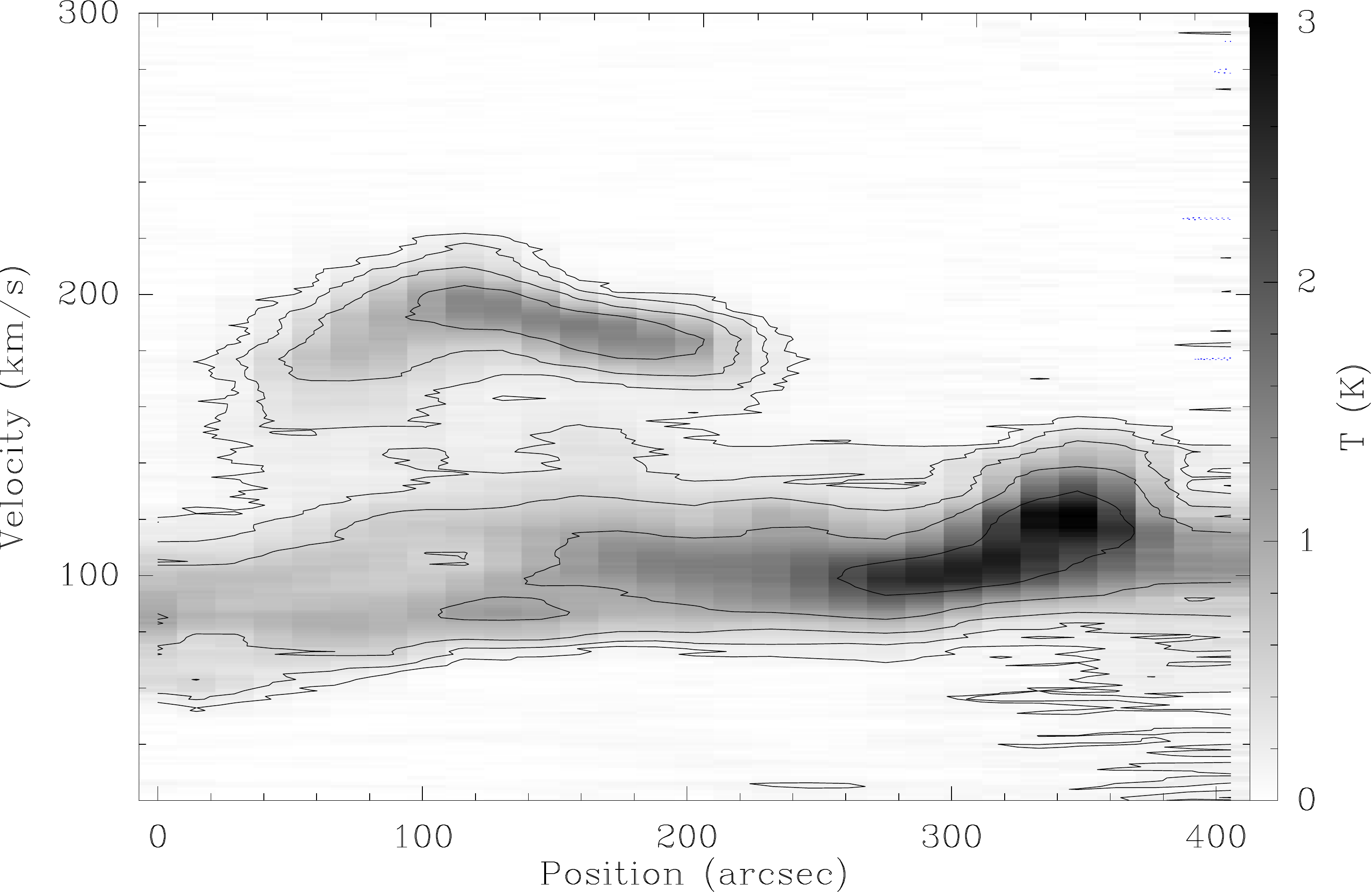}
    \begin{tikzpicture}[remember picture,overlay]
        \draw (1.3,12.52) node {a)};
    \end{tikzpicture}  
    \includegraphics[width=0.51\textwidth]{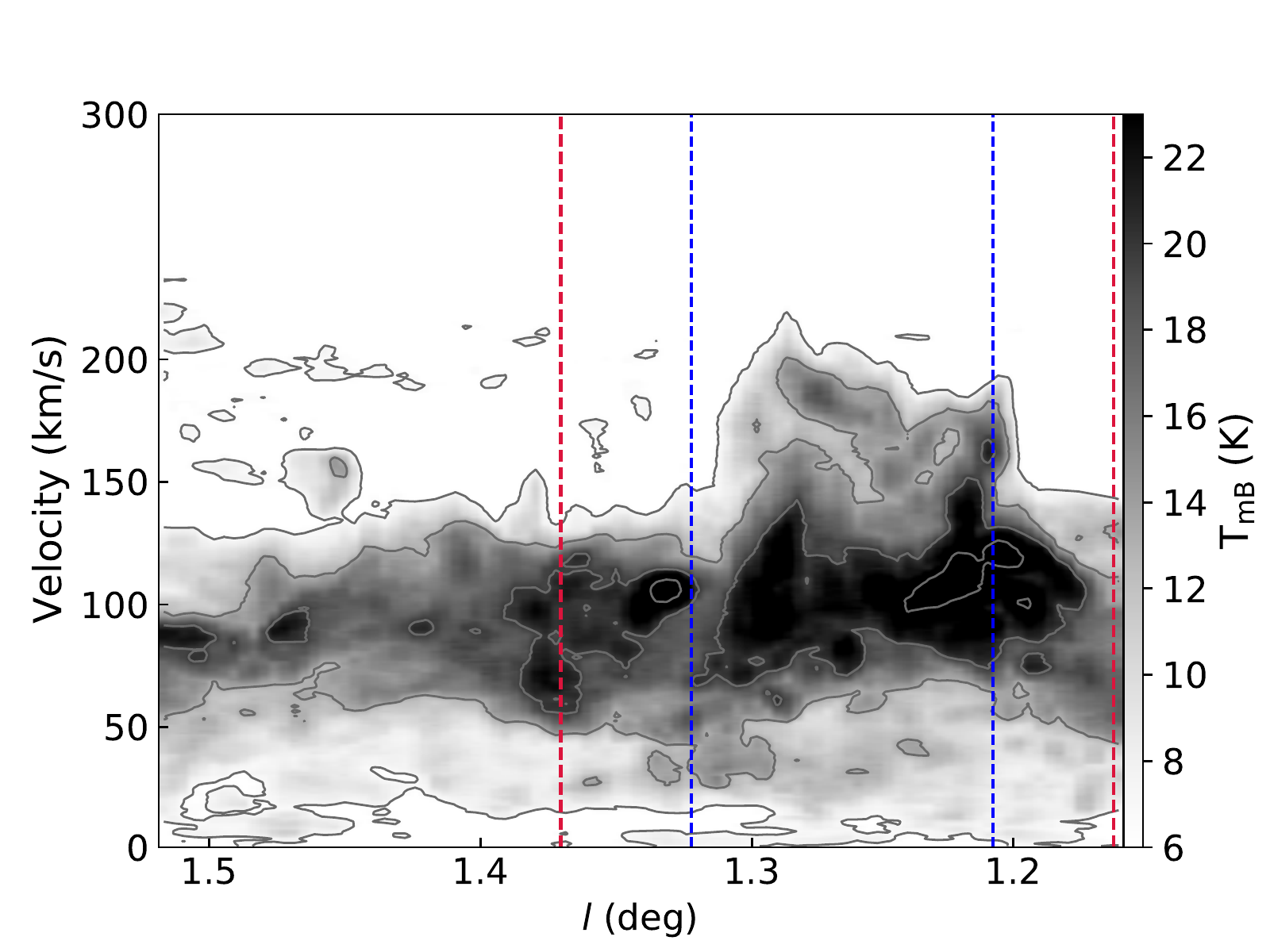}
    \begin{tikzpicture}[remember picture,overlay]
        \draw [dashed, blue] (5.1,13.3) -- (1.05,14.7);
        \draw [dashed, blue] (7.4,13.3) -- (8.15,14.7);
        \draw (1.55,12.75) node {b)};
    \end{tikzpicture}  
    \includegraphics[width=0.51\textwidth]{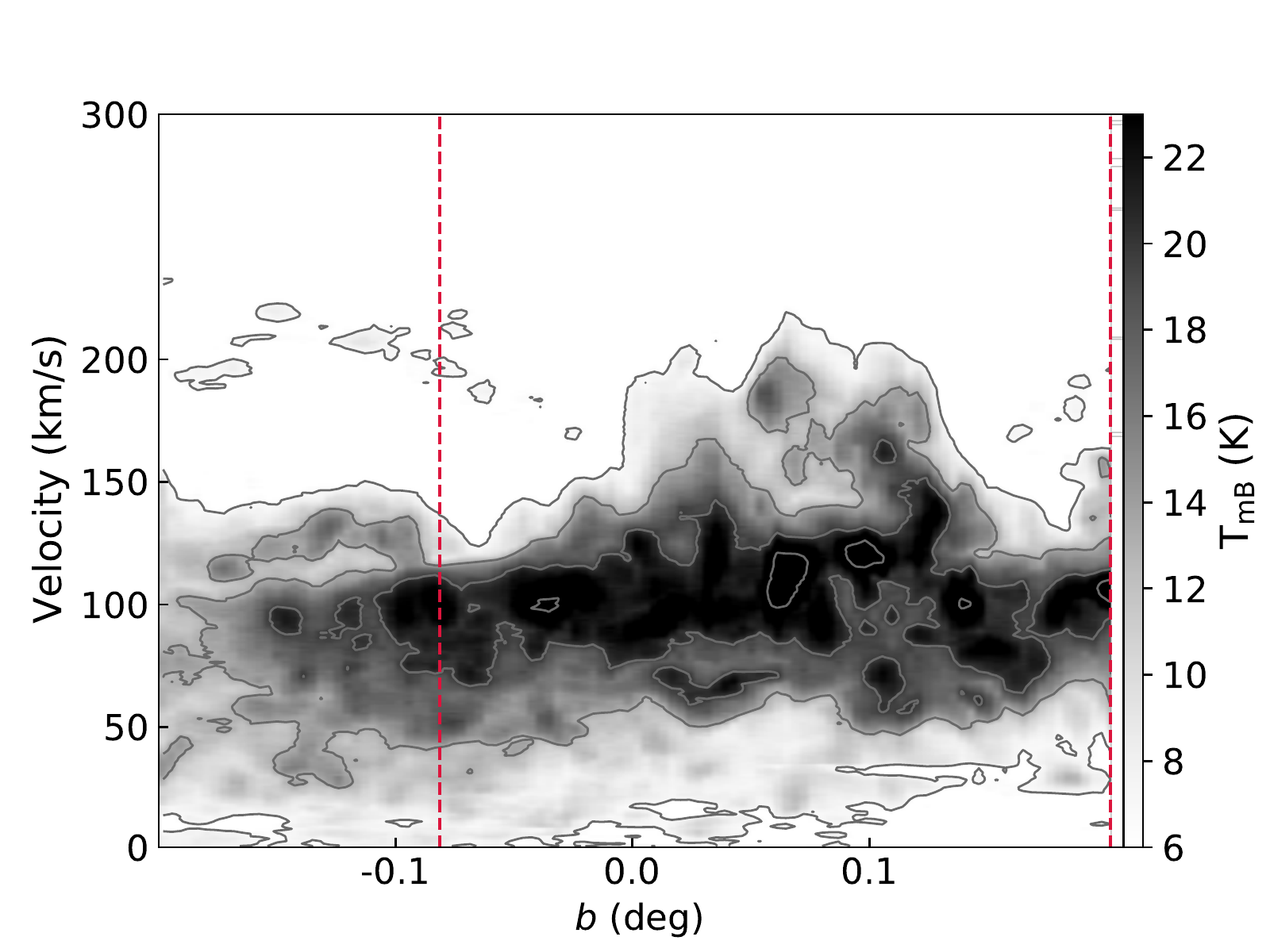}
    \begin{tikzpicture}[remember picture,overlay]
        \draw (1.55,6.15) node {c)};
    \end{tikzpicture}  
    \caption{\textit{(a)} PV diagram of CS 2\,--\,1 emission towards G1.3 along the black arrow labelled D1 in Fig.\,\ref{fig:13x1x2}. The position axis originates from the start of the arrow and is completely covered by the length of it. The contour levels are -4$\sigma$, 4$\sigma$, 8$\sigma$, 16$\sigma$, 32$\sigma$, and 64$\sigma$ with $\sigma=0.03$\,K. \textit{(b)} Longitude-velocity diagram of CO 2\,--\,1 data observed towards G1.3, where we show the maximum intensity along the latitude axis per pixel. The contour steps start at 20$\sigma$ and then increase by the same value, with $\sigma=0.33$\,K. Blue dashed lines indicate the position axis covered in (a). \textit{(c)}  Same as (b), but it shows latitude versus velocity and the maximum intensity along the longitude axis per pixel. Dashed red lines in (b) and (c) indicate the complete longitude and latitude ranges observed in this work with the IRAM\,30\,m telescope, respectively. }
    \label{fig:13slice}
\end{figure}

\begin{figure*}
    \centering
    \includegraphics[width=\textwidth]{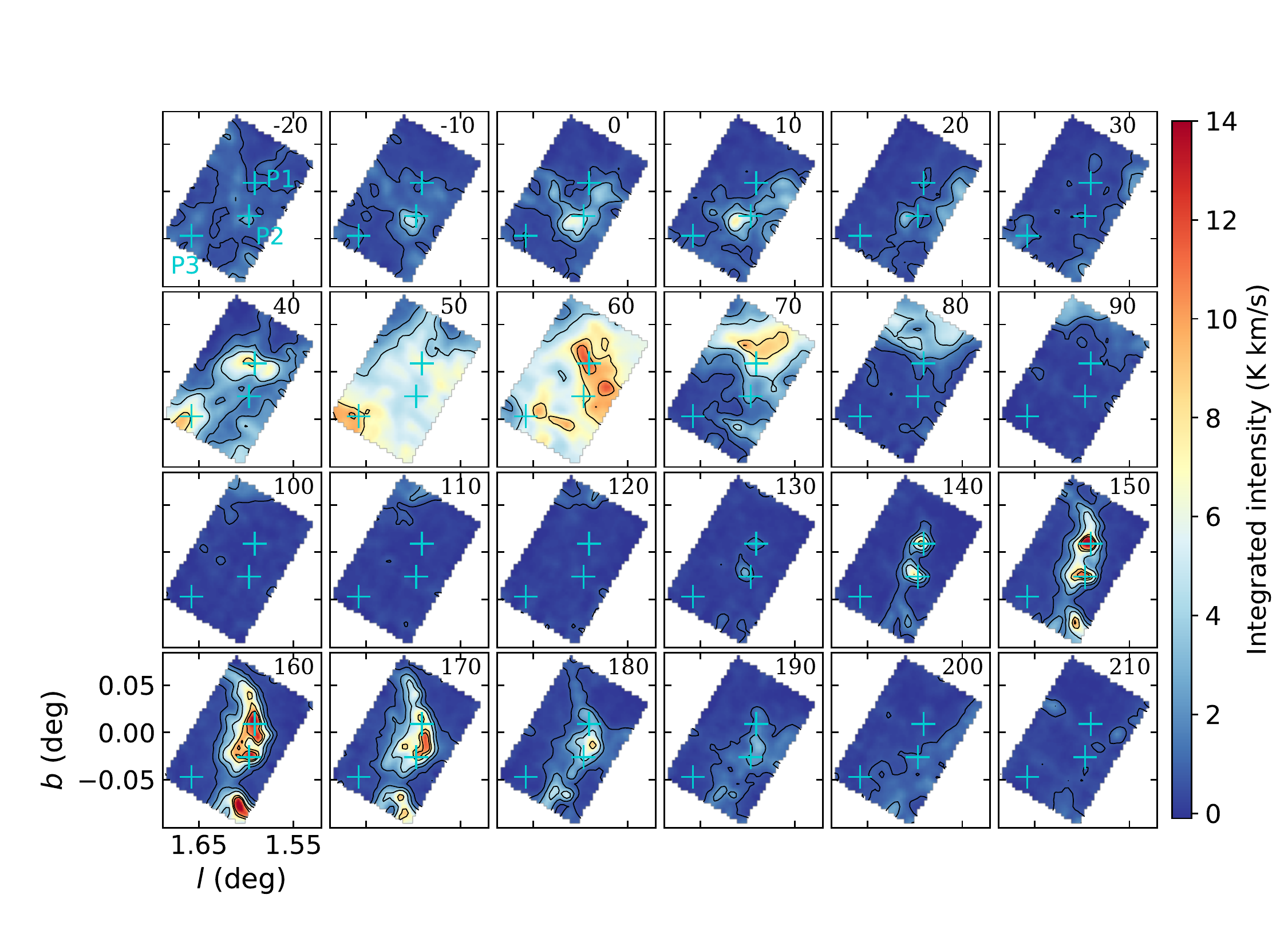}
    \caption{Velocity-channel maps of CS 2\,--\,1 emission towards G1.6. Intensities are integrated in velocity steps of 10\,km\,s$^{-1}$. The central velocity is displayed in km\,s$^{-1}$ in the upper right corner in each map. The contour steps are $3\sigma$, $9\sigma$, $18\sigma$, $36\sigma$, and $45\sigma$ with $\sigma=0.22$\,K\,km\,s$^{-1}$. Blue crosses labelled P1\,--\,P3 indicate positions selected for further analysis.
    }
    \label{fig:cmaps16}
\end{figure*}
\begin{figure}[htpb]
    \hspace{-0.5cm}
    \includegraphics[width=0.55\textwidth]{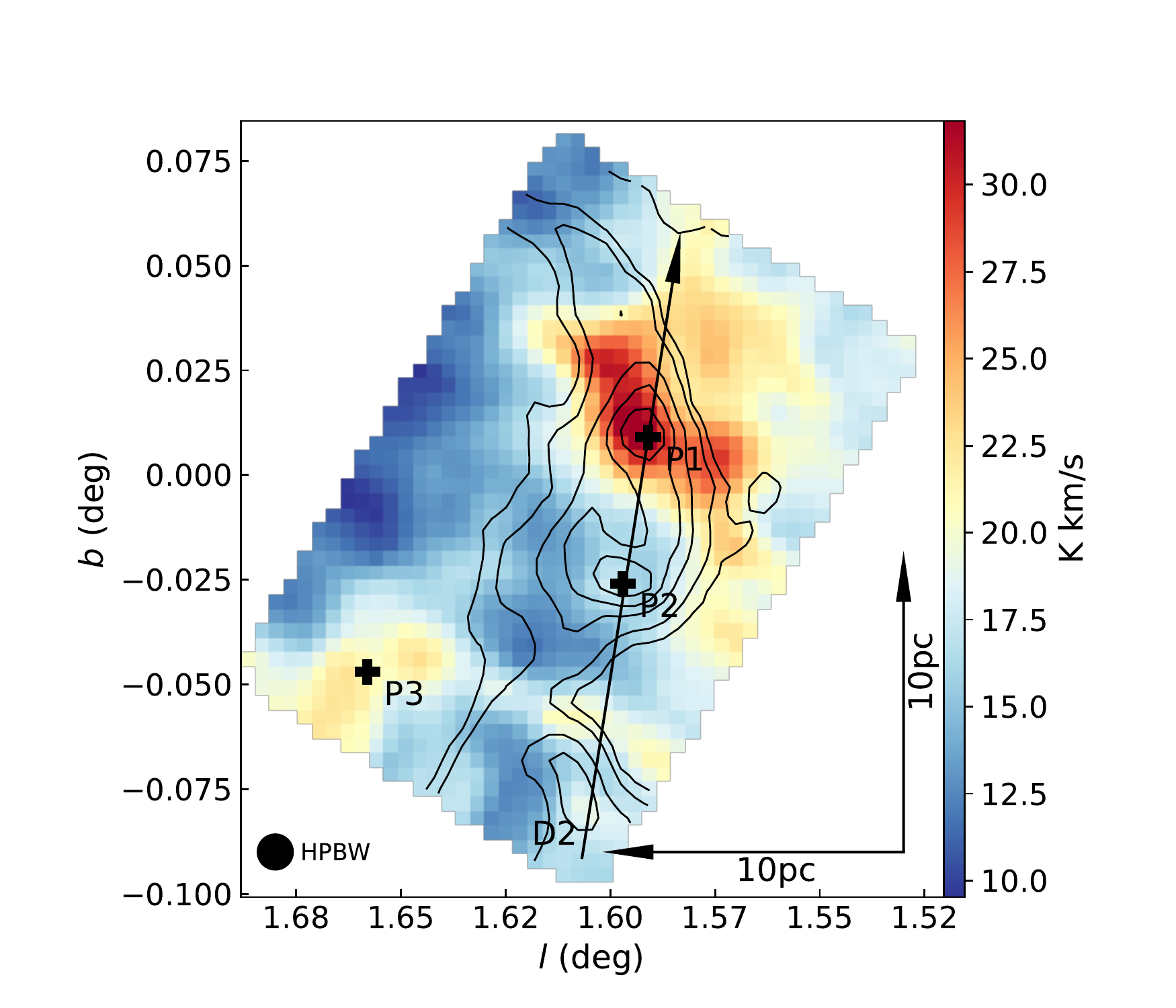}
    \caption{CS 2\,--\,1 intensities towards G1.6 integrated from 25 to 112\,km\,s$^{-1}$ (low-velocity component) towards G1.6 are shown in colour-scale and from 112 to 240\,km\,s$^{-1}$ (high-velocity component) are shown with black contours. Contour levels are $8\sigma$, $12\sigma$, and then increase by $12\sigma$ with $\sigma=1.0$\,K\,km\,s$^{-1}$. The HPBW is shown in the bottom left corner. The black arrow labelled D2 indicates the position axis along which the PV diagram in Fig.\,\ref{fig:16slice} was taken. Black crosses labelled P1\,--\,P3 indicate positions selected for further analysis.}
    \label{fig:16x1x2}
\end{figure}
\begin{figure}
    \includegraphics[width=0.5\textwidth]{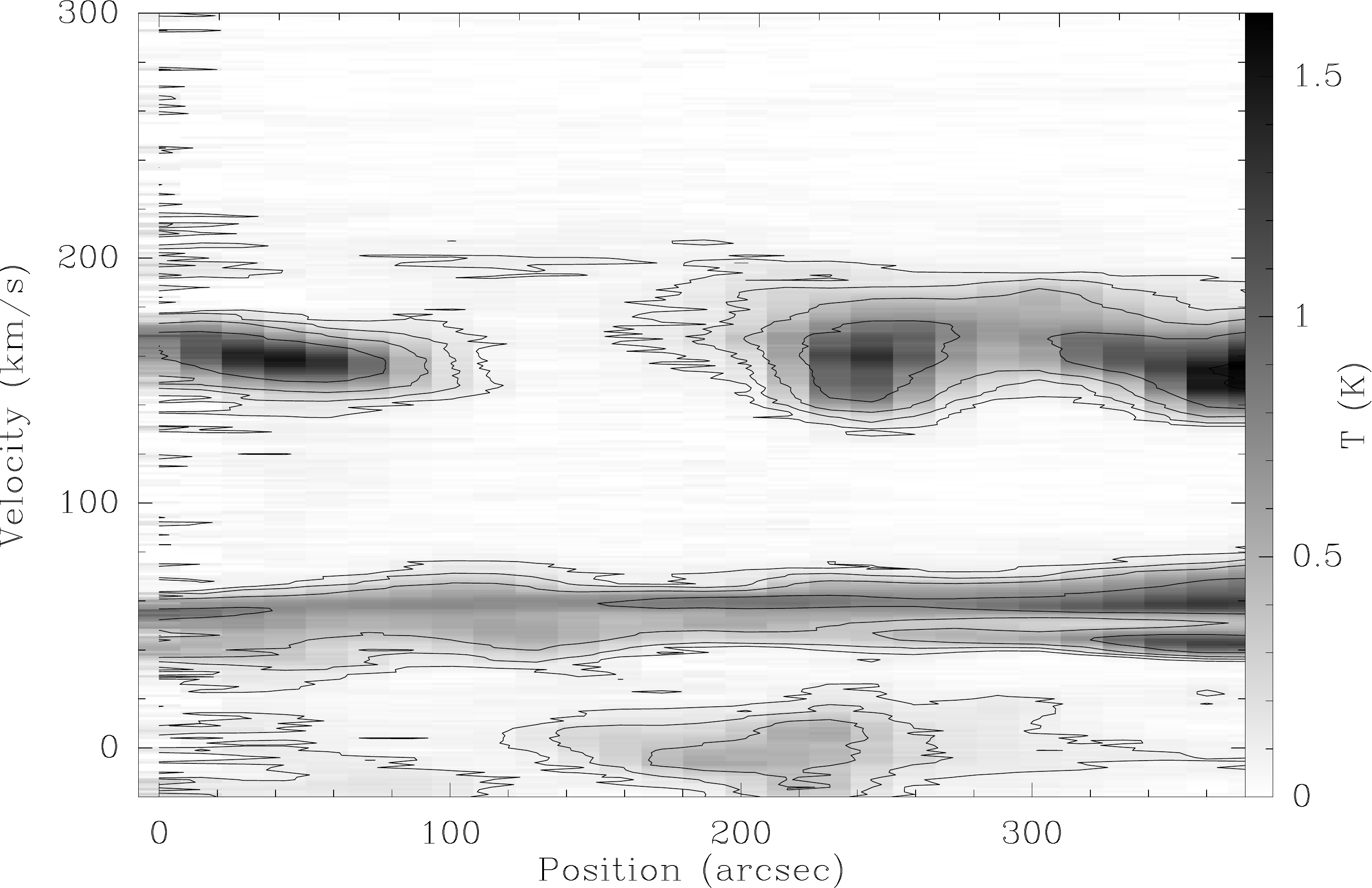}
    \includegraphics[width=0.51\textwidth]{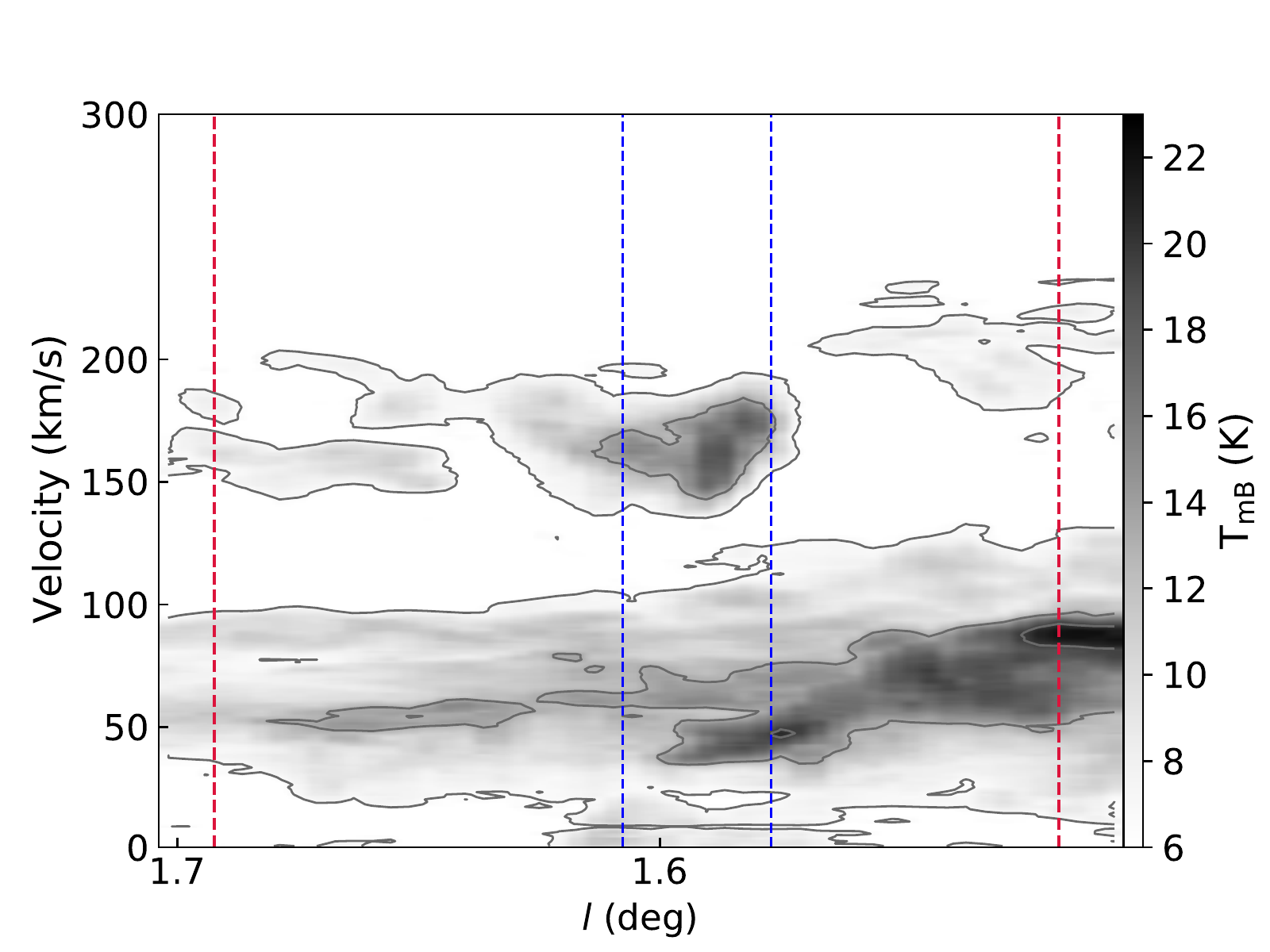}
    \includegraphics[width=0.51\textwidth]{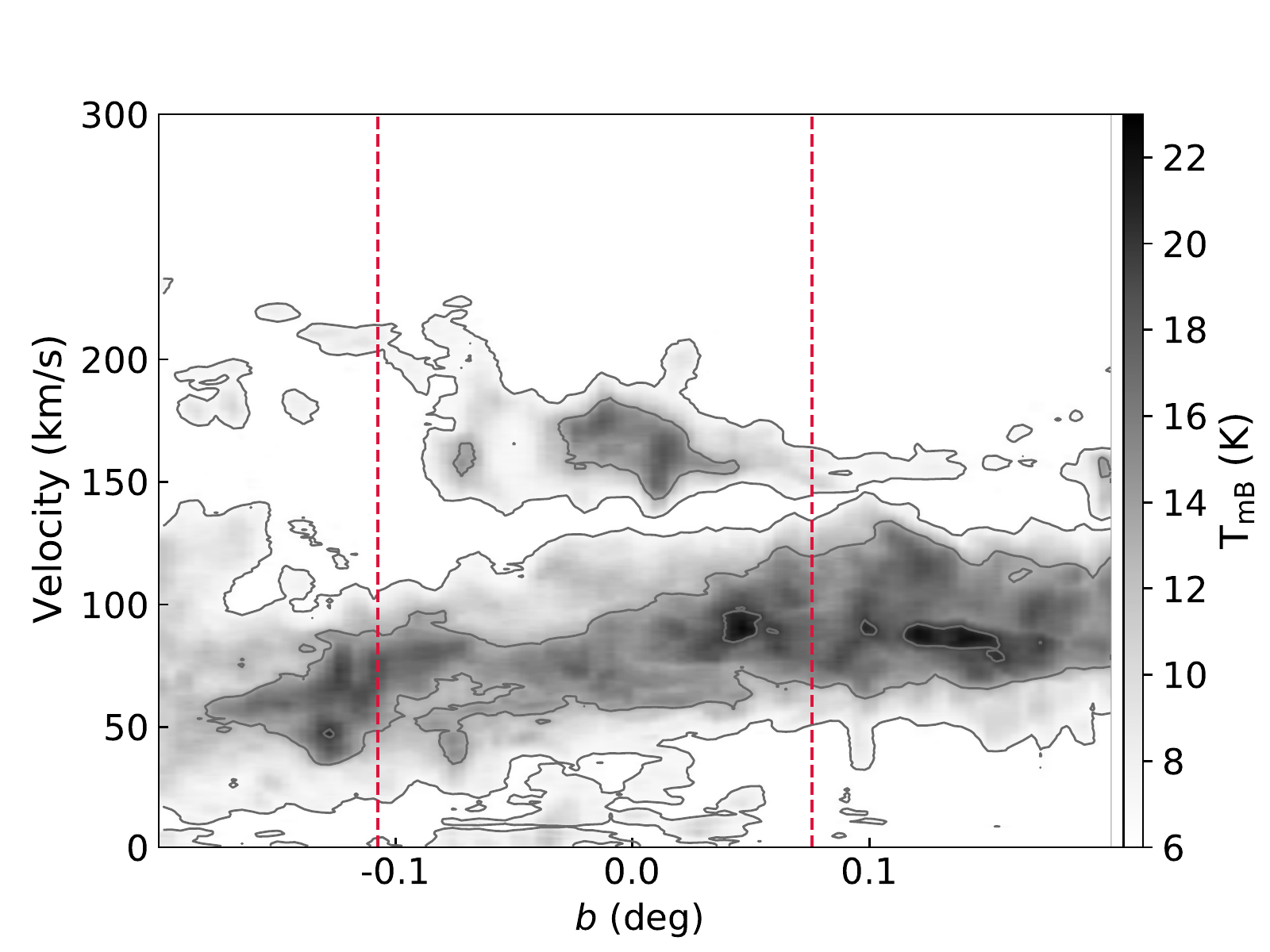}
    \begin{tikzpicture}[remember picture,overlay]
        \draw [dashed, blue] (4.55,13.7) -- (1.0,15.1);
        \draw [dashed, blue] (5.75,13.7) -- (8.3,15.1);
        \draw (1.5,13.2) node {b)};
        \draw (1.65,20.) node {a)};
        \draw (1.5,6.2) node {c)};
    \end{tikzpicture}  
    \caption{\textit{(a)} PV diagram of CS 2\,--\,1 emission towards G1.6 along the black arrow labelled D2 in Fig.\,\ref{fig:16x1x2}. The position axis originates from the start of the arrow and is completely covered by the length of it. The contour levels are -4$\sigma$, 4$\sigma$, 8$\sigma$, 16$\sigma$, 32$\sigma$, and 64$\sigma$ with $\sigma=0.02$\,K. \textit{(b)} Longitude-velocity diagram of CO 2\,--\,1 data observed towards G1.6, where we show the maximum intensity along the latitude axis per pixel. The contour steps start at 20$\sigma$ and then increase by the same value, with $\sigma=0.33$\,K. Blue dashed lines indicate the positions covered in (a). \textit{(c)}  Same as (b), but it shows latitude versus velocity and the maximum intensity along the longitude axis per pixel. Dashed red lines in (b) and (c) indicate the complete longitude and latitude ranges observed in this work with the IRAM\,30\,m telescope, respectively.}
    \label{fig:16slice}
\end{figure}
\begin{figure*}[h!]
    \includegraphics[width=0.35\textwidth]{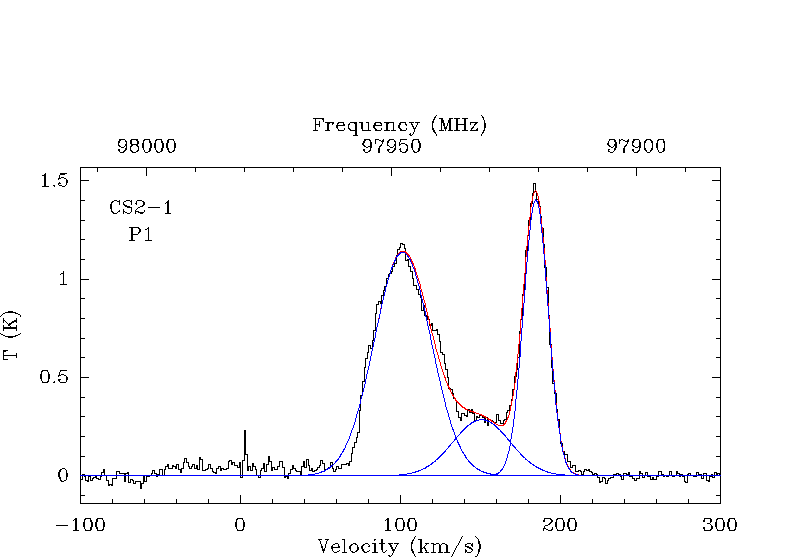}\hspace{-0.4cm}
    \includegraphics[width=0.35\textwidth]{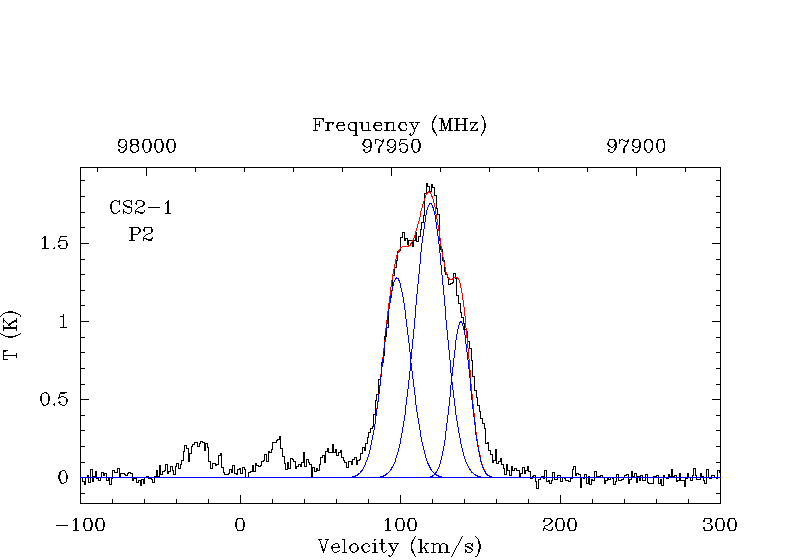}\hspace{-0.4cm}
    \includegraphics[width=0.35\textwidth]{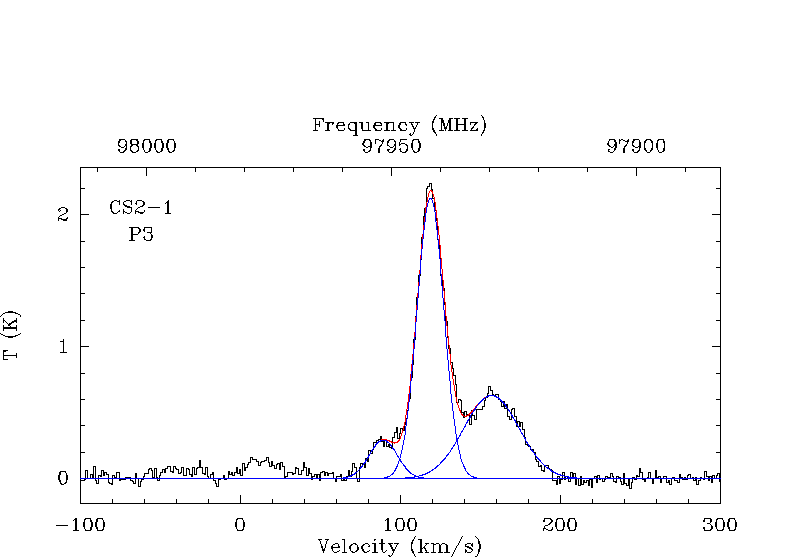}\hspace{-0.4cm}
    \caption{CS 2\,--\,1 spectra at positions P1\,--\,P3 in G1.3 together with Gaussian profile fitting results. The single components are shown in blue, the sum of all components in red. The intensity is given in main-beam temperatures.}
    \label{fig:CS13}
\end{figure*}
\begin{figure*}[h!]
    \includegraphics[width=0.35\textwidth]{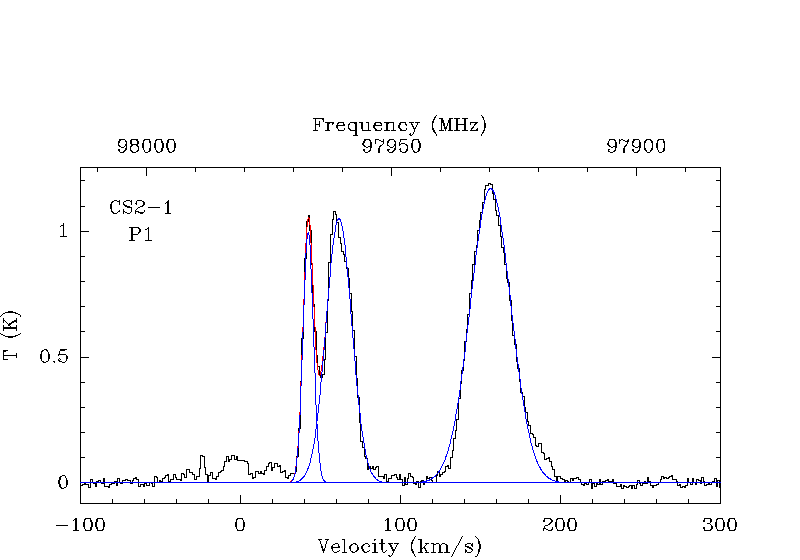}\hspace{-0.4cm}
    \includegraphics[width=0.35\textwidth]{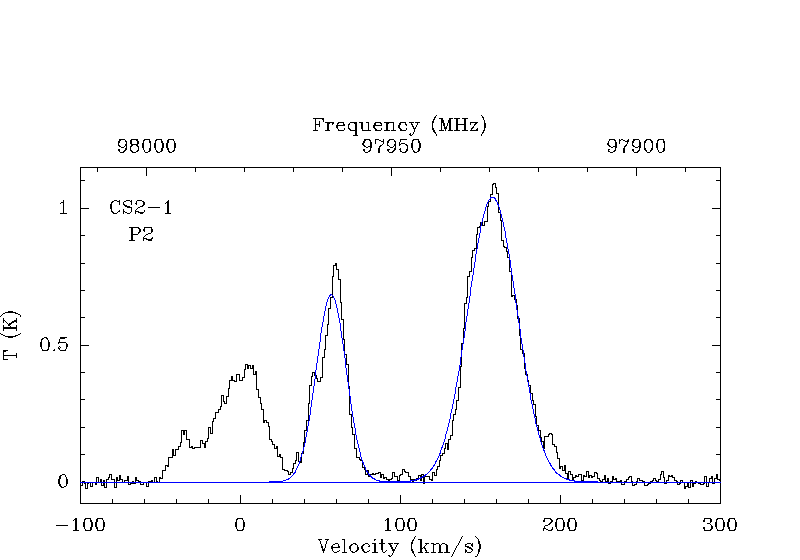}\hspace{-0.4cm}
    \includegraphics[width=0.35\textwidth]{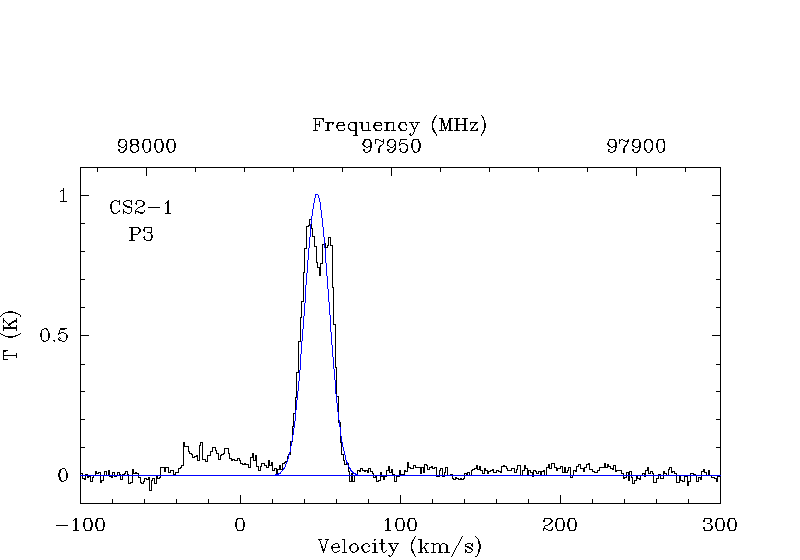}\hspace{-0.4cm}
    \caption{Same as Fig.\,\ref{fig:CS13}, but for positions in G1.6}
    \label{fig:CS16}
\end{figure*}

\subsection{Line selection}\label{ss:lines}
A set of molecules covered by the line survey 
was selected that either trace high densities, temperatures, or shock chemistry. Table\,\ref{tab:lines} lists the selected molecular transitions along with their spectroscopic properties that we take from the CDMS \citep[Cologne Database for Molecular Spectroscopy,][]{CDMS} or the JPL catalogues \citep[Jet Propulsion Laboratory,][]{Pickett98}.
Shock tracing molecules are usually identified by their very existence, that is their detection in shocked and their absence (or low intensity) in quiescent gas. This is because they are known to reside on the surfaces of dust grains in cold and quiescent environments, or in the case of SiO, the Si is part of the grains. When a shock wave passes, molecules that reside on the dust grain surfaces can be released into the gas phase due to desorption  processes or, in the case of SiO, by grain destruction (`sputtering') and hence, be observed \citep[see, e.g. ][]{Miura17}.
We observe silicon monoxide (SiO), isocyanic acid (HNCO), sulfur monoxide (SO), carbonyl sulfide (OCS), and methanol (CH$_{3}$OH) \citep[e.g.][]{Flower2012,kelly17,Guzman18}, which are proposed to trace shock-like events.


Because of its asymmetric structure, methanol can undergo numerous transitions many of them detectable under the physical conditions found in the CMZ \citep[see e.g.][\citeyear{Leurini16}]{Leurini04}, with a relatively large number of them covered and detected in our survey.  Only three transitions of CH$_3$OH's A-type symmetry species are considered for the following analysis. They are selected because they are not blended with other lines and do
not show strong deviations from LTE, like maser action or enhanced absorption.

Typical density tracers are carbon monosulfide (CS), diazenylium (N$_2$H$^{+}$), formylium (HCO$^{+}$), hydrogen cyanide (HCN), hydrogen isocyanide (HNC), para-formaldehyde (p-H$_{2}$CO), cyanoacetylene (HC$_{3}$N), and HNCO. We list critical densities of molecular transitions at gas kinetic temperatures of 100\,K in Table\,\ref{tab:lines} as a reference, which we computed using $n_\mathrm{crit}=\frac{A_\mathrm{ul}}{C_\mathrm{ul}}$, where $A_\mathrm{ul}$ is the Einstein A coefficient of the transition and $C_\mathrm{ul}$ is the collisional rate coefficient. However, for optically thick molecular emission, the effective critical density will be
significantly reduced compared to the values quoted in Table\,\ref{tab:lines} \citep[see also][]{Shirley15}.

To derive fractional abundances of molecules, H$_2$ column densities have to be derived. Because H$_2$ has no permanent electric dipole moment it can basically not be observed \citep[only hardly from space at high temperatures,][]{rodriguez01,Mills17}. 
CO is generally used as a proxy for H$_2$. 
Because the optical depth of CO lines is generally extremely high, we use this molecule's 
optically thin(ner) isotopologue $^{13}$CO.

The symmetric top rotor methyl cyanide, CH$_{3}$CN, was used to derive kinetic temperatures and H$_2$ volume densities of the gas. For each value of its main angular momentum quantum number $J$, CH$_{3}$CN has several transitions with $0 < K \leq J-1$, where $K$ is the projection of $J$ on the C$_3$ symmetry axis. These span a wide range of upper level energies and, being close in frequency, can be observed simultaneously \citep{Boucher80}. We cover transitions with $J=5_K-4_K$ and $6_K-5_K$. 

\subsection{Morphology}\label{ss:morphology}
\subsubsection{G1.3}
Figure\,\ref{fig:cmaps13} shows velocity-channel maps of CS 2\,--\,1 emission in the velocity range from $-$10 to 220\,km\,s$^{-1}$. They reveal diffuse and rather weak emission at velocities $<$\,80\,km\,s$^{-1}$. At higher velocities, the morphology changes and the emission becomes more intense with the intensity peaking at velocities 90\,--\,120\,km\,s$^{-1}$. The emission is clumpy in the sense that there are several intensity peaks surrounded by more diffuse and extended emission across the map in this velocity range. Because these intensity peaks appear at various velocities, the emission in G1.3 appears to be `fluffy' (puffed-up) in the position-position-velocity (PPV) cube. 
The most prominent emission  seems to follow an elongated structure that goes from south\footnote{Analogous to the usage in the equatorial coordinate system, we use north (south) for higher (lower) Galacic latitute and east (west) for higher (lower) longitude.}
to north-west, where it 
wraps around the western part of an emission cavity in the south (see map at 100\,km\,s$^{-1}$ at $\sim(1\overset{\circ}{.}27,0\overset{\circ}{.}0)$) and the eastern part of another cavity further north (see map at 100\,--\,110\,km\,s$^{-1}$ at $\sim(1\overset{\circ}{.}23,0\overset{\circ}{.}07)$). The emission shells around these `holes' of emission were reported by \citet{Oka2001} and \citet{Tanaka07}. 
At velocities $\gtrsim$150\,km\,s$^{-1}$, the emission decreases again and emerges from a more compact structure with intensity peaks at different locations in the map than found for the lower-velocity gas.

Figure\,\ref{fig:13x1x2} shows CS 2\,--\,1 integrated intensity (moment zero, M$_0$) maps. Intensities were integrated over the low- and intermediate-velocity components from 35 to 150\,km\,s$^{-1}$, which are shown in colour-scale, and over the high-velocity component from 150 to 220\,km\,s$^{-1}$,  which are shown with black contours. Emission at velocities below 35\,km\,s$^{-1}$ can most probably be associated with local gas residing along the line of sight \citep[e.g.][]{Riquelme10}. The M$_0$ maps reveal the different morphologies of the low- and high-velocity clouds with shells primarily observed in the lower-velocity gas. We show shells S1--S3, which correspond to shells A, C, and C1 in \citet[][]{Tanaka07}. The remaining shells are not highlighted, however, may also be identified in the morphology in Fig.\,\ref{fig:13x1x2}.
The authors also observed a shell B, which is not covered in the field of view of our observations. 
Based on the overlay of lower- and higher-velocity gas, we identify spatial coincidences indicating that the gas components might be connected. The western edge of the compact high-velocity emission follows the elongated structure at one side (A1).
In addition, the highest intensity region at high velocities appears to be partially embedded in the elongated structure along A2.

Figure\,\ref{fig:13slice}a shows a position-velocity (PV) diagram of CS\,2\,--\,1 emission, which was taken along arrow D1 indicated in Fig.\,\ref{fig:13x1x2}. The arrow covers exactly the range of positions in the PV diagram.  
We clearly identify two strong velocity components, which are connected through an `emission bridge' at intermediate velocities. To ensure that what is observed here indeed is an emission bridge and not another cloud moving at 150\,km\,s$^{-1}$ that just appears between the low- and high-velocity component in projection, we investigated how the observed morphology is embedded in the larger-scale gas distribution. Using the  CO 2\,--\,1 data cubes whose emission covers the regions of G1.3 and G1.6 on larger scales,  
we created longitude-velocity and latitude-velocity diagrams of maximum intensity per pixel along the respective spatial axis. They are shown in Figs.\,\ref{fig:13slice}b and c, respectively. Based on these, it becomes clear that the intense intermediate-velocity component in G1.3 coincides solely with high-velocity gas at 180\,--\,200\,km\,s$^{-1}$, which supports our presumption of an emission bridge. Such a feature is indicative of a scenario involving cloud-cloud interaction \citep[e.g.][]{Haworth15} that is discussed in more detail in Sect.\,\ref{ss:CCC}.

\subsubsection{G1.6}
The velocity-channel maps in Fig.\,\ref{fig:cmaps16} show the distribution of CS 2\,--\,1 emission in the velocity range from $-$20 to 210\,km\,s$^{-1}$ in G1.6. At negative velocities, extended weak emission can probably be assigned to local gas. There is also extended emission between 0 and 20\,km\,s$^{-1}$, which could be associated with G1.6, because its detection across three maps suggests a large line width of this component. 
\citet{Salii02} observed this component in shocked methanol emission, which led them to associate it with the cloud complex, however, since this velocity range could also show some contribution of local gas we do not include it in the further analysis. 
Highest intensities at low velocities are observed in the maps of 40\,--\,80\,km\,s$^{-1}$. The emission is clumpy with intensity peaks randomly distributed in this velocity range meaning that we cannot identify a clear connection of emissions in neighbouring maps. This may suggest that we observe several smaller clouds with various systemic velocities and dispersions.
Channel maps between 90 and 130\,km\,s$^{-1}$ lack emission almost completely. At 140\,km\,s$^{-1}$ the high-velocity gas component starts to emit, peaks at 150\,--\,160\,km\,s$^{-1}$, and remains visible up to 190\,km\,s$^{-1}$. The morphology shows an elongated structure extending from north to south \citep[e.g.][]{Gardner87,Salii02,Menten09}. The emission is again clumpy with several intensity peaks, however, the morphology suggests that the emission in each map can be associated with one cloud. 
Overall, the emission in the PPV cube is more distinct and less fluffy than what is observed in G1.3.

Figure\,\ref{fig:16x1x2} shows M$_0$ maps of CS 2\,--\,1 emission in G1.6. We integrated intensities over the low-velocity component from 25 to 112\,km\,s$^{-1}$, which is shown in colour-scale and over the high-velocity component from 112 to 240\,km\,s$^{-1}$ shown with black contours.
The most conspicuous spatial coincidence presents the almost exact congruence of the highest intensity peaks of both velocity components at $\sim$(1$\overset{\circ}{.}$59, 0$\overset{\circ}{.}$01). However, the other peak at high velocities has no counterpart at low velocities. 

Figure\,\ref{fig:16slice}a shows a PV diagram of CS 2\,--\,1 emission taken along the arrow labelled D2 in Fig.\,\ref{fig:16x1x2}. 
Between $\sim$40\,--\,60\,km\,s$^{-1}$, emission extends along the whole position axis. The low- and high-velocity components do not seem to be connected in the intermediate-velocity range as it is the case in G1.3, suggesting a spatial separation of the two. In the PV diagram the component at $\sim$0\,km\,s$^{-1}$ appears as broad emission feature, which might be connected to the low-velocity component if it was spatially associated with the cloud complex. 
The comparison with the CO\,2\,--\,1 data shown in Figs.\,\ref{fig:16slice}b and c also suggest that there is no connection between the low- and high-velocity gas on larger scales, at least not at a detection level of 20$\sigma$ with $\sigma=0.33$\,K as is the case for G1.3.

\subsubsection{Position \& component selection}\label{s:positions}
We selected three positions in each cloud complex to determine physical and chemical properties. In G1.3 positions P1 and P2 show CS 2\,--\,1 intensity peaks at high and intermediate velocities (see channel maps at 180 and 120\,km\,s$^{-1}$ in Fig.\,\ref{fig:cmaps13}, respectively, and Fig.\,\ref{fig:13x1x2}) and P3 has moderate intensity (see channel map 140\,km\,s$^{-1}$ and Fig.\,\ref{fig:13x1x2}). In G1.6 positions P1 and P2 correspond to intensity peaks at high velocities (see channel map 160\,km\,s$^{-1}$ in Fig.\,\ref{fig:cmaps16} and Fig.\,\ref{fig:16x1x2}) while the intensity at P3 is dominant at lower velocities (see channel map 50\,km\,s$^{-1}$). 
All selected positions are listed in Table\,\ref{tab:positions}. 
We averaged the intensities of nine neighbouring spectra by using the \texttt{average} command in CLASS, where the position of the blue cross in the figures indicates the centre. 
\begin{table}[]
    \caption{Galactic coordinates of selected positions in G1.3 and G1.6.}
    \centering
    \begin{tabular}{lcrr}
     \hline\hline
         Source & Position & $l$ [$^\circ$] & $b$ [$^\circ$]  \\
         \hline\\[-.3cm]
         G1.3 & P1 & 1.277 & 0.054 \\
         & P2 & 1.229 & 0.054 \\
         & P3 & 1.221 & 0.084 \\
         G1.6 & P1 & 1.591 & 0.009 \\
         & P2 & 1.597 & $-$0.026 \\
         & P3 & 1.658 & $-$0.047 \\
    \hline\hline 
    \end{tabular}
    \label{tab:positions}
\end{table}

In Figs.\,\ref{fig:CS13} and \ref{fig:CS16} we show the CS 2\,--\,1 spectra for positions P1\,--\,P3 in G1.3 and G1.6, respectively. CS 2\,--\,1 emission is strong, however, does not severely suffer from opacity. 
Each position (except for G1.6 P3) reveals multiple spectral lines corresponding to multiple gas components at different velocities. Because we want to compare the physical and chemical properties of different gas components, we fitted one-dimensional (1D) Gaussian profiles to velocity components that were subsequently used for modelling. In some cases, the different components are difficult to disentangle. 
Therefore, we fitted as few Gaussian profiles as possible to a (possibly blended) spectral line and only considered emission lines in the velocity ranges of $\sim$70\,--\,200\,km\,s$^{-1}$ for G1.3 and 40\,--\,200\,km\,s$^{-1}$ for G1.6. The number of transitions to fit was selected by visual inspection. The line fitting was performed using the \texttt{lines} and \texttt{minimize} commands in CLASS, with which all lines in the spectra at one positions can be fitted at once. The fit results are shown in Figs.\,\ref{fig:CS13} and \ref{fig:CS16} for G1.3 and G1.6, respectively.
Deviations from Gaussian profiles may occur as a consequence of higher optical depth or because of actually blended velocity components or lines. We number the components with increasing central velocity (c1, c2, c3). The list of all components is shown in Table\,\ref{tab:cs-gauss}.

A spectral line is regarded as detection, when the signal-to-noise $(S/N)$ ratio, which is determined by computing the ratio of the modelled peak temperature to the rms value $\sigma$ of the spectrum, is larger than 3. 
When this is not the case, for features for
which weak emission is detected, we smoothed the spectra to a lower spectral resolution to increase the $S/N$ ratio.

\begin{table}[]
\caption{Results of the 1D Gaussian fitting for CS\, 2--1 spectra.}
\centering
\begin{tabular}[t]{rrrr}
    \hline\hline\\[-0.3cm]
    Component & $\varv_\mathrm{lsr}$\tablefootmark{a} & $\Delta\varv$\tablefootmark{b} & $T_{\rm peak}$\tablefootmark{c} \\
    & (km\,s$^{-1}$) & (km\,s$^{-1}$) & K \\
    \hline\\[-.3cm]
     G1.3 P1/c1 & 101.6 & 41.8 & 1.14 \\
     P1/c2 & 151.0 & 42.0 & 0.29 \\
     P1/c3 & 184.9 & 18.1 & 1.40 \\
     P2/c1 & 98.0 & 10.0 & 1.28 \\
     P2/c2 & 119.0 & 22.0 & 1.76 \\
     P2/c3 & 138.0 & 14.0 & 1.00 \\
     P3/c1 & 89.7 & 18.7 & 0.30 \\
     P3/c2 & 119.9 & 21.6 & 1.95 \\
     P3/c3 & 158.7 & 36.6 & 0.66 \\
     G1.6 P1/c1 & 42.6 & 7.6 & 0.99 \\
     P1/c2 & 61.8 & 18.9 & 1.05 \\
     P1/c3 & 156.6 & 29.8 & 1.17 \\
     P2/c1 & 57.0 & 22.0 & 0.68 \\
     P2/c2 & 157.8 & 36.1 & 1.04 \\
     P3/c1 & 48.0 & 18.0 & 0.99 \\
    \hline\hline 
\end{tabular}
\tablefoot{\tablefoottext{a}{Source velocity at rest frequency.}\tablefoottext{b}{Full width at half maximum.}\tablefoottext{c}{Peak intensity.}}
\label{tab:cs-gauss}
\end{table}

\subsection{Non-LTE modelling with RADEX}\label{ss:lvg}

The first step was to derive kinetic temperatures from CH$_3$CN. For each velocity component that we identified and fitted in the CH$_3$CN spectra, we performed non-LTE modelling using the radiative transfer code RADEX \citep[][]{vanderTak07}. We assumed the geometrical approach of an expanding spherical shell \citep[or Large Velocity Gradient (LVG) approximation,][]{Sobolev60}. Collisional rate coefficients were taken from the LAMDA data base \citep{Schoier05}. We did not find significant differences in the derived temperature values between the components ($\sim$60--100\,K, see Table\,\ref{tab:ch3cn-radex}) and, therefore, used a fixed value of 75\,K as input for further modelling to derive column densities of the molecules that are marked with an asterisk in Table\,\ref{tab:lines}. The details on the temperature derivation can be found in App.\,\ref{app:ch3cn}. 

With the temperature derived from CH$_3$CN as a fixed parameter, we run RADEX for a $40\times34$ grid of H$_2$ number densities and column densities and compare observed and modelled peak intensities. 
If data for three transitions of a molecule are available, we can apply the reduced $\chi^2$ method described in Sect.\,\ref{app:ch3cn}. 
If data are only available for one or two transitions, this method cannot be applied because of an insufficient number of degrees of freedom. Nonetheless, for two transitions, a solution can  still be found at the intersection of the lines.

H$_2$ volume densities are constrained by the range derived from CH$_3$CN (see Table\,\ref{tab:ch3cn-radex}). If this is not possible, because either only lower limits for the number density could be  derived or the CH$_3$CN emission could  not be fitted for this component, we determined the volume density using HCN or HCO$^+$ data as these two species trace similar density ranges as CH$_3$CN. 
If neither of these three species provides a volume density range, we fixed it by using the minimum and maximum value obtained from CH$_3$CN in the respective cloud complex. In most cases, we can then extract column densities of the molecules. In rare cases, the intersection of lines lies completely outside the restricted H$_2$ volume density range. In these cases, we read off column densities from this intersection disregarding the volume density.

For all components, the comparison of modelled and observed peak intensities for $^{13}$CO shows the solution at significantly lower H$_2$ volume densities of $\sim$10$^3$\,cm$^{-3}$. Given that we only use the lowest $J\,-$\,rotational transitions of the molecule, this is not surprising as these can be detected in the lower-density gas of the regions. However, only observations of higher $J\,-$\,rotational transitions may confirm the presence of multiple density layers along one line of sight. 

Also, the density solutions for HNCO generally lie at lower values than the densities constrained by CH$_3$CN, HCN, or HCO$^+$, and are similar to the values attained from $^{13}$CO. Moreover, because the contours of modelled and observed peak intensities for the two detected transitions ($J=4-3$ and $5-4$) overlap for most parts, a constraint on the column density is ambiguous. Therefore, we only give lower limits on the column density for this molecule.


\begin{table}
    \centering
    \caption{H$_2$ column densities derived from $^{13}$CO column densities.} 
    \begin{tabular}{rcccc}
         \hline\hline\\[-0.3cm]
         Component & $N(^{13}$CO$)$\tablefootmark{a} & $N($H$_2)\tablefootmark{b}$ & $N_D($H$_2)\tablefootmark{c}$ \\[0.1cm]
         & ($10^{16}$\,cm$^{-2}$) & ($10^{21}$\,cm$^{-2}$) & ($10^{21}$\,cm$^{-2}$) \\[0.1cm]
         \hline \\[-0.3cm]
         G1.3: P1/c1 & 6.3$_{-2.3}^{+18.8}$ & 25.3$_{-9.2}^{+75.4}$ & 25 \\[0.06cm]
               P1/c2 & 1.4$_{-0.4}^{+1.1}$ & 5.6$_{-1.6}^{+4.4}$ & 5 \\[0.06cm]
               P1/c3 & 1.6$_{-0.3}^{+0.9}$ & 6.4$_{-1.2}^{+3.6}$ & 10 \\[0.06cm]
               P2/c1 & 4.0$_{-1.5}^{+11.8}$ & 16.0$_{-6.0}^{+47.3}$ & \multirow{2}{*}{28} \\[0.06cm]
               P2/c2 & 1.6$_{-0.6}^{+0.6}$ & 6.4$_{-2.4}^{+2.4}$ & \\[0.06cm]
               P2/c3 & 1.6$_{-0.6}^{+0.6}$ & 6.4$_{-2.4}^{+2.4}$ & 12\\[0.06cm]
               P3/c1 & 1.0$_{-0.2}^{+2.5}$ & 4.0$_{-0.8}^{+10.0}$ & 4 \\[0.06cm]
               P3/c2 & 2.2$_{-0.8}^{+0.3}$ & 8.8$_{-3.2}^{+1.2}$ & 10 \\[0.06cm]
               P3/c3 & 2.5$_{-0.9}^{+1.0}$ & 10.0$_{-3.6}^{+1.6}$ & 8 \\[0.06cm]
         G1.6: P1/c1 & 0.6$_{-0.3}^{+0.4}$ & 2.4$_{-1.2}^{+1.6}$ & \multirow{2}{*}{44}\\[0.06cm]
               P1/c2 & $\geq$1.0(2.5) & $\geq$6.0(10.0) & \\[0.06cm]
               P1/c3 & 4.0$_{-1.7}^{+2.3}$ & 16.0$_{-6.8}^{+9.2}$ & 72 \\[0.06cm]
               P2/c1 & $\geq$0.9(2.0) & $\geq$4.4(8.0) & 16 \\[0.06cm]
               P2/c2 & 3.2$_{-1.6}^{+2.5}$ & 12.8$_{-6.4}^{+10.0}$ & 33 \\[0.06cm]
               P3/c1 & $\geq$1.4(2.2) & $\geq$3.2(8.8) & 47 \\[0.06cm]
               P3/c2 & 0.2$_{-0.1}^{+0.2}$ & 0.8$_{-0.4}^{+0.8}$ & 20 \\[0.06cm]
        \hline\hline
    \end{tabular}
    \label{tab:13co}
    \tablefoot{\tablefoottext{a}{$N(^{13}\mathrm{CO})$ derived from RADEX modelling.}\\\tablefoottext{b}{$N(\mathrm{H}_2)=N(^{13}\mathrm{CO})\times24\times1.67\times10^4$\,cm$^{-2}$.}\tablefoottext{c}{$N_D(\mathrm{H}_2)$ derived from dust surface density maps \citep{Guzman15}, where the surface density was divided by $2.8\,m_{\rm H}$ to obtain $N_D(\mathrm{H}_2)$.}\\
    Whenever only lower limits can be derived, the best-fit result is still shown in parentheses.  }
\end{table} 

To model HCN\,1--0, HCO$^+$\,1--0, CS\,2--1, and HNC\,1--0, we used their $^{13}$C isotopic substitutions when the main isotopologues showed signs of high optical depth or self-absorption. The $^{12}$C/$^{13}$C isotopic ratio is one of the best-studied in the ISM including the GC. It decreases from 80\,--\,90 in the circumsolar medium to 20\,--\,25 in the GC \citep{Wannier80,Wilson99,Humire20}. \citet{Langer90} reported a value of 24 for Sgr\,B2 and \citet{Riquelme10iso} derived values of 22\,--\,30 in the low-velocity component of G1.3. Therefore, we expected these values for the intensity ratios of lines of $^{12}$C- to $^{13}$C-bearing species with the same quantum numbers, if the emission in the main isotopologue lines were optically thin. 

Therefore, to derive the column density of the main isotopologue, we modelled the spectra of the $^{13}$C isotopic substitution in RADEX and multiply the column density values used in the grid for the $^{13}$C isotopolgue by 24 in order to mimic results of the same transition of the main isotopologue. These results together with higher $J\,-$\,transitions of the main isotopologue are then used to derive column densities.
Collisional rate coefficients of HN$^{13}$C are not provided in the LAMDA data base \citep{Schoier05}. Therefore, we were not able to model the $^{13}$C isotopologue and only used the line from the main isotopologue HNC. 


N$_2$H$^+$\,1--0, H$^{13}$CN\,1--0 (also HCN\,1--0 but only the $^{13}$C isotopologue was used during the analysis), HNCO\, 4--3 and 5--4, and H$^{13}$CO$^+$\,1--0 show hyperfine structure (HFS). For the latter two molecules, given the large observed line widths, the spacing between the transitions is negligible and does not affect our analysis. For the former two, we do not resolve the HFS, however, we notice a widening of the line width. Assuming that the emission is optically thin, we made use of the observed integrated intensity of the line at the observed line width and, based on this, estimated the peak intensity of the line if HFS was not present using a typical line width in this component that was observed for other molecules. This peak intensity is then compared to the modelled ones, where in the model HFS is not taken into account.

\subsection{H$_2$ column densities}
We use $^{13}$CO to derive H$_2$ column densities because CO emission is extremely optically thick. 
The column densities of $^{13}$CO determined with RADEX are listed in Table\,\ref{tab:13co}. 
$^{13}$CO column densities are in a range of $(1-6)\times 10^{16}$\,cm$^{-2}$ for G1.3 and $(0.2-4) \times 10^{16}$\,cm$^{-2}$ for G1.6. 
Assuming optically thin emission, the $^{13}$CO column densities are first converted to CO densities by multiplying with an $^{12}$C/$^{13}$C isotopic ratio of 24 (see Sect.\,\ref{ss:lvg}), and, subsequently, to H$_2$ column densities by multiplying with an H$_2$/CO abundance ratio of $1.67 \times 10^4$ \citep{Roueff21}. 
This H$_2$/CO abundance ratio presents an update of the commonly used value of $10^4$ in both the Galactic plane and centre \citep[e.g.][as references for the GC]{Huettemeister98,rodriguez01}. This value is proposed to be stable against enhanced cosmic ray fluxes \citep{Farquhar94,Huettemeister98}, which are shown to exist in the GC. H$_2$ column densities are found to be in the range of $(4-25) \times 10^{21}$\,cm$^{-2}$ and $(0.8-16) \times 10^{21}$\,cm$^{-2}$ in G1.3 and G1.6, respectively. 

\begin{figure*}[h!]
    \centering
    \includegraphics[width=\textwidth]{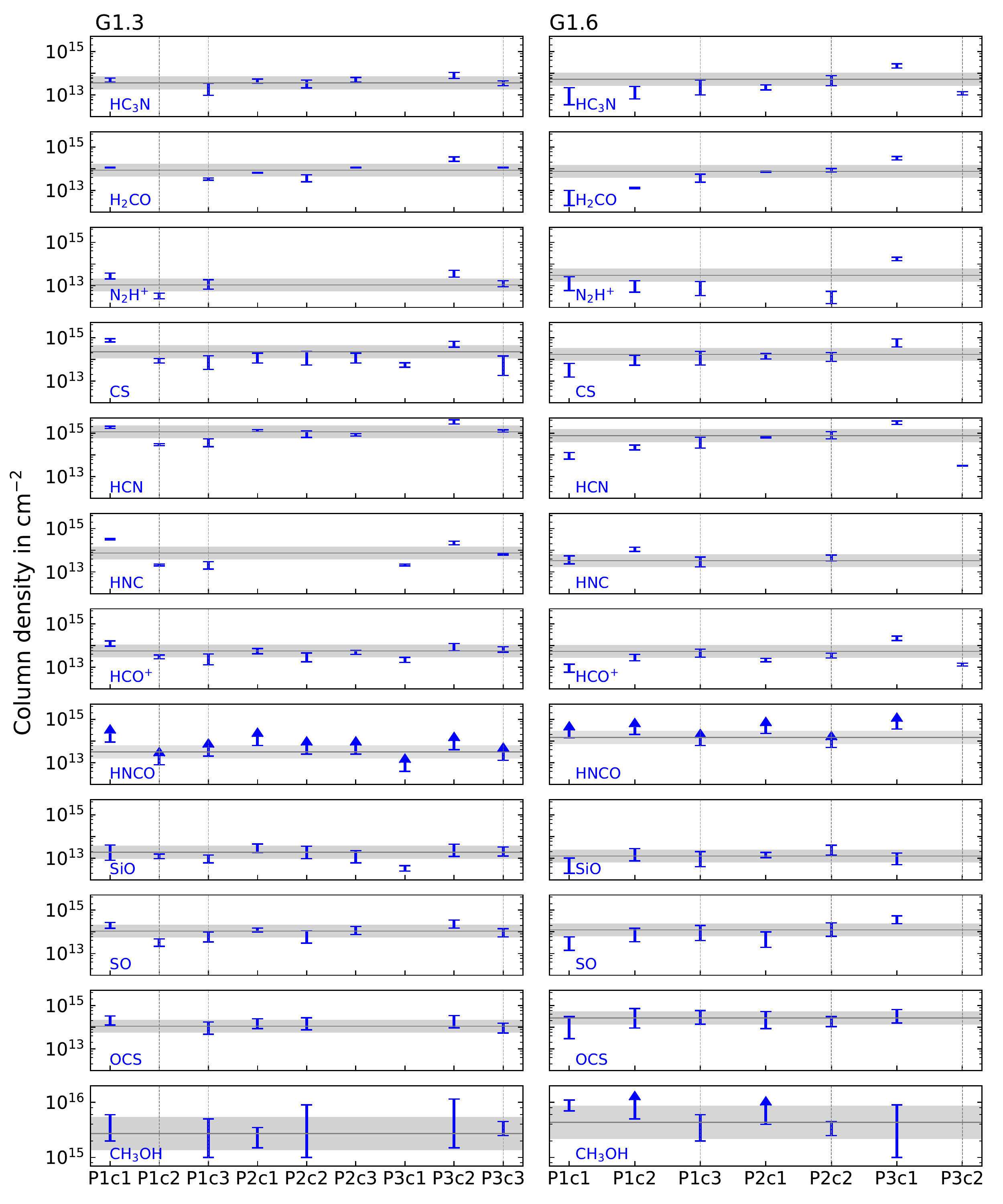}
    \caption{\textit{Left column:} Column densities in all components of G1.3 obtained from non-LTE modelling with RADEX. The horizontal line shows the average column density in the cloud complex, the shaded area indicates a deviation from the average by a factor 2. Vertical dashed lines indicate the high-velocity components. Arrows indicate lower limits. If no marker is shown, either the molecule is not detected in this component or the molecule has not been modelled for various reasons (see Table\,\ref{tab:Ncol} and text). \textit{Right column:} Same as on the left, but for G1.6.}
    \label{fig:Ncols}
\end{figure*}
\begin{figure*}[h!]
    \centering
    \includegraphics[width=\textwidth]{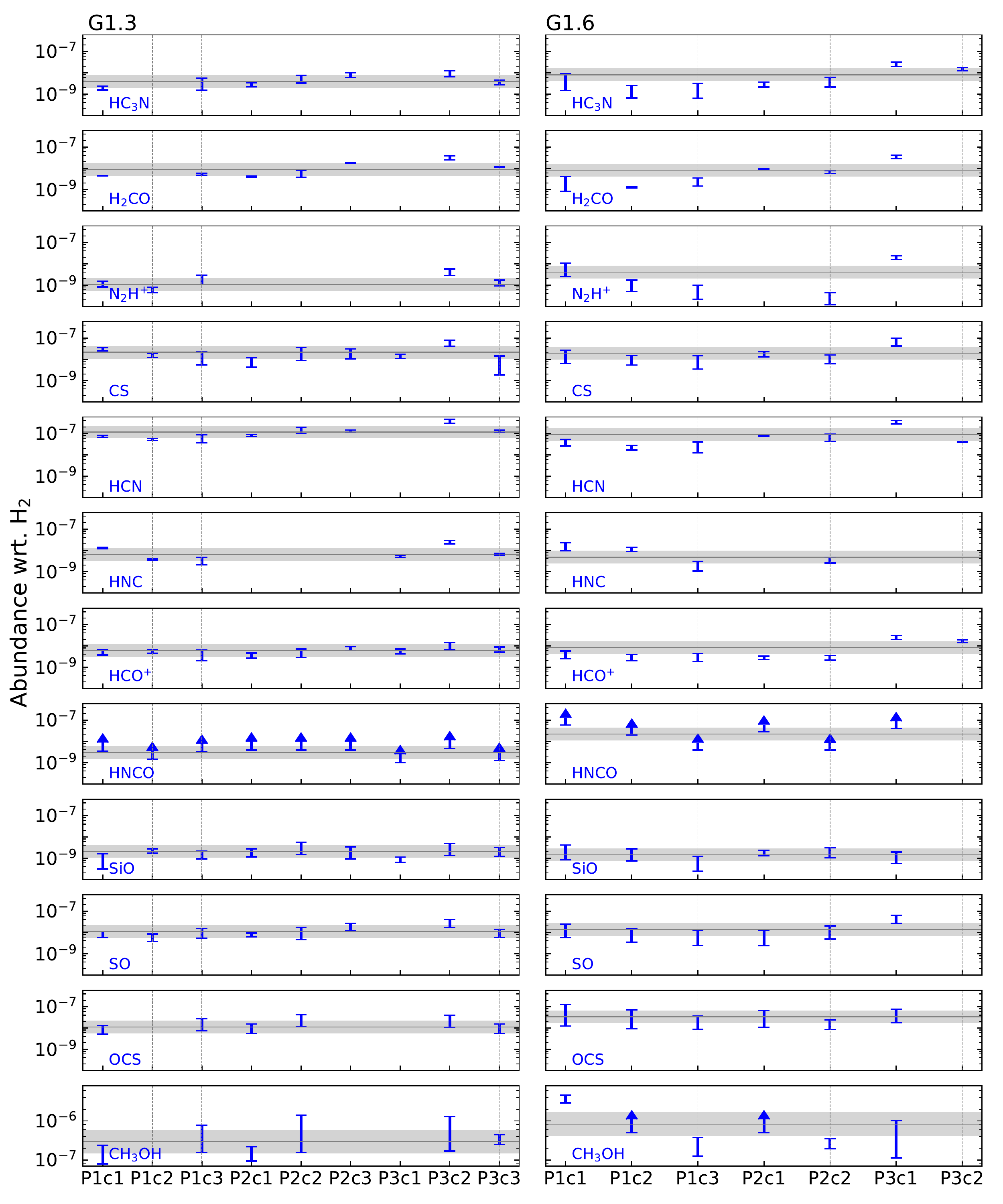}
    \caption{Same as Fig.\,\ref{fig:Ncols}, but, instead of column densities, abundances with respect to H$_2$ are shown.}
    \label{fig:Xmols}
\end{figure*}

Additionally, we estimated the H$_2$ column density from dust surface density (in units of g\,cm$^{-2}$) maps produced by \citet[][see their paper for details]{Guzman15}. Using data from the Herschel archive and from the APEX Telescope Large Survey of the Galaxy \citep[ATLASGAL,][]{Schuller09} covering wavelengths of 160\,--\,870\,$\mu$m, the authors derived dust temperatures $T_\mathrm{dust}$ and surface densities $\Sigma_\mathrm{dust}$ for more than 3000 molecular clouds including the whole CMZ region. We determine median values of $T_\mathrm{dust}$ and $\Sigma_\mathrm{dust}$ in a circle of 20\arcsec\,\,radius around our selected positions. On average, dust temperatures are slightly higher in G1.3 (with values of $\sim$\,21\,--\,23\,K) than in G1.6 (with $\sim$\,17\,--\,19\,K).
Dust surface densities were derived from integrated intensities, where the integration was done over the whole velocity range, that is, without consideration of different velocity components or local gas contributions. 
To derive H$_2$ column densities from these values, we needed an estimate on the contributions of each velocity component to the total surface density. 
Therefore, we simply related integrated intensities of one velocity component to the total integrated intensity, which considers all emission detected at the respective position. We use the $^{13}$CO 2\,--\,1 transitions for this rough estimation. 
Subsequently, dust surface densities are converted to H$_2$ column densities by dividing by $2.8 m_{\rm H}$, where $m_{\rm H}$ is the mass of an hydrogen atom and 2.8 is the mean molecular weight. The results are shown in Table\,\ref{tab:13co}. 
Because the low-velocity components of G1.3/P2 and G1.6/P1 are strongly blended, only one value of H$_2$ column density was derived there, respectively. In G1.3 H$_2$ column densities derived from dust and $^{13}$CO agree well for some components while for others the dust yields higher values by a factor 2\,--\,3. 
In G1.6 the column densities derived from dust vary over a larger range and are higher by more than one order of magnitude for some components. The results obtained from dust suggest overall higher H$_2$ column densities in G1.6 than in G1.3, in contrast to the results from $^{13}$CO. However, H$_2$ column densities derived from dust may be taken with caution as the derivation depends on dust properties such as the dust mass opacity and dust emissivity spectral index, both of which are highly uncertain parameters \citep[for their estimation on the dust properties see][]{Guzman15}. 
On the other hand, the $^{13}$CO emission may not necessarily be optically thin, which would result in an underestimation of the $^{13}$CO and H$_2$ column densities. However, opacity values of $<0.2$ determined during the RADEX modelling for the $J=1-0$ transitions suggest that the $^{13}$CO emission is optically thin at each position. Also, we cannot exclude some uncertainty in the conversion factors from $^{13}$CO to $^{12}$CO to H$_2$.  

\subsection{Molecular column densities \& fractional abundances}\label{ss:abund}


Column densities for all molecules and components that were obtained from the RADEX modelling are listed in Tables\,\ref{tab:Ncol} and \ref{tab:Ncol2} for G1.3 and G1.6, respectively. 
The column densities of each molecule in each component of G1.3 and G1.6 are shown in the left and right columns of Fig.\,\ref{fig:Ncols}, respectively. The horizontal line indicates the mean column density of a molecule in the cloud complex. The grey-shaded area indicates a deviation from the mean by a factor of 2. These mean values show that the difference between G1.3 and G1.6 for a respective molecule is at most a factor 4, in most cases even less, which indicates that the chemistry is generally not too different between both cloud complexes. 

Regarding G1.3 and G1.6 separately, there are no remarkable differences between the high-velocity (indicated with dashed vertical lines) and low-velocity components. However, molecules in component P3/c2 in G1.3 seem to have systematically higher column densities by factors of a few than in the other components, except for HCO$^+$, SiO, OCS, and CH$_3$OH. On the other hand, column densities of molecules in components P1/c2 and P3/c1 in G1.3, if they are detected at all, are lower than the average by factors of a few. Interestingly, however, an increase or a decrease of column densities in G1.3 
affects molecules in a similar manner, while in G1.6 column densities can be above the average for some molecules and below for others in one component. For example, in P1/c1  or P1/c2 in G1.6 the majority of molecules has a column density below the average except for methanol and, in P1/c2, HNC, which show higher column densities than average. Besides,  column densities of all density-tracer molecules and SO are higher than the average by a factor of a few in component P3/c1 in G1.6. Component P3/c2 in G1.6 is only detected in HCN, HC$_3$N, and HCO$^+$ emission and these molecules have only low column densities.

Using the H$_2$ column densities derived from $^{13}$CO, we determined fractional abundances for each molecule in each velocity component, where the minimum and maximum value of column density is divided by the most likely value of H$_2$ column density shown in Table\,\ref{tab:13co}, neglecting the uncertainty on the latter. The results for G1.3 and G1.6 are shown in the left and right columns of Fig.\,\ref{fig:Xmols}, respectively. 
Similar to the column densities, the abundances with respect to H$_2$ in both cloud complexes differ by less than a factor of 4 for all molecules on average, except for HNCO, for which the average in both complexes differs by a factor 7. Given that we only derive upper limits for this molecule, however, the difference could also be smaller. Regarding both cloud complexes separately, similar trends as seen for column density are evident: P3/c2 in G1.3 and P3/c1 in G1.6 show higher column densities. Components that showed column densities lower than  average do not stand out in case of abundances.

\section{Discussion}\label{s:discussion}

\subsection{Implications for cloud properties}

\subsubsection{Kinetic temperatures \& densities}\label{dss:temp}

\begin{table*}[t]
    \caption{Kinetic temperatures $T_\mathrm{kin}$ and H$_2$ number densities derived in this and previous studies for G1.3 and G1.6.}
    \centering
    \begin{tabular}{lrrlrrr}
        \hline\hline\\[-0.3cm] 
         Source & Position (l,b) & Resolution & Reference & $\varv_\mathrm{lsr}$ & $T_\mathrm{kin}$ & $n($H$_2)$ \\
         & ($^\circ$) & ($^{\prime\prime}$) & & (km\,s$^{-1}$) & (K) & (10$^4$\,cm$^{-2}$) \\\hline\\[-0.3cm]
         G1.3 & (1.31,-0.13) & $\sim$50 & \citet{Huettemeister98} & 80 & $>$100 & 0.2 \\
         & $\sim$(1.27,0.01) & $\sim$40 & \citet{Tanaka07} & $<$110 & 25 & $\sim$1.0 \\
         & & & & $>$110 & $\gtrsim$25 & $\gtrsim$1.0 \\
         & (1.28,0.07) & $\sim$40 & \citet{Riquelme13} & \multirow{2}{*}{100} & 40 & 6.0  \\
         & & & & & 100 & 3.0 \\
         & & & & \multirow{2}{*}{180} & 40 & 10.0 \\
         & & & & & 300 & 3.6 \\
         G1.6 & (1.59,0.015) & $\sim$50 & \citet{Salii02} & 50 & 25\,--\,55 & $<$1.0 \\
         & 2:(1.59,0.015) & & & 160 & 150\,--\,180 & $<$1.0 \\
         & \& 3:(1.64,-0.064) & $\sim$120 & \citet{Menten09} & 60 & 30 \& 60 & 7.0 \\
         & & & & 160 & 190 \& 16 & 3.8 \\
         \hline\hline
    \end{tabular}
    \label{tab:Refs}
\end{table*}

Based on the non-LTE modelling of CH$_3$CN emission spectra, reliable kinetic temperatures have been derived in a range of 60--100\,K in G1.3 and G1.6. A kinetic temperature of 175\,K has been found for component P3/c2 and 100\,K in the component P2/c2 in G1.3, however, this is likely the consequence of contamination by another velocity component that was not fitted in the CH$_3$CN spectra, but was detected for other molecules. No great differences are observed between the cloud complexes nor between the low- and high-velocity gas in either complex.
H$_2$ number densities derived from non-LTE modelling of CH$_3$CN emission spectra are on average a few 10$^4$\,cm$^{-3}$. Only components P1/c1 and P2/c1 in G1.6 tend to a slightly higher value above 10$^5$\,cm$^{-3}$ and P3/c1 in G1.6 tends to a value of less than 10$^4$\,cm$^{-3}$. 

Table\,\ref{tab:Refs} lists kinetic temperatures and H$_2$ number densities derived for G1.3 and G1.6 in previous studies. Although a position with a different velocity component was studied by \citet{Huettemeister98} we still show these early results.
H$_2$ number densities generally agree with earlier results. 
\citet{Riquelme13} for G1.3 and \citet{Menten09} for G1.6 showed that the low- and high-velocity component each have a low- and a high-temperature as well as low- and high-density components. Model results of $^{13}$CO suggests the presence of an even lower density regime at $\sim$10$^3$\,cm$^{-3}$, while in some components SiO, SO, and CS point to a slightly higher density than seen for CH$_3$CN. However, based on our results, we cannot certainly confirm the presence of different density regimes in one velocity component.

Values for the kinetic temperatures that are similar to the values derived by us, have been reported in previous studies for G1.3 and G1.6 (see Table\,\ref{tab:Refs}) and are typical for clouds in the CMZ \citep[e.g.][]{Huettemeister98, Ginsburg16, Immer16}. However, in G1.3 kinetic temperatures of $\gtrsim$300\,K have been determined \citep[][]{Riquelme13}. The authors used ammonia (NH$_3$), which is another commonly used thermometer for interstellar gas similar to CH$_3$CN.
Because NH$_3$ has a lower critical density than CH$_3$CN, it might trace a more diffuse gas component, which could have higher temperatures than the denser gas traced by CH$_3$CN and may explain why we do not see these extreme temperature with the latter molecule.  However, \citet{Menten09} found a maximum kinetic temperature of $\sim$200\,K in G1.6 based on observations of methanol, which generally traces similar densities as CH$_3$CN.
As can be seen from Fig.\,\ref{fig:ch3cn-spectra}, the CH$_3$CN emission is fairly weak, especially in G1.3, which made the line fitting challenging and even impossible for some components, also because the blending of lines and velocity components is difficult to distinguish. A proper analysis of methanol or sensitive observations in ammonia emission may help to investigate possible temperature differences between the high- and low-velocity gas, between positions, and between the cloud complexes. 


\subsubsection{H$_2$ column density \& molecular fractional abundances}\label{dss:abund}
H$_2$ column densities were derived from $^{13}$CO column densities, which were obtained from the LVG modelling of multiple transitions of the molecule.
In this process we assumed that the $^{13}$CO emission is optically thin, which must not always be the case and therefore, can lead to an underestimation of H$_2$ column densities.  Our results are lower than H$_2$ column densities derived from $^{13}$CO in G1.3 by \citet{Riquelme13} by an order of magnitude on average. However, the authors only used the $J=1-0$ transition of $^{13}$CO, which may introduce a higher uncertainty on the H$_2$ column density, eventually. 
H$_2$ column densities that we derived from dust are smaller by factors of a few only.  As already mentioned, dust properties are generally highly uncertain. Furthermore, it is known that the gas is decoupled from the dust as temperatures are generally significantly lower than gas kinetic temperatures 
\citep{Lis99,Rodriguez-Fernandez02,Molinari11}. Therefore, dust may not be suitable to derive H$_2$ column densities in these regions. The results estimated from the dust compare, however, to those derived from dust observations at 1.3\,mm within the CMZoom survey \citep{CMZoom20}, where the total H$_2$ column densities are generally higher in G1.6 \citep[few 10$^{22}$ up to $\sim$10$^{23}$\,cm$^{-2}$,][]{Hatchfield20} than in G1.3.
Underestimating H$_2$ column densities would lead to an overestimation of molecular abundances relative to H$_2$. On the other hand, the size of the emitting region of a molecule and, thus, the beam filling factor is fairly uncertain. We may be overestimating the size of this region by assuming a beam filling factor of 1, which would lead to an underestimation of the column densities and abundances. 

The chemistry indicated by the molecular fractional abundances seems to generally be similar between G1.3 and G1.6. Moreover, small deviations from the average seen in a few components do not seem to depend on the cloud's velocity. 
In G1.6 some molecules show higher abundances than average, while others show lower values. 
This may indicate a greater variation of cloud properties in G1.6.

We compare the molecular abundances with values found in other sources of the CMZ, which are summarised in \citet{Riquelme18}. These authors also show abundances for G1.3 and G1.6, however, without differentiating between velocity components, that is their abundances show only an average of one position in both complexes, which explains minor differences seen for the molecules.
Of all the sources listed in Table\,5 in \citet[][]{Riquelme18}, results for G$+$0.693 and G$-$0.11 \citep[taken from][]{Armijos15} are closest to our findings. Notably, abundances of shock tracers such as SiO, SO, OCS, and HNCO are at the upper end of the range spanned by the GMCs mentioned in this table. 
Like G1.3 and G1.6, G$+$0.693 and G$-$0.11 do not show active star formation \citep[although G$+$0.693 was recently proposed to possibly be on the verge of star formation,][]{Colzi22} and their chemistry is proposed to be driven by low-velocity shocks. Moreover, G$+$0.693 has been proposed as a site of cloud-cloud collision that drives shocks into the gas. In contrast to these last two sources, G1.3 and G1.6 do not show evidence for complex organic chemistry beyond CH$_3$CN and CH$_3$OH, which is likely triggered by shocks in G$+$0.693 and G$-$0.11 as well.  
Nonetheless, shocks seem to likely be the driving source of the chemistry in G1.3 and G1.6 like in other GC GMCs. 

\begin{figure}[]
    \centering
    \includegraphics[width=0.5\textwidth]{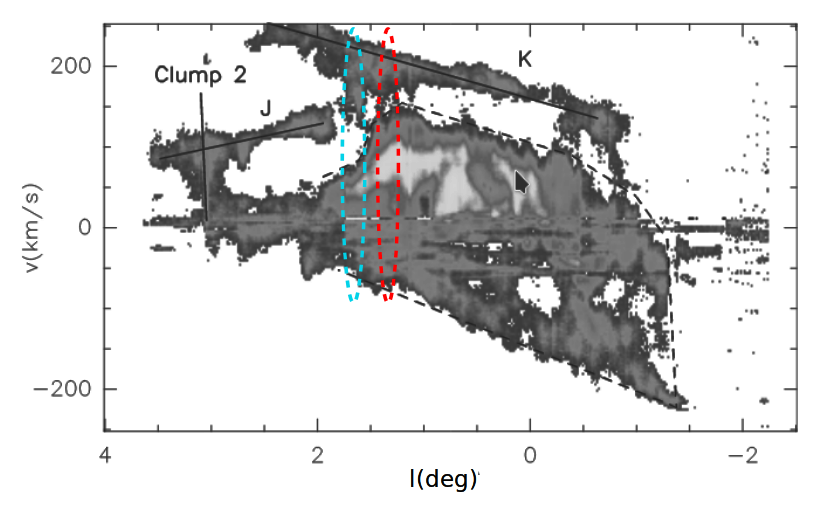}
    \includegraphics[width=.3\textwidth]{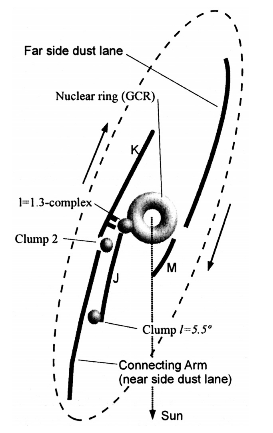}\\[1.cm]
    \includegraphics[width=.5\textwidth]{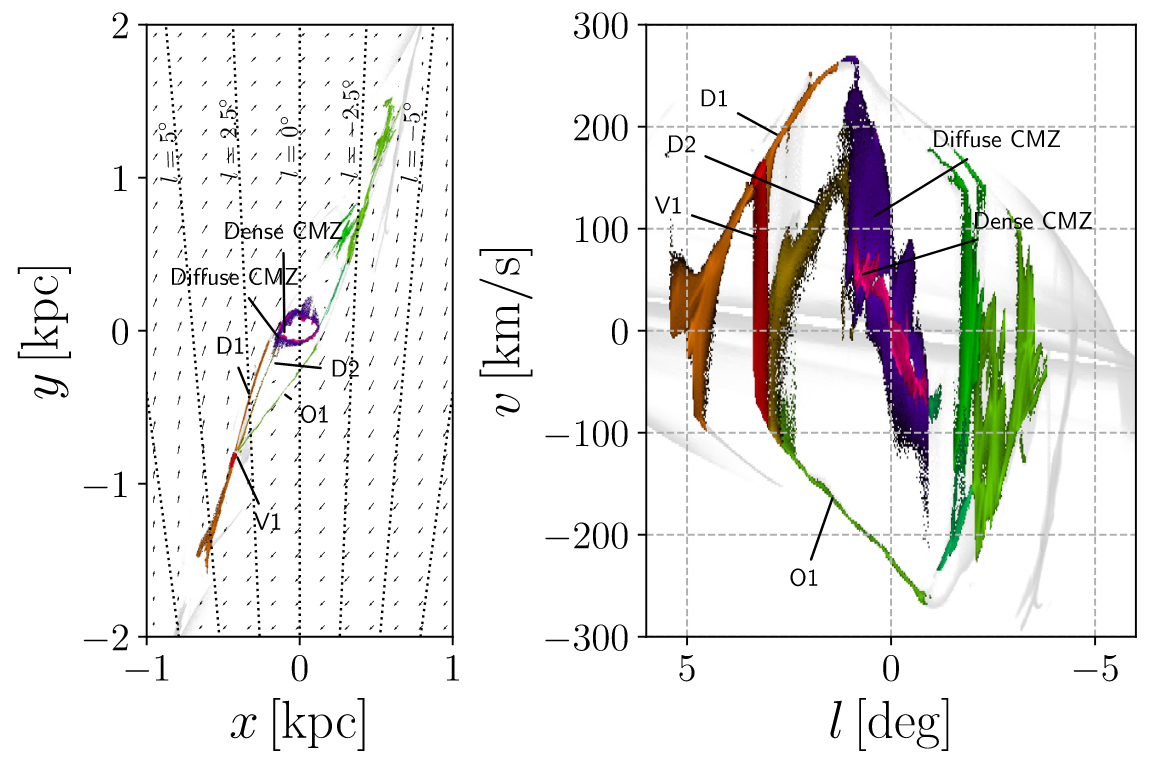}
    \begin{tikzpicture}[remember picture,overlay]
        \draw (2.4,5.1) -- (3.,6.);
        \draw (3.12,6.1) node {\tiny V2};
        \draw (-2.36,3.88) -- (-2.85,4.);
        \draw (-3.1,4.) node {\tiny V2};
    \end{tikzpicture}  
    \caption{\textit{Top panel:} Longitude-velocity diagram of CO 1–0 emission in the CMZ \citep[taken from][]{Rodriguez08} using data of \citet{Bally87}. Intensities were integrated over all negative Galactic latitudes. Solid lines indicate features such as Clump\,2 and dust lanes K and J \citep{Rodriguez06}. \textit{Middle panel:} Illustration of a possible face-on view of the GC \citep[taken from][]{Rodriguez06}. \textit{Bottom panels:} Recent simulations of the gas kinematics in the CMZ by \citet{Sormani19}.}
    \label{fig:GCkin}
\end{figure}

\subsection{Implications for the CMZ}\label{ss:CCC}
In this section, we combine the results on the cloud properties and the results on the morphology of G1.3 and G1.6 in order to place the two cloud complexes in the picture of the CMZ.
Based on the molecular abundances, we concluded that turbulence most likely drives the chemistry in G1.6 and G1.3. Turbulence  is also responsible for the high kinetic temperatures \citep[e.g.][]{Ginsburg16}.
The main result obtained from the analysis of the morphology in Sect.\,\ref{ss:morphology} is the connection of low- ($\sim$100\,--\,120\,km\,s$^{-1}$) and high-velocity ($\sim$180\,km\,s$^{-1}$) gas by an emission bridge at intermediate velocities ($\sim$150\,km\,s$^{-1}$) in G1.3. Such a connection of gas components is not identified in G1.6. 
An emission bridge presents observational evidence for the interaction of at least two clouds that move at different velocities as shown by numerical simulations and observations \citep[e.g.][]{Haworth15,Haworth15a,Fukui21}.  
In the case of G1.3, the projected velocity separation (i.e. disregarding inclination), between the participating clouds would be $\sim$80\,km\,s$^{-1}$, which is much higher than the typical separation of a few km\,s$^{-1}$ \citep[][and references therein]{Sano21}. In their Fig.\,9a, \citet{Enokiya21a} show velocity separations of 45 possible cloud-cloud interaction regions, where only six of them have higher values than 15\,km\,s$^{-1}$ and the largest value is at 24\,km\,s$^{-1}$, which corresponds to the possible cloud-cloud collision (CCC) in the Giant Molecular Loops (GMLs), which is another region in the GC. 
According to their observations, the velocity separation could be as high as $\sim$60\,km\,s$^{-1}$ if the angle between collision axis and line of sight were 45$^\circ$ \citep{Enokiya21a}. Because inclination is a highly uncertain parameter the authors assumed a lower limit for the velocity separation of 24\,km\,s$^{-1}$ based on the average line width of observed spectral lines.
Therefore, depending on the viewing angle at which we observe a possible CCC in G1.3, it may present an extreme example. If a CCC indeed happened at this velocity separation, one might expect a more different chemistry for G1.3 compared to other GMCs in the CMZ. However, our findings are comparable to what is seen in sources that experience only low-velocity shocks. It remains uncertain whether solely inclination can explain this scenario. 

An interaction of clouds in G1.3 is further supported by the observed fluffy, shell-like structure of the gas.
Given the absence of massive star formation in G1.3 \citep[and G1.6,][]{Menten09}, which otherwise is associated with this kind of morphology, a cloud-cloud collision presents the most possible explanation. 
Early theoretical studies by \citet{Habe92} and \cite{Anath10} and, most recently, \citet{Sakre22} discussed the impact of a collision between clouds of different sizes or density distribution or both on the morphology of such a region. Upon collision, the smaller, denser part of the colliding cloud is able to displace the material of the other cloud, which leaves behind a hole. Therefore, the morphology of a post-collision cloud complex may present complementary structures of an emission hole and a dense clump accounting for it. This has been observed in various Galactic plane clouds \citep[e.g.][]{Fukui18,Fukui18.2,Sano21}, but also in the GC \citep[][]{Enokiya21a,Enokiya21b,Zeng20}. However, depending on the timescale, these complementary structures might have already moved apart substantially and are not found at the same position. 
Given the larger number of emission cavities and the clumpiness of emission that we observe in the lower- and higher-velocity clouds in G1.3 
we cannot discern which clump might have caused which hole. 
Additionally, given the probably high degree of turbulence in G1.3 it is possible that complementary structures might have been substantially displaced or even fused. 

A reason for the possible occurrence of a CCC may be associated with the location of G1.3 at the edge of the CMZ and, therefore, its potential to be a region where material is accreted into the CMZ. 
The top panel in Fig.\,\ref{fig:GCkin} shows an LV diagram in CO 1\,--\,0 emission, which is integrated over negative latitudes \citep{Rodriguez08}, of the whole CMZ with its characteristic parallelogram shape. The approximate locations of G1.3 and G1.6 are indicated by red and blue ellipses, respectively. Based on this, it is evident that the gas in G1.3 extends to highest velocities of the CMZ representing the tip of the parallelogram. Additionally, it seems to be connected with the dust lane K \citep[the connecting arm in the middle panel of Fig.\,\ref{fig:GCkin},][]{Rodriguez06,Rodriguez08} at even higher velocities. Because the K lane is most prominent at negative latitude at the position of G1.3 (and G1.6) \citep[cf. Fig.\,4 in][]{Rodriguez06} and our observations of G1.3 only extend to $b\sim-0.05^\circ$ we cannot investigate a possible interaction of the two based on our observations. Nonetheless, if G1.3 would indeed be a region where material was accreted into the CMZ, this dust lane could be a reasonable supplier.

Although spatial coincidences between the low- and high-velocity gas in G1.6 are evident they do not seem to be connected by an emission bridge at intermediate velocities as can be seen in the PV diagram in Fig.\,\ref{fig:16slice}. This suggests that the high-velocity component is spatially separated from the low-velocity components. This may be supported by the fact that the molecular abundances in velocity components seem to be more independent of each other than in G1.3, that is, abundances of some molecules may be higher than average, while those of other molecules are lower in one component, while for another component at the same position all abundances can be average. However, given the overall similarity of the chemistry in G1.3 and G1.6, especially regarding the shock tracers such as SiO, we would expect that an event such as a CCC would also be responsible for the chemistry in G1.6.

Recent numerical calculations by \citet[][]{Sormani19,Sormani20} and \citet{Tress20} are able to explain the 
existence of G1.3 (also G1.6 and other features such as Clump\,2, see Fig.\,\ref{fig:GCkin}) at their  location in the CMZ,  purely by kinematical considerations, that is, not taking into account physical and chemical properties of single clouds in detail. 
According to these simulations, extended velocity features (EVFs), as the authors call the observed gas with high velocity dispersion in the CMZ, are a natural consequence of the barred gravitational potential. In particular, EFV V2 closely resembles G1.3 regarding its location, as can be seen in the bottom panel of Fig.\,\ref{fig:GCkin} and the PV diagram \citep[see Fig.\,7 in][]{Sormani19}, a fact that has already been pointed out by the authors. Accordingly, G1.3 appears as a result of accretion of gas originating from dust lane D1 (same as dust lane K) to the CMZ, where it interacts with the residing lower-velocity gas. 
This scenario may be supported by the findings of \citet{Riquelme10iso}, who reported on the different $^{12}$C/$^{13}$C isotopic ratios for the low- and high-velocity components. High ratios untypical for the CMZ were determined for the gas at high velocities, suggesting an origin from outside the CMZ, while typical ratios of $\sim$24 were found for the lower-velocity gas. The insufficient sensitivity of our observations prevented us from confirming these results. 

For G1.6, we rather propose a coincidental superposition of low- and high-velocity gas components, that is, although observed along the same line of sight clouds at low and high velocities are actually spatially separated. In the context of the simulations of \citet{Sormani19}, the high-velocity gas may be attributed to their dust lane D2 (same as dust lane J, see Fig.\,\ref{fig:GCkin}), while the low-velocity gas may be associated with gas located in the dense CMZ or it may correspond to gas that moved along the far side dust lane (or O1) and instead of plunging onto the CMZ at negative longitudes it overshot and was sprayed in between the dust lanes, where it decelerated. The movement along a dust lane or the abrupt deceleration may then be able to evoke the shock chemistry.

Our proposed scenarios in association with the models by \citet{Sormani19,Sormani20} and \citep{Tress20} can be supported by observations of galactic centres of other galaxies. Equivalents of the near and far side dust lanes attributed to the Milky Way's CMZ have been observed in, for example the centres of M\,83 \citep[][]{Harada19,Callanan21}, NGC\,3504 \citep[][]{Wu21}, and, recently, in a couple of more nearby galaxies thanks to the high angular resolution of the PHANGS survey \citep[Physics at High Angular resolution in Nearby GalaxieS, e.g.][]{Lee22}. The observed kinematic structure of the dust lanes and their connection to the CMZs of these galaxies resembles the predictions obtained from the simulations (cf. bottom right panel in Fig.\,\ref{fig:GCkin}). 

In addition to a proposed interaction with material originating from the dust lanes, \citet[][]{Sormani19} proposed that external perturbations can lead to the interactions of clouds, that is, as a consequence of the complex kinematics of the CMZ, it is possible that clouds at highly different speeds may be (nearly) co-located along the same line of sight and only a small perturbation, for example from an expanding supernova shock front, can cause a bridging feature between them.
An emission bridge provoked by such an event may, however, be weak, such that we would not be able to detect it with our observations or in molecular emission at all. Therefore, it remains questionable whether such an event would be able to trigger the rich chemistry that we see also at intermediate velocities in G1.3. However, in this case, the presence of an emission bridge in G1.6 is not completely ruled out as low-excitation CO emission is pervasive in the CMZ and including the low-intensity emission in Fig.\,\ref{fig:16slice}b and c would lead to a connection between the velocity components that is not seen for less abundant species. 

In contrast to our proposed scenario, the most recent work by \citet[][]{Tsujimoto21} tries to explain the shell-like structure in G1.3 based on the idea of a nascent molecular superbubble, which was first introduced by \citet{Tanaka07}. 
Based on archival CO 1\,--\,0 \citep{Tokuyma19} and 3\,--\,2 data \citep[][]{Eden20} the authors identified 11 shells, for which they derived masses of $10^{4-5}\,M_\sun$ and kinetic energies of $10^{49.5-51.8}$\,erg. They attributed these high energies to multiple supernovae explosions of members of  a young and extremely massive star cluster ($M_\mathrm{cl}\approx10^{7.5}\,M_\sun$). However, there are several caveats concerning this scenario such as the absence of H{\small II} regions and supernova remnants \citep[as traced by observations at cm wavelength,][]{Nguyen21}, and an infrared luminosity much below the expectations. 
We estimated the mass of the high-velocity component from the CO 2\,--\,1 data by using integrated intensities in the velocity range of 175\,--\,220\,km\,s$^{-1}$. Assuming a temperature of 75\,K, this yields a mass of 10$^{7.7}\,M_\sun$ and translates to a kinetic energy of $10^{54.5}$\,erg assuming a relative velocity of the low- and high-velocity gas of 80\,km\,s$^{-1}$, which is much higher than any value derived by \citet[][]{Tsujimoto21}. This would be the kinetic energy involved in the process of the high-velocity gas plunging into the low-velocity gas. As described, this interaction causes holes or shells in the lower-velocity cloud, however, whether these would expand as stated by \citet[][see also \citeauthor{Tanaka07} 2007]{Tsujimoto21} is uncertain. Moreover, if such a massive event indeed took place in G1.3, one might except the chemistry to stand out more from other GC GMCs, which, as we saw, is not the case. Maybe observations at higher angular resolution and better sensitivity are able to detect differences between these two complexes and other GC sources.

\section{Conclusion}\label{s:conclusion}

We mapped G1.3 and G1.6 in the complete 3\,mm spectral window with the IRAM\,30m telescope and over significant portions of the submillimetre range (between 0.65 and 1.38\,mm) with the APEX telescope. Based on a selected sample of 14 molecular species, we studied the morphology of the cloud complexes (Sect.\,\ref{ss:morphology}) and derived physical conditions and chemical composition at three positions in each complex (Sects.\,\ref{ss:lvg}--\ref{ss:abund}), in which a variety of gas components was detected at different peak velocities. 
Our main results are the following:

\begin{enumerate}
    \item The distribution of gas in G1.3 provides observational evidence for interaction between clouds. We identify an emission bridge at intermediate velocities in the PV diagrams connecting the high- and low-velocity gas components along with an overall fluffy morphology with a shell-like structure (Figs.\,\ref{fig:cmaps13}--\ref{fig:13slice}).
    \item There is no emission bridge identified in G1.6 suggesting a spatial separation of low- and high-velocity gas components (Fig.\,\ref{fig:16slice}).
    \item In both cloud complexes, kinetic temperatures and H$_2$ volume densities, derived from non-LTE modelling of CH$_3$CN emission, are in a range of $\sim$60--100\,K and $\sim$10$^4$--10$^5$\,cm$^{-3}$, respectively, which are overall in agreement with previous studies (Sect.\,\ref{dss:temp}). 
    \item The chemistry, as traced by molecular fractional abundances, is similar in G1.3 and G1.6 (Figs.\,\ref{fig:Ncols}--\ref{fig:Xmols}) in general and compares with the chemistry of other GC molecular clouds, especially, G$+$0.693 and G$-$0.11, whose chemistry is likely driven by low-velocity shocks (Sect.\,\ref{dss:abund}). 
    
\end{enumerate}
Based on our results and by comparing them to recent simulations by \citet{Sormani19} in Sect.\,\ref{ss:CCC}, we propose that G1.3 may present the region in which high-velocity gas from the near-side dust lane is accreted to the CMZ, where it interacts with pre-existing gas that moves at lower velocity. In contrast, at the location of G1.6, high-velocity gas moving along a dust lane and low-velocity gas, that either overshot from the far-side dust lane or resides in the CMZ, are observed coincidentally along the same line of sight and do not interact (yet).  

\begin{acknowledgements}
We thank Andr\'es E. Guzm\'an for providing the H$_2$ surface density map covering the CMZ for us. We would also like to thank Mattia Sormani for helpful discussions about their simulations of the CMZ kinematics and Rainer Mauersberger for useful comments that helped to improve the manuscript. D.R. acknowledges support from the Collaborative Research Council\,956, subproject A5, funded by the Deutsche Forschungsgemeinschaft (DFG). T.P.\ gratefully acknowledges support by the National Science Foundation under grants No.\ AST-2009842 and AST-2108989. The work of J.K.\ is in part supported by the National Science Foundation under grant No.\ AST-1909097. This publication makes use of data acquired with the IRAM\,30m\,telescope and the Atacama Pathfinder Experiment (APEX). IRAM is supported by INSU/CNRS (France), MPG (Germany) and IGN (Spain). APEX is a collaboration between the Max-Planck-Institut für Radioastronomie, the European Southern Observatory, and the Onsala Space Observatory. The authors would like to thank the developers of the many Python libraries, made available as open-source software, in particular this research has made use of the NumPy \citep[][]{numpy}, AstroPy \citep{astropy}, and matplotlib \citep{matplotlib} packages.

\end{acknowledgements}

\bibliographystyle{aa} 
\bibliography{refs.bib} 

\begin{thebibliography}{119}
\expandafter\ifx\csname natexlab\endcsname\relax\def\natexlab#1{#1}\fi

\bibitem[{{Anathpindika}(2010)}]{Anath10}
{Anathpindika}, S.~V. 2010, \mnras, 405, 1431

\bibitem[{{Ao} {et~al.}(2013){Ao}, {Henkel}, {Menten}, {Requena-Torres},
  {Stanke}, {Mauersberger}, {Aalto}, {M{\"u}hle}, \& {Mangum}}]{Ao13}
{Ao}, Y., {Henkel}, C., {Menten}, K.~M., {et~al.} 2013, \aap, 550, A135

\bibitem[{{Armijos-Abenda{\~n}o} {et~al.}(2015){Armijos-Abenda{\~n}o},
  {Mart{\'i}n-Pintado}, {Requena-Torres}, {Mart{\'i}n}, \&
  {Rodr{\'i}guez-Franco}}]{Armijos15}
{Armijos-Abenda{\~n}o}, J., {Mart{\'i}n-Pintado}, J., {Requena-Torres}, M.~A.,
  {Mart{\'i}n}, S., \& {Rodr{\'i}guez-Franco}, A. 2015, \mnras, 446, 3842

\bibitem[{{Astropy Collaboration} {et~al.}(2018){Astropy Collaboration},
  {Price-Whelan}, {Sip{\H{o}}cz}, {G{\"u}nther}, {Lim}, {Crawford}, {Conseil},
  {Shupe}, {Craig}, {Dencheva}, {Ginsburg}, {VanderPlas}, {Bradley},
  {P{\'e}rez-Su{\'a}rez}, {de Val-Borro}, {Aldcroft}, {Cruz}, {Robitaille},
  {Tollerud}, {Ardelean}, {Babej}, {Bach}, {Bachetti}, {Bakanov}, {Bamford},
  {Barentsen}, {Barmby}, {Baumbach}, {Berry}, {Biscani}, {Boquien}, {Bostroem},
  {Bouma}, {Brammer}, {Bray}, {Breytenbach}, {Buddelmeijer}, {Burke},
  {Calderone}, {Cano Rodr{\'i}guez}, {Cara}, {Cardoso}, {Cheedella}, {Copin},
  {Corrales}, {Crichton}, {D'Avella}, {Deil}, {Depagne}, {Dietrich}, {Donath},
  {Droettboom}, {Earl}, {Erben}, {Fabbro}, {Ferreira}, {Finethy}, {Fox},
  {Garrison}, {Gibbons}, {Goldstein}, {Gommers}, {Greco}, {Greenfield},
  {Groener}, {Grollier}, {Hagen}, {Hirst}, {Homeier}, {Horton}, {Hosseinzadeh},
  {Hu}, {Hunkeler}, {Ivezi{\'c}}, {Jain}, {Jenness}, {Kanarek}, {Kendrew},
  {Kern}, {Kerzendorf}, {Khvalko}, {King}, {Kirkby}, {Kulkarni}, {Kumar},
  {Lee}, {Lenz}, {Littlefair}, {Ma}, {Macleod}, {Mastropietro}, {McCully},
  {Montagnac}, {Morris}, {Mueller}, {Mumford}, {Muna}, {Murphy}, {Nelson},
  {Nguyen}, {Ninan}, {N{\"o}the}, {Ogaz}, {Oh}, {Parejko}, {Parley}, {Pascual},
  {Patil}, {Patil}, {Plunkett}, {Prochaska}, {Rastogi}, {Reddy Janga},
  {Sabater}, {Sakurikar}, {Seifert}, {Sherbert}, {Sherwood-Taylor}, {Shih},
  {Sick}, {Silbiger}, {Singanamalla}, {Singer}, {Sladen}, {Sooley},
  {Sornarajah}, {Streicher}, {Teuben}, {Thomas}, {Tremblay}, {Turner},
  {Terr{\'o}n}, {van Kerkwijk}, {de la Vega}, {Watkins}, {Weaver}, {Whitmore},
  {Woillez}, {Zabalza}, \& {Astropy Contributors}}]{astropy}
{Astropy Collaboration}, {Price-Whelan}, A.~M., {Sip{\H{o}}cz}, B.~M., {et~al.}
  2018, \aj, 156, 123

\bibitem[{{Baars} {et~al.}(1987){Baars}, {Hooghoudt}, {Mezger}, \& {de
  Jonge}}]{Baars87}
{Baars}, J.~W.~M., {Hooghoudt}, B.~G., {Mezger}, P.~G., \& {de Jonge}, M.~J.
  1987, \aap, 175, 319

\bibitem[{{Bally}{} {et~al.}(2010){Bally}{}, {Aguirre}, {Battersby}, {Bradley},
  {Cyganowski}, {Dowell}, {Drosback}, {Dunham}, {Evans}, {Ginsburg}, {Glenn},
  {Harvey}, {Mills}, {Merello}, {Rosolowsky}, {Schlingman}, {Shirley},
  {Stringfellow}, {Walawender}, \& {Williams}}]{Bally10}
{Bally}{}, J., {Aguirre}, J., {Battersby}, C., {et~al.} 2010, \apj, 721, 137

\bibitem[{{Bally} {et~al.}(1987){Bally}, {Stark}, {Wilson}, \&
  {Henkel}}]{Bally87}
{Bally}, J., {Stark}, A.~A., {Wilson}, R.~W., \& {Henkel}, C. 1987, \apjs, 65,
  13

\bibitem[{{Bally} {et~al.}(1988){Bally}, {Stark}, {Wilson}, \&
  {Henkel}}]{Bally88}
{Bally}, J., {Stark}, A.~A., {Wilson}, R.~W., \& {Henkel}, C. 1988, \apj, 324,
  223

\bibitem[{{Battersby} {et~al.}(2020){Battersby}, {Keto}, {Walker}, {Barnes},
  {Callanan}, {Ginsburg}, {Hatchfield}, {Henshaw}, {Kauffmann}, {Kruijssen},
  {Longmore}, {Lu}, {Mills}, {Pillai}, {Zhang}, {Bally}, {Butterfield},
  {Contreras}, {Ho}, {Ott}, {Patel}, \& {Tolls}}]{CMZoom20}
{Battersby}, C., {Keto}, E., {Walker}, D., {et~al.} 2020, \apjs, 249, 35

\bibitem[{{Binney}{} {et~al.}(1991){Binney}{}, {Gerhard}, {Stark}, {Bally}, \&
  {Uchida}}]{Binney91}
{Binney}{}, J., {Gerhard}, O.~E., {Stark}, A.~A., {Bally}, J., \& {Uchida},
  K.~I. 1991, \mnras, 252, 210

\bibitem[{{Boucher} {et~al.}(1980){Boucher}, {Burie}, {Bauer}, {Dubrulle}, \&
  {Demaison}}]{Boucher80}
{Boucher}, D., {Burie}, J., {Bauer}, A., {Dubrulle}, A., \& {Demaison}, J.
  1980, Journal of Physical and Chemical Reference Data, 9, 659

\bibitem[{{Callanan} {et~al.}(2021){Callanan}, {Longmore}, {Kruijssen},
  {Schruba}, {Ginsburg}, {Krumholz}, {Bastian}, {Alves}, {Henshaw}, {Knapen},
  \& {Chevance}}]{Callanan21}
{Callanan}, D., {Longmore}, S.~N., {Kruijssen}, J.~M.~D., {et~al.} 2021,
  \mnras, 505, 4310

\bibitem[{{Carter} {et~al.}(2012){Carter}, {Lazareff}, {Maier}, {Chenu},
  {Fontana}, {Bortolotti}, {Boucher}, {Navarrini}, {Blanchet}, {Greve}, {John},
  {Kramer}, {Morel}, {Navarro}, {Pe{\~n}alver}, {Schuster}, \& {Thum}}]{EMIR12}
{Carter}, M., {Lazareff}, B., {Maier}, D., {et~al.} 2012, \aap, 538, A89

\bibitem[{{Clark} {et~al.}(2013){Clark}, {Glover}, {Ragan}, {Shetty}, \&
  {Klessen}}]{Clark13}
{Clark}, P.~C., {Glover}, S. C.~O., {Ragan}, S.~E., {Shetty}, R., \& {Klessen},
  R.~S. 2013, \apjl, 768, L34

\bibitem[{{Colzi} {et~al.}(2022){Colzi}, {Mart{\'i}n-Pintado}, {Rivilla},
  {Jim{\'e}nez-Serra}, {Zeng}, {Rodr{\'i}guez-Almeida}, {Rico-Villas},
  {Mart{\'i}n}, \& {Requena-Torres}}]{Colzi22}
{Colzi}, L., {Mart{\'i}n-Pintado}, J., {Rivilla}, V.~M., {et~al.} 2022, \apjl,
  926, L22

\bibitem[{{Contopoulos}{} \& {Mertzanides}(1977)}]{Contopoulos77}
{Contopoulos}{}, G. \& {Mertzanides}, C. 1977, \aap, 61, 477

\bibitem[{{Eden} {et~al.}(2020){Eden}, {Moore}, {Currie}, {Rigby},
  {Rosolowsky}, {Su}, {Kim}, {Parsons}, {Morata}, {Chen}, {Minamidani}, {Park},
  {Ragan}, {Urquhart}, {Rani}, {Tahani}, {Billington}, {Deb}, {Figura},
  {Fujiyoshi}, {Joncas}, {Liao}, {Liu}, {Ma}, {Tuan-Anh}, {Yun}, {Zhang},
  {Zhu}, {Henshaw}, {Longmore}, {Kobayashi}, {Thompson}, {Ao},
  {Campbell-White}, {Ching}, {Chung}, {Duarte-Cabral}, {Fich}, {Gao}, {Graves},
  {Jiang}, {Kemper}, {Kuan}, {Kwon}, {Lee}, {Lee}, {Liu}, {Pe{\~n}aloza},
  {Peretto}, {Phuong}, {Pineda}, {Plume}, {Puspitaningrum}, {Samal}, {Soam},
  {Sun}, {Tang}, {Traficante}, {White}, {Yan}, {Yang}, {Yuan}, {Yue}, {Bemis},
  {Brunt}, {Chen}, {Cho}, {Clark}, {Cyganowski}, {Friberg}, {Fuller}, {Han},
  {Hoare}, {Izumi}, {Kim}, {Kim}, {Kim}, {Koch}, {Kuno}, {Lacialle}, {Lai},
  {Lee}, {Lee}, {Li}, {Liu}, {Mairs}, {Pan}, {Qian}, {Scicluna}, {Shi}, {Shi},
  {Srinivasan}, {Tan}, {Thomas}, {Torii}, {Trejo}, {Umemoto}, {Violino},
  {Wallstr{\"o}m}, {Wang}, {Wu}, {Yuan}, {Zhang}, {Zhang}, {Zhou}, \&
  {Zhou}}]{Eden20}
{Eden}, D.~J., {Moore}, T.~J.~T., {Currie}, M.~J., {et~al.} 2020, \mnras, 498,
  5936

\bibitem[{{Endres} {et~al.}(2016){Endres}, {Schlemmer}, {Schilke}, {Stutzki},
  \& {M{\"u}ller}}]{CDMS}
{Endres}, C.~P., {Schlemmer}, S., {Schilke}, P., {Stutzki}, J., \&
  {M{\"u}ller}, H. S.~P. 2016, Journal of Molecular Spectroscopy, 327, 95

\bibitem[{{Enokiya} {et~al.}(2021{\natexlab{a}}){Enokiya}, {Ohama}, {Yamada},
  {Sano}, {Fujita}, {Hayashi}, {Tsutsumi}, {Torii}, {Nishimura}, {Konishi},
  {Yamamoto}, {Tachihara}, {Hasegawa}, {Kimura}, {Ogawa}, \&
  {Fukui}}]{Enokiya21a}
{Enokiya}, R., {Ohama}, A., {Yamada}, R., {et~al.} 2021{\natexlab{a}}, \pasj,
  73, S256

\bibitem[{{Enokiya} {et~al.}(2021{\natexlab{b}}){Enokiya}, {Torii}, \&
  {Fukui}}]{Enokiya21b}
{Enokiya}, R., {Torii}, K., \& {Fukui}, Y. 2021{\natexlab{b}}, \pasj, 73, S75

\bibitem[{{Farquhar} {et~al.}(1994){Farquhar}, {Millar}, \&
  {Herbst}}]{Farquhar94}
{Farquhar}, P.~R.~A., {Millar}, T.~J., \& {Herbst}, E. 1994, \mnras, 269, 641

\bibitem[{{Figer} {et~al.}(1999){Figer}, {McLean}, \& {Morris}}]{Figer99}
{Figer}, D.~F., {McLean}, I.~S., \& {Morris}, M. 1999, \apj, 514, 202

\bibitem[{{Figer} {et~al.}(2002){Figer}, {Najarro}, {Gilmore}, {Morris}, {Kim},
  {Serabyn}, {McLean}, {Gilbert}, {Graham}, {Larkin}, {Levenson}, \&
  {Teplitz}}]{Figer02}
{Figer}, D.~F., {Najarro}, F., {Gilmore}, D., {et~al.} 2002, \apj, 581, 258

\bibitem[{{Flower} \& {Pineau des For{\^e}ts}(2012)}]{Flower2012}
{Flower}, D.~R. \& {Pineau des For{\^e}ts}, G. 2012, \mnras, 421, 2786

\bibitem[{{Fukui} {et~al.}(2021){Fukui}, {Habe}, {Inoue}, {Enokiya}, \&
  {Tachihara}}]{Fukui21}
{Fukui}, Y., {Habe}, A., {Inoue}, T., {Enokiya}, R., \& {Tachihara}, K. 2021,
  \pasj, 73, S1

\bibitem[{{Fukui} {et~al.}(2018{\natexlab{a}}){Fukui}, {Kohno}, {Yokoyama},
  {Nishimura}, {Torii}, {Hattori}, {Sano}, {Ohama}, {Yamamoto}, \&
  {Tachihara}}]{Fukui18}
{Fukui}, Y., {Kohno}, M., {Yokoyama}, K., {et~al.} 2018{\natexlab{a}}, \pasj,
  70, S44

\bibitem[{{Fukui} {et~al.}(2018{\natexlab{b}}){Fukui}, {Torii}, {Hattori},
  {Nishimura}, {Ohama}, {Shimajiri}, {Shima}, {Habe}, {Sano}, {Kohno},
  {Yamamoto}, {Tachihara}, \& {Onishi}}]{Fukui18.2}
{Fukui}, Y., {Torii}, K., {Hattori}, Y., {et~al.} 2018{\natexlab{b}}, \apj,
  859, 166

\bibitem[{Gardner{} {et~al.}(1985)Gardner{}, Whiteoak, Forster, Peters, \&
  Kuiper}]{Gardner85}
Gardner{}, F., Whiteoak, J., Forster, J., Peters, W., \& Kuiper, T. 1985,
  Publications of the Astronomical Society of Australia, 6

\bibitem[{{Gardner}{} \& {Boes}(1987)}]{Gardner87}
{Gardner}{}, F.~F. \& {Boes}, F. 1987, Proceedings of the Astronomical Society
  of Australia, 7, 185

\bibitem[{{Ginsburg} {et~al.}(2016){Ginsburg}, {Henkel}, {Ao}, {Riquelme},
  {Kauffmann}, {Pillai}, {Mills}, {Requena-Torres}, {Immer}, {Testi}, {Ott},
  {Bally}, {Battersby}, {Darling}, {Aalto}, {Stanke}, {Kendrew}, {Kruijssen},
  {Longmore}, {Dale}, {Guesten}, \& {Menten}}]{Ginsburg16}
{Ginsburg}, A., {Henkel}, C., {Ao}, Y., {et~al.} 2016, \aap, 586, A50

\bibitem[{{Gravity Collaboration}(2019)}]{gravity19}
{Gravity Collaboration}. 2019, \aap, 625, L10

\bibitem[{{Gravity Collaboration} {et~al.}(2021){Gravity Collaboration},
  {Abuter}, {Amorim}, {Baub{\"o}ck}, {Berger}, {Bonnet}, {Brandner},
  {Cl{\'e}net}, {Davies}, {de Zeeuw}, {Dexter}, {Dallilar}, {Drescher},
  {Eckart}, {Eisenhauer}, {F{\"o}rster Schreiber}, {Garcia}, {Gao}, {Gendron},
  {Genzel}, {Gillessen}, {Habibi}, {Haubois}, {Hei{\ss}el}, {Henning},
  {Hippler}, {Horrobin}, {Jim{\'e}nez-Rosales}, {Jochum}, {Jocou}, {Kaufer},
  {Kervella}, {Lacour}, {Lapeyr{\`e}re}, {Le Bouquin}, {L{\'e}na}, {Lutz},
  {Nowak}, {Ott}, {Paumard}, {Perraut}, {Perrin}, {Pfuhl}, {Rabien},
  {Rodr{\'\i}guez-Coira}, {Shangguan}, {Shimizu}, {Scheithauer}, {Stadler},
  {Straub}, {Straubmeier}, {Sturm}, {Tacconi}, {Vincent}, {von Fellenberg},
  {Waisberg}, {Widmann}, {Wieprecht}, {Wiezorrek}, {Woillez}, {Yazici},
  {Young}, \& {Zins}}]{gravity21}
{Gravity Collaboration}, {Abuter}, R., {Amorim}, A., {et~al.} 2021, \aap, 647,
  A59

\bibitem[{{G{\"u}sten} {et~al.}(2006){G{\"u}sten}, {Nyman}, {Schilke},
  {Menten}, {Cesarsky}, \& {Booth}}]{Gusten06}
{G{\"u}sten}, R., {Nyman}, L.~{\r{A}}., {Schilke}, P., {et~al.} 2006, \aap,
  454, L13

\bibitem[{{G{\"u}sten}{} \& {Philipp}(2004)}]{Gusten04}
{G{\"u}sten}{}, R. \& {Philipp}, S.~D. 2004, in The Dense Interstellar Medium
  in Galaxies, ed. S.~{Pfalzner}, C.~{Kramer}, C.~{Staubmeier}, \&
  A.~{Heithausen}, Vol.~91, 253

\bibitem[{{Guzm{\'a}n}{} {et~al.}(2018){Guzm{\'a}n}{}, {Guzm{\'a}n}, {Garay},
  {Bronfman}, \& {Hechenleitner}}]{Guzman18}
{Guzm{\'a}n}{}, A.~E., {Guzm{\'a}n}, V.~V., {Garay}, G., {Bronfman}, L., \&
  {Hechenleitner}, F. 2018, \apjs, 236, 45

\bibitem[{{Guzm{\'a}n} {et~al.}(2015){Guzm{\'a}n}, {Sanhueza}, {Contreras},
  {Smith}, {Jackson}, {Hoq}, \& {Rathborne}}]{Guzman15}
{Guzm{\'a}n}, A.~E., {Sanhueza}, P., {Contreras}, Y., {et~al.} 2015, \apj, 815,
  130

\bibitem[{{Habe} \& {Ohta}(1992)}]{Habe92}
{Habe}, A. \& {Ohta}, K. 1992, \pasj, 44, 203

\bibitem[{{Harada} {et~al.}(2019){Harada}, {Sakamoto}, {Mart{\'\i}n},
  {Watanabe}, {Aladro}, {Riquelme}, \& {Hirota}}]{Harada19}
{Harada}, N., {Sakamoto}, K., {Mart{\'\i}n}, S., {et~al.} 2019, \apj, 884, 100

\bibitem[{{Harris} {et~al.}(2020){Harris}, {Millman}, {van der Walt},
  {Gommers}, {Virtanen}, {Cournapeau}, {Wieser}, {Taylor}, {Berg}, {Smith},
  {Kern}, {Picus}, {Hoyer}, {van Kerkwijk}, {Brett}, {Haldane}, {del R{\'i}o},
  {Wiebe}, {Peterson}, {G{\'e}rard-Marchant}, {Sheppard}, {Reddy}, {Weckesser},
  {Abbasi}, {Gohlke}, \& {Oliphant}}]{numpy}
{Harris}, C.~R., {Millman}, K.~J., {van der Walt}, S.~J., {et~al.} 2020, \nat,
  585, 357

\bibitem[{{Haschick}{} \& {Baan}(1993)}]{Haschick93}
{Haschick}{}, A.~D. \& {Baan}, W.~A. 1993, \apj, 410, 663

\bibitem[{{Hatchfield} {et~al.}(2020){Hatchfield}, {Battersby}, {Keto},
  {Walker}, {Barnes}, {Callanan}, {Ginsburg}, {Henshaw}, {Kauffmann},
  {Kruijssen}, {Longmore}, {Lu}, {Mills}, {Pillai}, {Zhang}, {Bally},
  {Butterfield}, {Contreras}, {Ho}, {Ott}, {Patel}, \& {Tolls}}]{Hatchfield20}
{Hatchfield}, H.~P., {Battersby}, C., {Keto}, E., {et~al.} 2020, \apjs, 251, 14

\bibitem[{{Haworth} {et~al.}(2015{\natexlab{a}}){Haworth}, {Shima}, {Tasker},
  {Fukui}, {Torii}, {Dale}, {Takahira}, \& {Habe}}]{Haworth15}
{Haworth}, T.~J., {Shima}, K., {Tasker}, E.~J., {et~al.} 2015{\natexlab{a}},
  \mnras, 454, 1634

\bibitem[{{Haworth} {et~al.}(2015{\natexlab{b}}){Haworth}, {Tasker}, {Fukui},
  {Torii}, {Dale}, {Shima}, {Takahira}, {Habe}, \& {Hasegawa}}]{Haworth15a}
{Haworth}, T.~J., {Tasker}, E.~J., {Fukui}, Y., {et~al.} 2015{\natexlab{b}},
  \mnras, 450, 10

\bibitem[{{Humire} {et~al.}(2020){Humire}, {Thiel}, {Henkel}, {Belloche},
  {Loison}, {Pillai}, {Riquelme}, {Wakelam}, {Langer},
  {Hern{\'a}ndez-G{\'o}mez}, {Mauersberger}, \& {Menten}}]{Humire20}
{Humire}, P.~K., {Thiel}, V., {Henkel}, C., {et~al.} 2020, \aap, 642, A222

\bibitem[{{Hunter}(2007)}]{matplotlib}
{Hunter}, J.~D. 2007, Computing in Science and Engineering, 9, 90

\bibitem[{{H{\"u}ttemeister}{} {et~al.}(1998){H{\"u}ttemeister}{}, {Dahmen},
  {Mauersberger}, {Henkel}, {Wilson}, \& {Mart\'in-Pintado}}]{Huettemeister98}
{H{\"u}ttemeister}{}, S., {Dahmen}, G., {Mauersberger}, R., {et~al.} 1998,
  \aap, 334, 646

\bibitem[{{H{\"u}ttemeister} {et~al.}(1993){H{\"u}ttemeister}, {Wilson},
  {Bania}, \& {Mart\'in-Pintado}}]{Huttemeister93}
{H{\"u}ttemeister}, S., {Wilson}, T.~L., {Bania}, T.~M., \& {Mart\'in-Pintado},
  J. 1993, \aap, 280, 255

\bibitem[{{Immer} {et~al.}(2016){Immer}, {Kauffmann}, {Pillai}, {Ginsburg}, \&
  {Menten}}]{Immer16}
{Immer}, K., {Kauffmann}, J., {Pillai}, T., {Ginsburg}, A., \& {Menten}, K.~M.
  2016, \aap, 595, A94

\bibitem[{{Immer} {et~al.}(2012){Immer}, {Schuller}, {Omont}, \&
  {Menten}}]{Immer2012}
{Immer}, K., {Schuller}, F., {Omont}, A., \& {Menten}, K.~M. 2012, \aap, 537,
  A121

\bibitem[{{Kaifu} {et~al.}(1972){Kaifu}, {Kato}, \& {Iguchi}}]{Kaifu72}
{Kaifu}, N., {Kato}, T., \& {Iguchi}, T. 1972, Nature Physical Science, 238,
  105

\bibitem[{{Kauffmann} {et~al.}(2017{\natexlab{a}}){Kauffmann}, {Pillai},
  {Zhang}, {Menten}, {Goldsmith}, {Lu}, \& {Guzm{\'a}n}}]{Kaufmann17a}
{Kauffmann}, J., {Pillai}, T., {Zhang}, Q., {et~al.} 2017{\natexlab{a}}, \aap,
  603, A89

\bibitem[{{Kauffmann} {et~al.}(2017{\natexlab{b}}){Kauffmann}, {Pillai},
  {Zhang}, {Menten}, {Goldsmith}, {Lu}, {Guzm{\'a}n}, \&
  {Schmiedeke}}]{Kauffmann17b}
{Kauffmann}, J., {Pillai}, T., {Zhang}, Q., {et~al.} 2017{\natexlab{b}}, \aap,
  603, A90

\bibitem[{Kelly {et~al.}(2017)Kelly, Viti, Garc{\'i}a-Burillo, Fuente, Usero,
  Krips, \& Neri}]{kelly17}
Kelly, G., Viti, S., Garc{\'i}a-Burillo, S., {et~al.} 2017, \aap, 597, A11

\bibitem[{{Klein} {et~al.}(2012){Klein}, {Hochg{\"u}rtel}, {Kr{\"a}mer},
  {Bell}, {Meyer}, \& {G{\"u}sten}}]{Klein12}
{Klein}, B., {Hochg{\"u}rtel}, S., {Kr{\"a}mer}, I., {et~al.} 2012, \aap, 542,
  L3

\bibitem[{{Klein} {et~al.}(2014){Klein}, {Ciechanowicz}, {Leinz}, {Heyminck},
  {Gusten}, {Kasemann}, {Wunsch}, {Maier}, \& {Sekimoto}}]{Klein14}
{Klein}, T., {Ciechanowicz}, M., {Leinz}, C., {et~al.} 2014, IEEE Transactions
  on Terahertz Science and Technology, 4, 588

\bibitem[{{Kramer}(1997)}]{iram-cali}
{Kramer}, C. 1997, Calibration of spectral line data at the IRAM 30m radio
  telescope

\bibitem[{{Kramer}(2016)}]{kramer16}
{Kramer}, C. 2016, IRAM 30m Telescope -- Observing Capabilities and
  Organisation, Tech. rep., IRAM

\bibitem[{{Kruijssen} {et~al.}(2014){Kruijssen}, {Dale}, \&
  {Longmore}}]{Kruijssen15}
{Kruijssen}, J.~D., {Dale}, J.~E., \& {Longmore}, S.~N. 2014, \mnras, 447, 1059

\bibitem[{{Kruijssen}{} {et~al.}(2014){Kruijssen}{}, {Longmore}, {Elmegreen},
  {Murray}, {Bally}, {Testi}, \& {Kennicutt}}]{Kruijssen14}
{Kruijssen}{}, J.~M.~D., {Longmore}, S.~N., {Elmegreen}, B.~G., {et~al.} 2014,
  \mnras, 440, 3370

\bibitem[{{Langer} \& {Penzias}(1990)}]{Langer90}
{Langer}, W.~D. \& {Penzias}, A.~A. 1990, \apj, 357, 477

\bibitem[{{Lapkin} {et~al.}(2008){Lapkin}, {Nystr{\"o}m}, {Desmaris}, {Dochev},
  {Vassilev}, {Monje}, {Meledin}, {Henke}, {Strandberg}, {Sundin}, {Fredrixon},
  {Ferm}, \& {Belitsky}}]{shefi08}
{Lapkin}, I., {Nystr{\"o}m}, O., {Desmaris}, V., {et~al.} 2008, in Ninteenth
  International Symposium on Space Terahertz Technology, ed. W.~{Wild}, 351

\bibitem[{{Lee} {et~al.}(2022){Lee}, {Whitmore}, {Thilker}, {Deger}, {Larson},
  {Ubeda}, {Anand}, {Boquien}, {Chandar}, {Dale}, {Emsellem}, {Leroy},
  {Rosolowsky}, {Schinnerer}, {Schmidt}, {Lilly}, {Turner}, {Van Dyk}, {White},
  {Barnes}, {Belfiore}, {Bigiel}, {Blanc}, {Cao}, {Chevance}, {Congiu},
  {Egorov}, {Glover}, {Grasha}, {Groves}, {Henshaw}, {Hughes}, {Klessen},
  {Koch}, {Kreckel}, {Kruijssen}, {Liu}, {Lopez}, {Mayker}, {Meidt}, {Murphy},
  {Pan}, {Pety}, {Querejeta}, {Razza}, {Saito}, {S{\'a}nchez-Bl{\'a}zquez},
  {Santoro}, {Sardone}, {Scheuermann}, {Schruba}, {Sun}, {Usero}, {Watkins}, \&
  {Williams}}]{Lee22}
{Lee}, J.~C., {Whitmore}, B.~C., {Thilker}, D.~A., {et~al.} 2022, \apjs, 258,
  10

\bibitem[{{Leurini}{} {et~al.}(2016){Leurini}{}, {Menten}, \&
  {Walmsley}}]{Leurini16}
{Leurini}{}, S., {Menten}, K.~M., \& {Walmsley}, C.~M. 2016, \aap, 592, A31

\bibitem[{{Leurini} {et~al.}(2004){Leurini}, {Schilke}, {Menten}, {Flower},
  {Pottage}, \& {Xu}}]{Leurini04}
{Leurini}, S., {Schilke}, P., {Menten}, K.~M., {et~al.} 2004, \aap, 422, 573

\bibitem[{{Lis} {et~al.}(1999){Lis}, {Menten}, \& {Zylka}}]{Lis99}
{Lis}, D.~C., {Menten}, K.~M., \& {Zylka}, R. 1999, \apj, 527, 856

\bibitem[{{Longmore} {et~al.}(2013){Longmore}, {Bally}, {Testi}, {Purcell},
  {Walsh}, {Bressert}, {Pestalozzi}, {Molinari}, {Ott}, {Cortese}, {Battersby},
  {Murray}, {Lee}, {Kruijssen}, {Schisano}, \& {Elia}}]{Longmore13}
{Longmore}, S.~N., {Bally}, J., {Testi}, L., {et~al.} 2013, \mnras, 429, 987

\bibitem[{{Mart{\'i}n-Pintado}{} {et~al.}(1997){Mart{\'i}n-Pintado}{}, {de
  Vicente}, {Fuente}, \& {Planesas}}]{Martin-Pintado97}
{Mart{\'i}n-Pintado}{}, J., {de Vicente}, P., {Fuente}, A., \& {Planesas}, P.
  1997, \apjl, 482, L45

\bibitem[{{Menten}{} {et~al.}(2009){Menten}{}, {Wilson}, {Leurini}, \&
  {Schilke}}]{Menten09}
{Menten}{}, K.~M., {Wilson}, R.~W., {Leurini}, S., \& {Schilke}, P. 2009, \apj,
  692, 47

\bibitem[{{Mills} {et~al.}(2017){Mills}, {Togi}, \& {Kaufman}}]{Mills17}
{Mills}, E. A.~C., {Togi}, A., \& {Kaufman}, M. 2017, \apj, 850, 192

\bibitem[{{Miura} {et~al.}(2017){Miura}, {Yamamoto}, {Nomura}, {Nakamoto},
  {Tanaka}, {Tanaka}, \& {Nagasawa}}]{Miura17}
{Miura}, H., {Yamamoto}, T., {Nomura}, H., {et~al.} 2017, \apj, 839, 47

\bibitem[{{Molinari} {et~al.}(2011){Molinari}, {Bally}, {Noriega-Crespo},
  {Compi{\`e}gne}, {Bernard}, {Paradis}, {Martin}, {Testi}, {Barlow}, {Moore},
  {Plume}, {Swinyard}, {Zavagno}, {Calzoletti}, {Di Giorgio}, {Elia},
  {Faustini}, {Natoli}, {Pestalozzi}, {Pezzuto}, {Piacentini}, {Polenta},
  {Polychroni}, {Schisano}, {Traficante}, {Veneziani}, {Battersby}, {Burton},
  {Carey}, {Fukui}, {Li}, {Lord}, {Morgan}, {Motte}, {Schuller},
  {Stringfellow}, {Tan}, {Thompson}, {Ward-Thompson}, {White}, \&
  {Umana}}]{Molinari11}
{Molinari}, S., {Bally}, J., {Noriega-Crespo}, A., {et~al.} 2011, \apjl, 735,
  L33

\bibitem[{{Molinari} {et~al.}(2016){Molinari}, {Schisano}, {Elia},
  {Pestalozzi}, {Traficante}, {Pezzuto}, {Swinyard}, {Noriega-Crespo}, {Bally},
  {Moore}, {Plume}, {Zavagno}, {di Giorgio}, {Liu}, {Pilbratt}, {Mottram},
  {Russeil}, {Piazzo}, {Veneziani}, {Benedettini}, {Calzoletti}, {Faustini},
  {Natoli}, {Piacentini}, {Merello}, {Palmese}, {Del Grande}, {Polychroni},
  {Rygl}, {Polenta}, {Barlow}, {Bernard}, {Martin}, {Testi}, {Ali},
  {Andr{\'e}}, {Beltr{\'a}n}, {Billot}, {Carey}, {Cesaroni}, {Compi{\`e}gne},
  {Eden}, {Fukui}, {Garcia-Lario}, {Hoare}, {Huang}, {Joncas}, {Lim}, {Lord},
  {Martinavarro-Armengol}, {Motte}, {Paladini}, {Paradis}, {Peretto},
  {Robitaille}, {Schilke}, {Schneider}, {Schulz}, {Sibthorpe}, {Strafella},
  {Thompson}, {Umana}, {Ward-Thompson}, \& {Wyrowski}}]{Molinari16}
{Molinari}, S., {Schisano}, E., {Elia}, D., {et~al.} 2016, \aap, 591, A149

\bibitem[{{Morris}{} \& {Serabyn}(1996)}]{Morris96}
{Morris}{}, M. \& {Serabyn}, E. 1996, \araa, 34, 645

\bibitem[{{Mulder}{} \& {Liem}(1986)}]{Mulder86}
{Mulder}{}, W.~A. \& {Liem}, B.~T. 1986, \aap, 157, 148

\bibitem[{{Nguyen} {et~al.}(2021{\natexlab{a}}){Nguyen}, {Rugel}, {Menten},
  {Brunthaler}, {Dzib}, {Yang}, {Kauffmann}, {Pillai}, {Nandakumar},
  {Schultheis}, {Urquhart}, {Dokara}, {Gong}, {Medina}, {Ortiz-Le{\'o}n},
  {Reich}, {Wyrowski}, {Beuther}, {Cotton}, {Csengeri}, {Pandian}, \&
  {Roy}}]{Nguyen2021}
{Nguyen}, H., {Rugel}, M.~R., {Menten}, K.~M., {et~al.} 2021{\natexlab{a}},
  \aap, 651, A88

\bibitem[{{Nguyen} {et~al.}(2021{\natexlab{b}}){Nguyen}, {Rugel}, {Menten},
  {Brunthaler}, {Dzib}, {Yang}, {Kauffmann}, {Pillai}, {Nandakumar},
  {Schultheis}, {Urquhart}, {Dokara}, {Gong}, {Medina}, {Ortiz-Le{\'o}n},
  {Reich}, {Wyrowski}, {Beuther}, {Cotton}, {Csengeri}, {Pandian}, \&
  {Roy}}]{Nguyen21}
{Nguyen}, H., {Rugel}, M.~R., {Menten}, K.~M., {et~al.} 2021{\natexlab{b}},
  \aap, 651, A88

\bibitem[{{Oka} \& {Geballe}(2022)}]{Oka22}
{Oka}, T. \& {Geballe}, T.~R. 2022, \apj, 927, 97

\bibitem[{{Oka}{} {et~al.}(2001{\natexlab{a}}){Oka}{}, {Hasegawa}, {Sato},
  {Tsuboi}, \& {Miyazaki}}]{Oka2001}
{Oka}{}, T., {Hasegawa}, T., {Sato}, F., {Tsuboi}, M., \& {Miyazaki}, A.
  2001{\natexlab{a}}, \pasj, 53, 787

\bibitem[{{Oka}{} {et~al.}(2001{\natexlab{b}}){Oka}{}, {Hasegawa}, {Sato},
  {Tsuboi}, {Miyazaki}, \& {Sugimoto}}]{Oka01}
{Oka}{}, T., {Hasegawa}, T., {Sato}, F., {et~al.} 2001{\natexlab{b}}, \apj,
  562, 348

\bibitem[{{Ott} {et~al.}(2014){Ott}, {Wei{\ss}}, {Staveley-Smith}, {Henkel}, \&
  {Meier}}]{ott14}
{Ott}, J., {Wei{\ss}}, A., {Staveley-Smith}, L., {Henkel}, C., \& {Meier},
  D.~S. 2014, \apj, 785, 55

\bibitem[{{Pickett} {et~al.}(1998){Pickett}, {Poynter}, {Cohen}, {Delitsky},
  {Pearson}, \& {M{\"u}ller}}]{Pickett98}
{Pickett}, H.~M., {Poynter}, R.~L., {Cohen}, E.~A., {et~al.} 1998, \jqsrt, 60,
  883

\bibitem[{{Polehampton} {et~al.}(2019){Polehampton}, {Hafok}, \&
  {Muders}}]{apex-cali}
{Polehampton}, E., {Hafok}, H., \& {Muders}, D. 2019, APEX Calibration and Data
  -- Reduction Manual

\bibitem[{{Press} {et~al.}(1992){Press}, {Teukolsky}, {Vetterling}, \&
  {Flannery}}]{Press92}
{Press}, W.~H., {Teukolsky}, S.~A., {Vetterling}, W.~T., \& {Flannery}, B.~P.
  1992, {Numerical recipes in FORTRAN. The art of scientific computing}
  (Cambridge: University Press)

\bibitem[{{Ridley} {et~al.}(2017){Ridley}, {Sormani}, {Tre{\ss}}, {Magorrian},
  \& {Klessen}}]{Ridley17}
{Ridley}, M. G.~L., {Sormani}, M.~C., {Tre{\ss}}, R.~G., {Magorrian}, J., \&
  {Klessen}, R.~S. 2017, \mnras, 469, 2251

\bibitem[{{Riquelme}{} {et~al.}(2013){Riquelme}{}, Amo-Baladr{\'o}n,
  Mart{\'i}n-Pintado, Mauersberger, Mart{\'i}n, \& Bronfman}]{Riquelme13}
{Riquelme}{}, D., Amo-Baladr{\'o}n, M., Mart{\'i}n-Pintado, J., {et~al.} 2013,
  \aap, 549, A36

\bibitem[{{Riquelme}{}{} {et~al.}(2010){Riquelme}{}{}, {Amo-Baladr{\'o}n},
  {Mart{\'i}n-Pintado}, {Mauersberger}, {Mart{\'i}n}, \&
  {Bronfman}}]{Riquelme10iso}
{Riquelme}{}{}, D., {Amo-Baladr{\'o}n}, M.~A., {Mart{\'i}n-Pintado}, J.,
  {et~al.} 2010, \aap, 523, A51

\bibitem[{Riquelme{}{} {et~al.}(2018)Riquelme{}{}, {Amo-Baladr{\'o}n},
  {Mart{\'i}n-Pintado}, {Mauersberger}, {Mart{\'i}n}, {Burton}, {Cunningham},
  {Jones}, Menten, Bronfman, \& G{\"u}sten}]{Riquelme18}
Riquelme{}{}, D., {Amo-Baladr{\'o}n}, M.~A., {Mart{\'i}n-Pintado}, J., {et~al.}
  2018, \aap, 613, A42

\bibitem[{Riquelme{} {et~al.}(2010)Riquelme{}, {Bronfman}, {Mauersberger},
  {May}, \& {Wilson}}]{Riquelme10}
Riquelme{}, D., {Bronfman}, L., {Mauersberger}, R., {May}, J., \& {Wilson}, T.
  2010, \aap, 523, A45

\bibitem[{{Rodr\'iguez-Fern\'andez} \& {Combes}(2008)}]{Rodriguez08}
{Rodr\'iguez-Fern\'andez}, N.~J. \& {Combes}, F. 2008, \aap, 489, 115

\bibitem[{{Rodr{\'i}guez-Fern\'andez}
  {et~al.}(2006){Rodr{\'i}guez-Fern\'andez}, {Combes}, {Mart\'in-Pintado},
  {Wilson}, \& {Apponi}}]{Rodriguez06}
{Rodr{\'i}guez-Fern\'andez}, N.~J., {Combes}, F., {Mart\'in-Pintado}, J.,
  {Wilson}, T.~L., \& {Apponi}, A. 2006, \aap, 455, 963

\bibitem[{{Rodr{\'i}guez-Fern{\'a}ndez}
  {et~al.}(2002){Rodr{\'i}guez-Fern{\'a}ndez}, {Mart{\'i}n-Pintado}, {de
  Vicente}, \& {Fuente}}]{Rodriguez-Fernandez02}
{Rodr{\'i}guez-Fern{\'a}ndez}, N.~J., {Mart{\'i}n-Pintado}, J., {de Vicente},
  P., \& {Fuente}, A. 2002, \apss, 281, 331

\bibitem[{{Rodr{\'i}guez-Fern{\'a}ndez}
  {et~al.}(2001){Rodr{\'i}guez-Fern{\'a}ndez}, {Mart{\'i}n-Pintado}, {Fuente},
  {de Vicente}, {Wilson}, \& {H{\"u}ttemeister}}]{rodriguez01}
{Rodr{\'i}guez-Fern{\'a}ndez}, N.~J., {Mart{\'i}n-Pintado}, J., {Fuente}, A.,
  {et~al.} 2001, \aap, 365, 174

\bibitem[{{Roueff} {et~al.}(2021){Roueff}, {Gerin}, {Gratier}, {Levrier},
  {Pety}, {Gaudel}, {Goicoechea}, {Orkisz}, {de Souza Magalhaes}, {Vono},
  {Bardeau}, {Bron}, {Chanussot}, {Chainais}, {Guzman}, {Hughes},
  {Kainulainen}, {Languignon}, {Le Bourlot}, {Le Petit}, {Liszt}, {Marchal},
  {Miville-Desch{\^e}nes}, {Peretto}, {Roueff}, \& {Sievers}}]{Roueff21}
{Roueff}, A., {Gerin}, M., {Gratier}, P., {et~al.} 2021, \aap, 645, A26

\bibitem[{{Ruze}(1966)}]{Ruze66}
{Ruze}, J. 1966, IEEE Proceedings, 54, 633

\bibitem[{{Sakre} {et~al.}(2022){Sakre}, {Habe}, {Pettitt}, {Okamoto},
  {Enokiya}, {Fukui}, \& {Hosokawa}}]{Sakre22}
{Sakre}, N., {Habe}, A., {Pettitt}, A.~R., {et~al.} 2022, arXiv e-prints,
  arXiv:2205.07057

\bibitem[{{Salii}{} {et~al.}(2002){Salii}{}, {Sobolev}, \&
  {Kalinina}}]{Salii02}
{Salii}{}, S.~V., {Sobolev}, A.~M., \& {Kalinina}, N.~D. 2002, Astronomy
  Reports, 46, 955

\bibitem[{{Sano} {et~al.}(2021){Sano}, {Tsuge}, {Tokuda}, {Muraoka},
  {Tachihara}, {Yamane}, {Kohno}, {Fujita}, {Enokiya}, {Rowell}, {Maxted},
  {Filipovi{\'c}}, {Knies}, {Sasaki}, {Onishi}, {Plucinsky}, \&
  {Fukui}}]{Sano21}
{Sano}, H., {Tsuge}, K., {Tokuda}, K., {et~al.} 2021, \pasj, 73, S62

\bibitem[{{Sawada} {et~al.}(2004){Sawada}, {Hasegawa}, {Handa}, \&
  {Cohen}}]{Sawada04}
{Sawada}, T., {Hasegawa}, T., {Handa}, T., \& {Cohen}, R.~J. 2004, \mnras, 349,
  1167

\bibitem[{{Schilke} {et~al.}(1997){Schilke}, {Walmsley}, {Pineau des F\^orets},
  \& {Flower}}]{Schilke1997}
{Schilke}, P., {Walmsley}, C.~M., {Pineau des F\^orets}, G., \& {Flower}, D.~R.
  1997, \aap, 321, 293

\bibitem[{{Sch{\"o}ier} {et~al.}(2005){Sch{\"o}ier}, {van der Tak}, {van
  Dishoeck}, \& {Black}}]{Schoier05}
{Sch{\"o}ier}, F.~L., {van der Tak}, F.~F.~S., {van Dishoeck}, E.~F., \&
  {Black}, J.~H. 2005, \aap, 432, 369

\bibitem[{{Schuller} {et~al.}(2009){Schuller}, {Menten}, {Contreras},
  {Wyrowski}, {Schilke}, {Bronfman}, {Henning}, {Walmsley}, {Beuther},
  {Bontemps}, {Cesaroni}, {Deharveng}, {Garay}, {Herpin}, {Lefloch}, {Linz},
  {Mardones}, {Minier}, {Molinari}, {Motte}, {Nyman}, {Reveret}, {Risacher},
  {Russeil}, {Schneider}, {Testi}, {Troost}, {Vasyunina}, {Wienen}, {Zavagno},
  {Kovacs}, {Kreysa}, {Siringo}, \& {Wei{\ss}}}]{Schuller09}
{Schuller}, F., {Menten}, K.~M., {Contreras}, Y., {et~al.} 2009, \aap, 504, 415

\bibitem[{{Scoville}(1972)}]{Scoville72}
{Scoville}, N.~Z. 1972, \apjl, 175, L127

\bibitem[{{Shirley}(2015)}]{Shirley15}
{Shirley}, Y.~L. 2015, \pasp, 127, 299

\bibitem[{{Sobolev}(1960)}]{Sobolev60}
{Sobolev}, V.~V. 1960, {Moving envelopes of stars} (Cambridge: Harvard
  University Press)

\bibitem[{{Sofue}{}(1995)}]{Sofue95}
{Sofue}{}, Y. 1995, \pasj, 47, 551

\bibitem[{{Sofue}(2022)}]{Sofue22}
{Sofue}, Y. 2022, \pasj, 74, L23

\bibitem[{{Sormani} {et~al.}(2015){Sormani}, {Binney}, \&
  {Magorrian}}]{Sormani2015}
{Sormani}, M.~C., {Binney}, J., \& {Magorrian}, J. 2015, \mnras, 449, 2421

\bibitem[{{Sormani} {et~al.}(2019){Sormani}, {Tre{\ss}}, {Glover}, {Klessen},
  {Barnes}, {Battersby}, {Clark}, {Hatchfield}, \& {Smith}}]{Sormani19}
{Sormani}, M.~C., {Tre{\ss}}, R.~G., {Glover}, S. C.~O., {et~al.} 2019, \mnras,
  488, 4663

\bibitem[{{Sormani} {et~al.}(2020){Sormani}, {Tress}, {Glover}, {Klessen},
  {Battersby}, {Clark}, {Hatchfield}, \& {Smith}}]{Sormani20}
{Sormani}, M.~C., {Tress}, R.~G., {Glover}, S. C.~O., {et~al.} 2020, \mnras,
  497, 5024

\bibitem[{{Tanaka}{} {et~al.}(2007){Tanaka}{}, {Kamegai}, {Nagai}, \&
  {Oka}}]{Tanaka07}
{Tanaka}{}, K., {Kamegai}, K., {Nagai}, M., \& {Oka}, T. 2007, \pasj, 59, 323

\bibitem[{{Tokuyama} {et~al.}(2019){Tokuyama}, {Oka}, {Takekawa}, {Iwata},
  {Tsujimoto}, {Yamada}, {Furusawa}, \& {Nomura}}]{Tokuyma19}
{Tokuyama}, S., {Oka}, T., {Takekawa}, S., {et~al.} 2019, \pasj, 71, S19

\bibitem[{{Tress} {et~al.}(2020){Tress}, {Sormani}, {Glover}, {Klessen},
  {Battersby}, {Clark}, {Hatchfield}, \& {Smith}}]{Tress20}
{Tress}, R.~G., {Sormani}, M.~C., {Glover}, S. C.~O., {et~al.} 2020, \mnras,
  499, 4455

\bibitem[{{Tsujimoto} {et~al.}(2021){Tsujimoto}, {Oka}, {Takekawa}, {Iwata},
  {Uruno}, {Yokozuka}, {Nakagawara}, {Watanabe}, {Kawakami}, {Nishiyama},
  {Kaneko}, {Kanno}, \& {Ogawa}}]{Tsujimoto21}
{Tsujimoto}, S., {Oka}, T., {Takekawa}, S., {et~al.} 2021, \apj, 910, 61

\bibitem[{{van der Tak}{} {et~al.}(2007){van der Tak}{}, {Black},
  {Sch{\"o}ier}, {Jansen}, \& {van Dishoeck}}]{vanderTak07}
{van der Tak}{}, F.~F.~S., {Black}, J.~H., {Sch{\"o}ier}, F.~L., {Jansen},
  D.~J., \& {van Dishoeck}, E.~F. 2007, \aap, 468, 627

\bibitem[{{Wannier}(1980)}]{Wannier80}
{Wannier}, P.~G. 1980, ARA\&A, 18, 399

\bibitem[{{Whiteoak}{} \& {Peng}(1989)}]{Whiteoak89}
{Whiteoak}{}, J.~B. \& {Peng}, R.~S. 1989, \mnras, 239, 677

\bibitem[{{Wilson}(1999)}]{Wilson99}
{Wilson}, T.~L. 1999, Reports on Progress in Physics, 62, 143

\bibitem[{{Wu} {et~al.}(2021){Wu}, {Trejo}, {Espada}, \& {Miyamoto}}]{Wu21}
{Wu}, Y.-T., {Trejo}, A., {Espada}, D., \& {Miyamoto}, Y. 2021, \mnras, 504,
  3111

\bibitem[{{Zeng} {et~al.}(2020){Zeng}, {Zhang}, {Jim{\'e}nez-Serra}, {Tercero},
  {Lu}, {Mart{\'i}n-Pintado}, {de Vicente}, {Rivilla}, \& {Li}}]{Zeng20}
{Zeng}, S., {Zhang}, Q., {Jim{\'e}nez-Serra}, I., {et~al.} 2020, \mnras, 497,
  4896

\end{thebibliography}


\begin{appendix}

\section{Spline interpolation}\label{s:spline}
The simplest way to correct for subtraction of off-source emission during calibration is to add on- and off-source intensities in each channel. This way, not only the lines detected in the off-position are added but also the noise of its spectrum. To avoid this, we used spline interpolation, with which we modelled the line features of the off-position while letting the intensity of channels without any features be zero. 
The observed features are divided into $n$ segments by $n+1$ points of intersection. Each section $n_i$ is then described by a polynomial function of order 3 
as
\begin{align}\label{eq:splinefct}
    n_i(\varv) = a_i + b_i(\varv - \varv_{i-1}) + c_i(\varv - \varv_{i-1})^2 + d_i(\varv - \varv_{i-1})^3,
\end{align}
where $\varv_{i-1}$ is the ending point of the preceding and the starting point of the subsequent segment. To solve this equation, initial conditions are assumed, which are:
\begin{align}
    n_i(\varv_{i-1}) &= n_{i+1}(\varv_{i-1}) \notag  \\
    \left. \frac{dn_i(\varv)}{d\varv}\right|_{\varv=\varv_{i-1}} &= \left. \frac{dn_{i+1}(\varv)}{d\varv}\right|_{\varv=\varv_{i-1}} \\
    \left. \frac{d^2n_i(\varv)}{d\varv^2}\right|_{\varv=\varv_i} &= \left. \frac{d^2n_{i+1}(\varv)}{d\varv^2}\right|_{\varv=\varv_i}. \notag
\end{align}
Based on these relations, expressions for the coefficients $a_i, b_i, c_i,$ and $d_i$ can be found in order to solve Eq.~\ref{eq:splinefct} for each segment. 
In practice, we used the function \texttt{UnivariateSpline}, which is implemented in python. This function fits a spline $spl(v)$ of degree $k$ (in our case: $k=3$) to the data. It selects the number of intersection points according to a given smoothing condition~$s$. The number of knots is increased until the smoothing condition is satisfied, which is
\begin{align}
    \sum_i w_i \cdot (Y(i)-spl(i))^2 \le s, 
\end{align}
where $Y(i)$ is the observed intensity at velocity/channel $i$, and $w_i$ is the weighting, which we set equal to 1 because we assumed equal weighting of all points. We chose the smoothing condition such that the difference $Y-spl$ was similar to the rms value in the observed spectrum. 

\section{Non-LTE modelling of CH$_3$CN}\label{app:ch3cn}

\begin{table}[]
\caption{Results of the 1D Gaussian fit for CH$_3$CN spectra.}
\centering
\begin{tabular}[t]{rrrrr}
    \hline\hline\\[-0.3cm]
    Component & Transition & $\varv_\mathrm{lsr}$\tablefootmark{a} & $\Delta\varv$\tablefootmark{b} & $T_{\rm int}$\tablefootmark{c} \\
    & & (km\,s$^{-1}$) & (km\,s$^{-1}$) & K\,km\,s$^{-1}$ \\
    \hline\\[-.3cm]
     G1.3 P1/c3 & 5$_0$--4$_0$ & \multirow{3}{*}{183.6} & \multirow{3}{*}{13.9} & 0.94\\
     & 5$_1$--4$_1$ & & & 0.94 \\
     & 5$_3$--4$_3$ & & & 0.49 \\
     & 6$_0$--5$_0$ & \multirow{3}{*}{183.0} & \multirow{3}{*}{14.0} & 0.77 \\
     & 6$_1$--5$_1$ & & & 0.77 \\
     & 6$_3$--5$_3$ & & & 0.47 \\
     P2/c2 & 5$_0$--4$_0$ & \multirow{3}{*}{112.0} & \multirow{3}{*}{28.0} & 2.54 \\
     & 5$_1$--4$_1$ & & & 2.54 \\
     & 5$_3$--4$_3$ & & & 2.23 \\
     & 6$_0$--5$_0$ & \multirow{3}{*}{117.0} & \multirow{3}{*}{29.9} & 1.88 \\
     & 6$_1$--5$_1$ & & & 1.88 \\
     & 6$_3$--5$_3$ & & & 1.57 \\
     P3/c2 & 5$_0$--4$_0$ & \multirow{3}{*}{118.8} & \multirow{3}{*}{19.5} & 2.21 \\
     & 5$_1$--4$_1$ & & & 2.21 \\
     & 5$_3$--4$_3$ & & & 2.85 \\
     & 6$_0$--5$_0$ & \multirow{3}{*}{119.5} & \multirow{3}{*}{15.3} & 1.36 \\
     & 6$_1$--5$_1$ & & & 1.36 \\
     & 6$_3$--5$_3$ & & & 1.96 \\
     G1.6 P1/c1 & 5$_0$--4$_0$ & \multirow{3}{*}{47.1} & \multirow{3}{*}{11.7} & 1.77 \\
     & 5$_1$--4$_1$ & & & 1.77 \\
     & 5$_3$--4$_3$ & & & 0.66 \\
     & 6$_0$--5$_0$ & \multirow{3}{*}{47.7} & \multirow{3}{*}{12.2} & 1.27 \\
     & 6$_1$--5$_1$ & & & 1.27 \\
     & 6$_3$--5$_3$ & & & 0.90 \\
     P1/c3 & 5$_0$--4$_0$ & \multirow{3}{*}{150.0} & \multirow{3}{*}{18.0} & 2.08 \\
     & 5$_1$--4$_1$ & & & 2.08 \\
     & 5$_3$--4$_3$ & & & 1.61 \\
     & 6$_0$--5$_0$ & \multirow{3}{*}{150.0} & \multirow{3}{*}{21.0} & 1.68 \\
     & 6$_1$--5$_1$ & & & 1.68 \\
     & 6$_3$--5$_3$ & & & 1.46 \\
     P2/c1 & 5$_0$--4$_0$ & \multirow{3}{*}{54.2} & \multirow{3}{*}{9.0} & 1.14 \\
     & 5$_1$--4$_1$ & & & 1.14 \\
     & 5$_3$--4$_3$ & & & 0.44 \\
     & 6$_0$--5$_0$ & \multirow{3}{*}{54.0} & \multirow{3}{*}{8.8} & 0.87 \\
     & 6$_1$--5$_1$ & & & 0.87 \\
     & 6$_3$--5$_3$ & & & 0.65 \\
     P2/c2 & 5$_0$--4$_0$ & \multirow{3}{*}{157.5} & \multirow{3}{*}{31.4} & 1.89 \\
     & 5$_1$--4$_1$ & & & 1.89 \\
     & 5$_3$--4$_3$ & & & 1.26 \\
     & 6$_0$--5$_0$ & \multirow{3}{*}{158.0} & \multirow{3}{*}{29.0} & 1.49 \\
     & 6$_1$--5$_1$ & & & 1.49 \\
     & 6$_3$--5$_3$ & & & 1.00 \\
     P3/c1 & 5$_0$--4$_0$ & \multirow{3}{*}{49.4} & \multirow{3}{*}{13.2} & 2.06 \\
     & 5$_1$--4$_1$ & & & 2.06 \\
     & 5$_3$--4$_3$ & & & 1.27 \\
     & 6$_0$--5$_0$ & \multirow{3}{*}{49.0} & \multirow{3}{*}{10.3} & 1.28 \\
     & 6$_1$--5$_1$ & & & 1.28 \\
     & 6$_3$--5$_3$ & & & 0.64 \\
    \hline\hline 
\end{tabular}
\tablefoot{\tablefoottext{a}{Source velocity at rest frequency.}\tablefoottext{b}{Full width at half maximum.}\tablefoottext{c}{Integrated intensity.}}
\label{tab:ch3cn-gauss}
\end{table}

\begin{table}
    \caption{RADEX results for CH$_3$CN.}
\centering
\begin{tabular}[t]{rrrrr}
    \hline\hline\\[-0.3cm]
    Component & $N$\tablefootmark{a} & $T_\mathrm{kin}$\tablefootmark{b} & $n({\rm H}_2)$\tablefootmark{c} & $\tau_{5_0-4_0}$\tablefootmark{d} \\
    & (10$^{13}$\,cm$^{-2}$) & (K) & (10$^{4}$\,cm$^{-3}$) & \\
    \hline\\[-.3cm]
     G1.3 P1/c3 & 0.6 & 65$^{+30}_{-10}$ & $>$4.0 & 0.010 \\[0.1cm]
     P2/c2 & 3.2 & 100$^{+30}_{-20}$ & 3.2$^{+3.1}_{-1.6}$ & 0.030 \\[0.1cm]
     P3/c2 & 5.0 & 175$^{+60}_{-35}$ & 1.6$^{+0.9}_{-0.8}$ & 0.050 \\[0.1cm]
     G1.6 P1/c1 & 1.3 & 70$^{+10}_{-10}$ & 12.6$^{+19.0}_{-6.3}$ & 0.020  \\[0.1cm]
     P1/c3 & 1.6 & 100$^{+15}_{-15}$ & 7.9$^{+4.6}_{-2.9}$ & 0.020 \\[0.1cm]
     P2/c1 & 1.0 & 70$^{+20}_{-10}$ & $>$31.6 & 0.002  \\[0.1cm]
     P2/c2 & 1.6 & 75$^{+20}_{-10}$ & 6.3$^{+6.3}_{-3.8}$ & 0.010 \\[0.1cm]
     P3/c1 & 10.0 & 60$^{+5}_{-5}$ & 0.8$^{+0.5}_{-0.5}$ & 0.200 \\[0.1cm]
    \hline\hline 
\end{tabular}
\tablefoot{\tablefoottext{a}{Best-fit column density for CH$_3$CN.}\tablefoottext{b}{Kinetic temperature determined by using $\chi^2$ minimisation (see Fig.\,\ref{fig:ch3cn-radex} and text).}\tablefoottext{c}{Number density determined in the same way as $T_{\rm kin}$.}\tablefoottext{d}{Opacity of the 5$_0$--4$_0$ transition at the position of minimum $\chi^2$.}}
\label{tab:ch3cn-radex}
\end{table}
\begin{figure*}
    \centering
    \includegraphics[width=.8\textwidth]{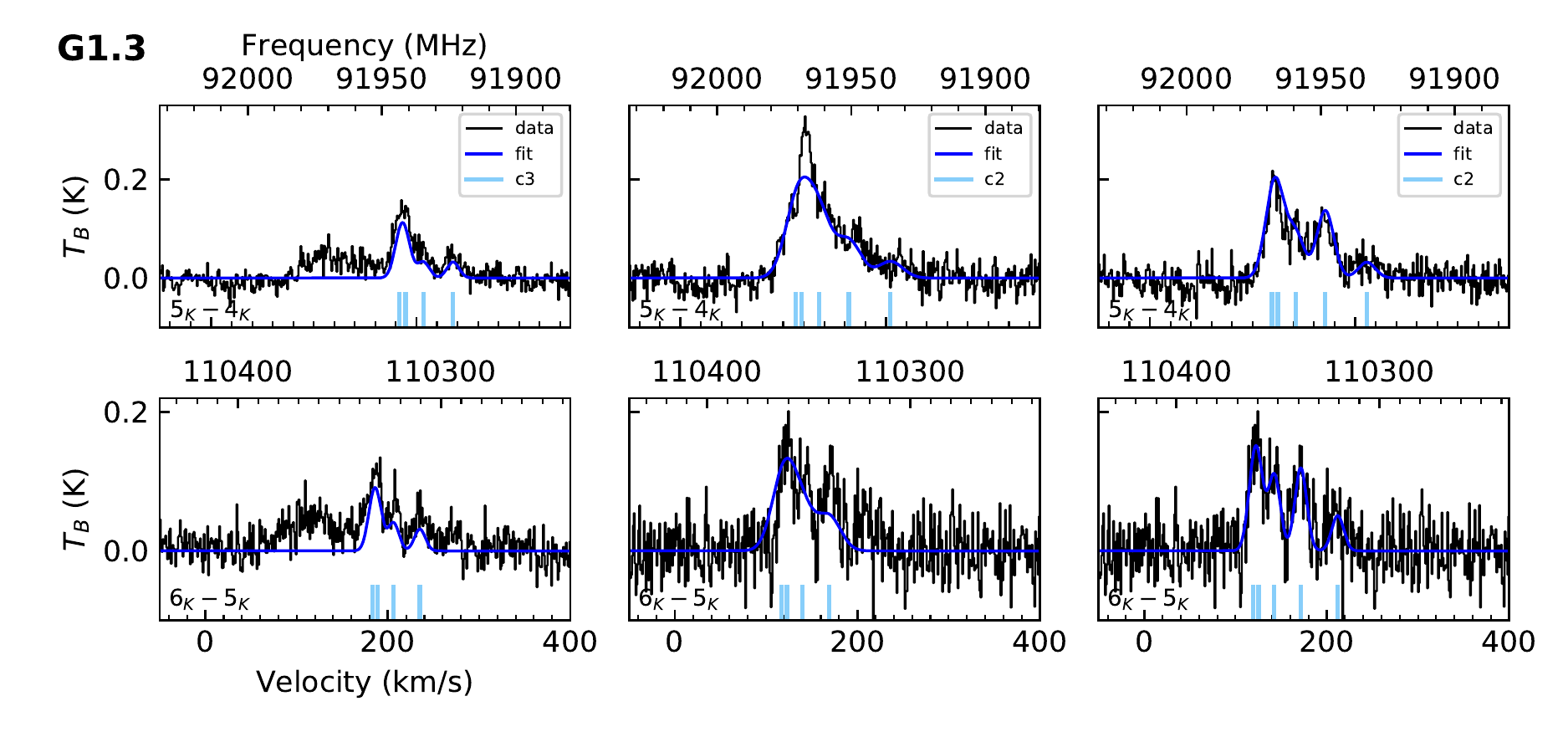}
    \includegraphics[width=.8\textwidth]{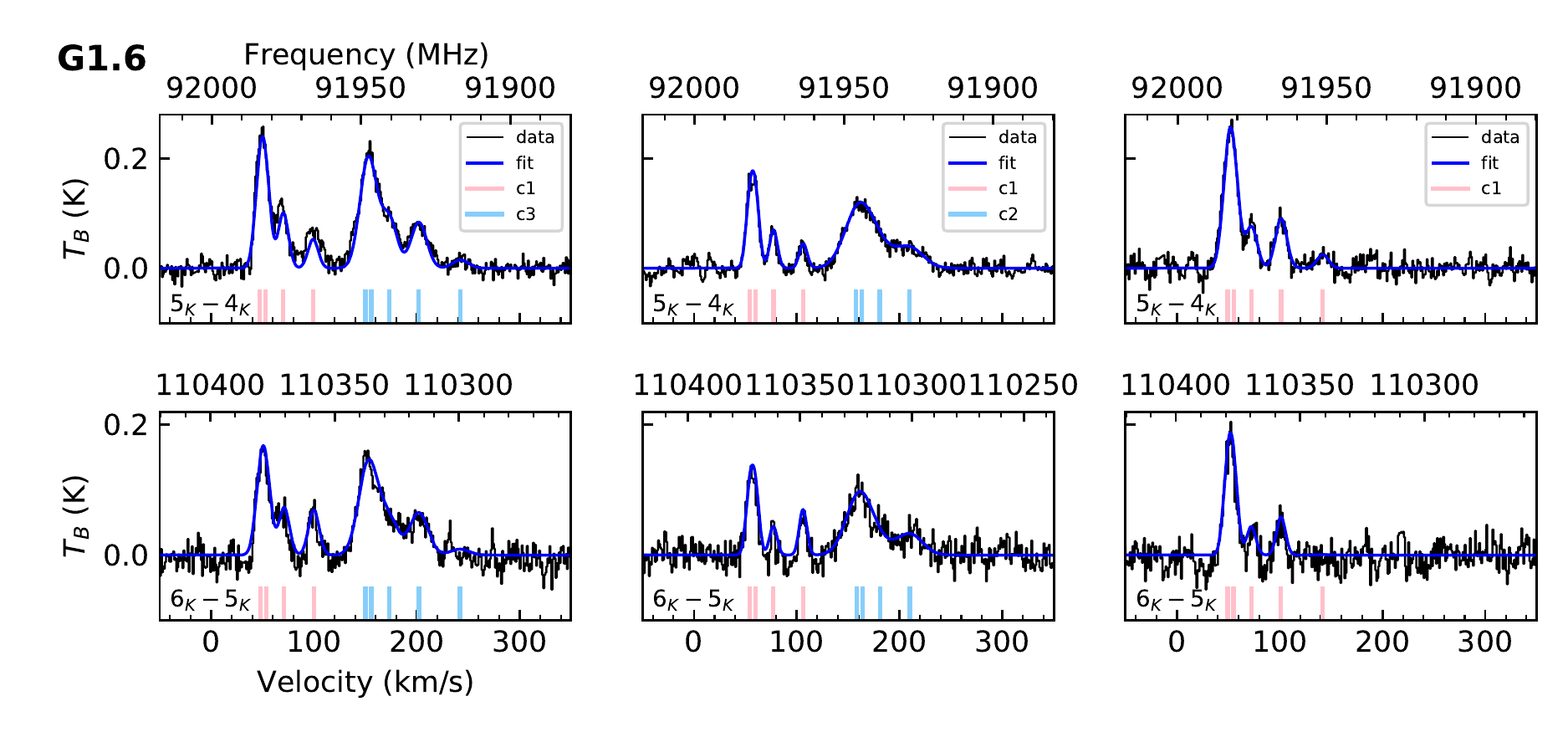}
    \caption{CH$_3$CN 5$_K$\,--\,4$_K$ and 6$_K$\,--\,5$_K$ spectra at positions P1 to P3 (\textit{left to right}) in G1.3 (\textit{top}) and G1.6 (\textit{bottom}), respectively. The blue curve shows the 1D Gaussian fit. Rest frequencies of all $K$ transitions for each fitted velocity component c$i$ with $i=(1,2)$ are indicated with vertical lines at the bottom.}
    \label{fig:ch3cn-spectra}
\end{figure*}
\begin{figure*}[h]
    \centering
    \includegraphics[width=.85\textwidth]{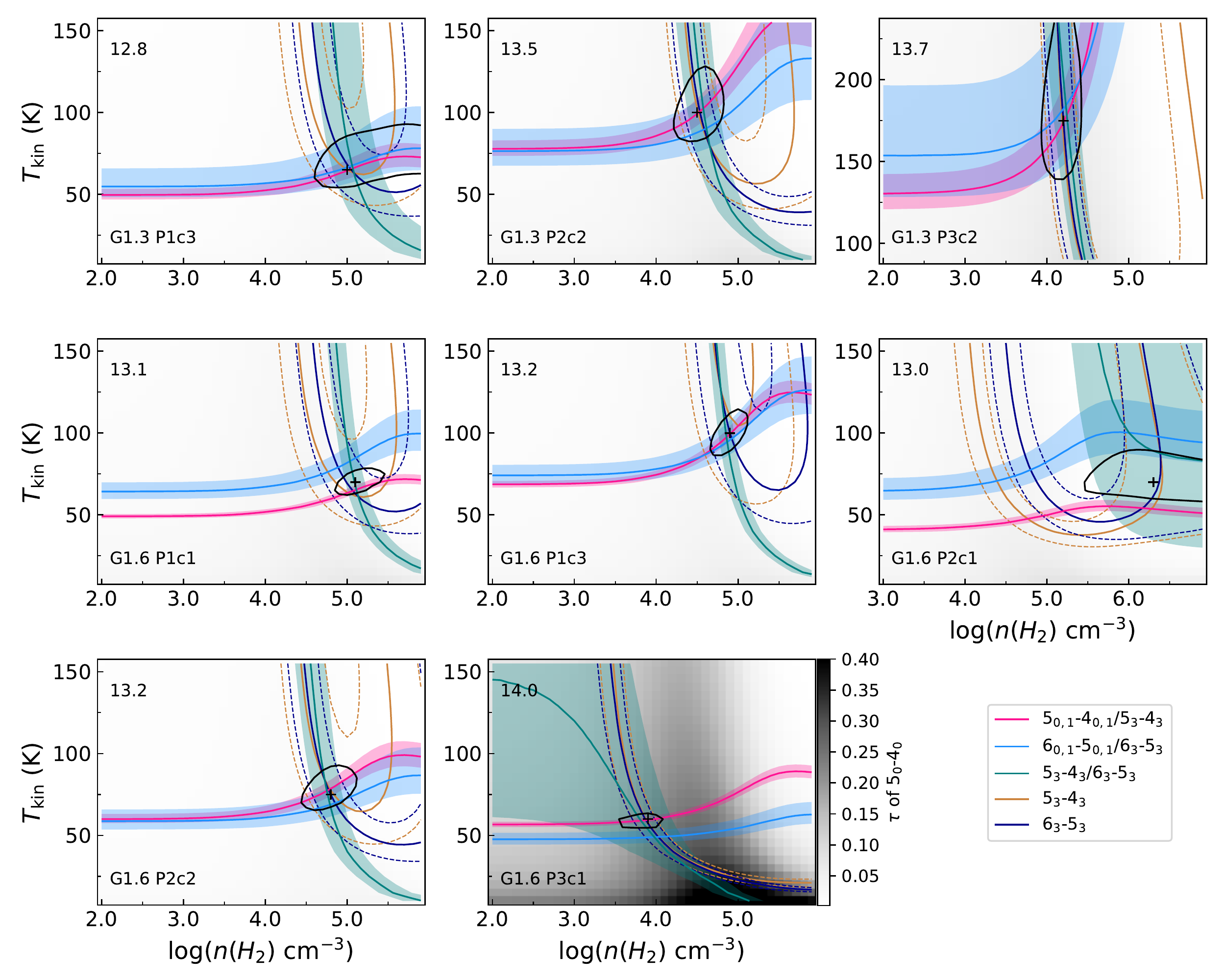}
    \caption{RADEX results for CH$_3$CN. Yellow and dark blue contours show where observed and modelled integrated intensities match, while teal, light blue and pink contours show the same but using ratios of integrated intensity. Dashed contours and shaded areas represent uncertainties, respectively. The grey scale shows the opacity of the 5$_0$--4$_0$ transition in all panels. The $K=0$ and $K=1$ lines cannot be separated due to blending and were, therefore, considered as one component. The black cross shows the most probable result for $T_{\rm kin}$ and $n({\rm H}_2)$ determined by minimizing $\chi^2$, that is, where $\Delta\chi^2= \chi^2-\chi^2_{\rm min}=0$. The closed black contours indicate a confidence level of 95.45\%. The respective source and velocity component are shown in the lower left, the best-fit decimal power of column density (in cm$^{-2}$) in the upper left corner.}
    \label{fig:ch3cn-radex}
\end{figure*}

To determine the full width at half maximum ($FWHM$) and the range over which to integrate intensities, we applied a 1D Gaussian profile to each transition using the HFS method of the \texttt{minimize} command in CLASS, which allows simultaneous fitting of different lines (and even blended ones). The rather noisy spectra shown in Fig.\,\ref{fig:ch3cn-spectra} and the strong blending of some components prevented us from fitting each component that was previously identified in the CS\,2--1 spectra shown in Figs.\,\ref{fig:CS13} and \ref{fig:CS16}. 
We show the fit for those components, for which it was possible, in Fig.\,\ref{fig:ch3cn-spectra} and summarise the results for source velocity at rest frequency, line width, and integrated intensities in Table\,\ref{tab:ch3cn-gauss}. 
We used only transitions that could be clearly identified and fitted in the spectra and that were, therefore, used during the non-LTE analysis with RADEX. These transitions are the $K=(0,1,3)$ lines of the $J=5_K-4_K$ and $6_K-5_K$ multiplets, respectively. The $K=0$ and 1 transitions were considered as one component as they are always blended and their relative individual contributions to the observed line are unknown. 
Therefore, we fitted the observed spectral feature 
such that both transitions contribute approximately equally to the total integrated intensity\footnote{This decision is based on  tests exploring the appropriate parameter space, in which it was found that the individual contributions of the $K=0$ and 1 line indeed always remained roughly equal.}. 

For each fitted component, RADEX was run  over  a $40\times30\times5$ grid of H$_2$ number densities $n({\rm H}_2)$, kinetic temperature $T_{\rm kin}$, and column density $N_X$, respectively. For each $N_X$ value, we investigated for which combination of $(n({\rm H}_2)$,$T_{\rm kin})$ the modelled integrated intensities best matched the observed ones. We compare the ratios of integrated intensities for $J=5_{0,1}-4_{0,1}$ to $J=5_3-4_3$, for $J=6_{0,1}-5_{0,1}$ to $J=6_3-5_3$, and for $J=5_3-4_3$ to $J=6_3-5_3$. 
The ratios are 
independent of column density indicating that the line emission is optically thin for the assumed range of column density of $10^{13}-10^{15}$\,cm$^{-2}$. Therefore, we also included the integrated intensities of the $K=3$ components of both $J\,-$\,multiplets, which are able to place a constraint on the column density. 
For velocity components in G1.3 and G1.6, for which a given $N_X$ does not lie within the intersection of the ratios, RADEX was rerun with a finer-tuned $N_X$ grid. The best-fit results are shown in Fig.\,\ref{fig:ch3cn-radex}. 
The uncertainty on the ratios (colour-shaded area) includes the 1$\sigma$ rms level of the respective spectrum and the uncertainty on the fit of integrated intensity provided by CLASS. The uncertainty on the integrated intensity (coloured dashed lines) includes the same contributions plus an uncertainty on the beam filling factor, where we allow for a deviation of 0.3 from unity.

The kinetic temperature and H$_2$ number density are determined by using a minimised $\chi^2$ method following section\,15.6 in \citet{Press92}, where $\chi^2$ is computed by\begin{align}
\chi^2 = \sum_{i=1}^N \frac{\left(T^{\rm mod}_{\rm int,i}-T^{\rm obs}_{\rm int,i}\right)^2}{\sigma_{\rm obs,i}^2},
\end{align}
where $T^{\rm mod}_{\rm int}$ is the modelled integrated intensity or the modelled ratio of the same for a given transition, $T^{\rm obs}_{\rm int}$ is the observed integrated intensity or ratio, and $\sigma_{\rm obs}$ the propagated error on the observed values. The sum considers all available transitions.
The black cross in each panel of Fig.\,\ref{fig:ch3cn-radex} indicates the most probable combination of $(n({\rm H}_2)$,$T_{\rm kin})$, that is the location of $\Delta\chi^2=\chi^2-\chi^2_{\rm min}= 0$. The black contour in each panel defines a confidence level of 95.45\% \citep[][]{Press92}. 
The results are summarised in Table\,\ref{tab:ch3cn-radex}. The uncertainties on $T_{\rm kin}$ and $n({\rm H}_2)$ correspond to minimum and maximum values covered by the black contour, respectively. For components P1/c3 in G1.3 and P2/c1 in G1.6, only lower limits for H$_2$ number density can be determined. 
Temperature values span the  range from 60 to 175\,K. The highest temperature of 175\,K, measured in component P3/c2 in G1.3, is most probably the result of a contamination of the $K=3$ transitions by the $K=(0,1)$ transitions of another velocity component that was identified in the CS spectra but could not be fitted properly in the CH$_3$CN spectra. 
Similarly, this may apply to component P2/c2 in G1.3 leading to a higher temperature of 100\,K. In the following we use the kinetic temperature derived from CH$_3$CN to derive column densities of other molecules. Because the temperatures generally agree within the error bars, except for the values for the two positions mentioned above, which are probably the result of blending of spectral lines, we decided to use an average of $T_{\rm kin}=75$\,K for all components during the subsequent RADEX modelling. We even use this value for components that could not be fitted in the CH$_3$CN spectra, and hence, did not yield temperature constraints, to obtain column densities for these components, too. 

H$_2$ number densities are at a few 10$^{4}$\,cm$^{-3}$ for all velocity components, except P1/c1 in G1.6, which possibly shows evidence for a slightly higher density of $\sim$10$^5$\,cm$^{-3}$, and P3/c1 in G1.6, which shows a lower value of $<$10$^{4}$\,cm$^{-3}$. 
Additionally, we show optical depth of the $J=5_{0}-4_{0}$ transitions obtained from RADEX in Table\,\ref{tab:ch3cn-radex}. 

\section{Additional tables: Column densities}
Tables\,\ref{tab:Ncol} and \ref{tab:Ncol2} list molecular column densities derived from non-LTE modelling with RADEX.
\begin{table*}
    \caption{Column densities for each component in G1.3 obtained from non-LTE modelling with RADEX.}
    \centering
    \begin{tabular}{rcccccc}
    \hline\hline \\[-0.3cm]
    & \multicolumn{5}{c}{$N_{\rm col} (10^{13}\,{\rm cm}^{-2})$} \\\cmidrule(lr){2-7}
    & HC$_3$N & H$_2$CO & N$_2$H$^+$ & CS & HCN & HNC  \\ \hline\\[-0.3cm]
     G1.3 P1/c1 & 1.0$\,-\,$8.9 & 0.2$\,-\,$22.4 & 0.9$\,-\,$5.0 & 12.6$\,-\,$141.3 & 22.4$\,-\,$354.8 & 2.2$\,-\,$63.1 \\
     P1/c2 & -- & -- & 0.1$\,-\,$0.6 & 2.0$\,-\,$15.8 & 3.2$\,-\,$56.2 & 0.2$\,-\,$4.0 \\
     P1/c3 & 1.3$\,-\,$3.2 & 0.4$\,-\,$6.3 & 0.6$\,-\,$2.0 & 5.6$\,-\,$12.6 & 15.8$\,-\,$63.1 & 0.8$\,-\,$3.5 \\
     P2/c1 & 1.0$\,-\,$7.9 & 0.3$\,-\,$12.6 & --$^*$ &  6.3$\,-\,$20.0 & 14.1$\,-\,$251.2 & --$^*$ \\
     P2/c2 & 1.4$\,-\,$5.6 & 1.4$\,-\,$6.3 & --$^*$ & 8.9$\,-\,$20.0 & 31.6$\,-\,$158.5 & --$^*$ \\
     P2/c3 & 1.3$\,-\,$8.9 & 0.4$\,-\,$22.4 & --$^*$ & 6.3$\,-\,$20.0 & 10.0$\,-\,$158.5 & --$^*$ \\
     P3/c1 & -- & -- & -- & 1.3$\,-\,$10.0 & --$^*$ & 0.2$\,-\,$4.0 \\
     P3/c2 & 2.5$\,-\,$14.1 & 6.3$\,-\,$50.1 & 1.3$\,-\,$6.3 & 15.8$\,-\,$89.1 & 75.9$\,-\,$602.9 & 4.0$\,-\,$39.8 \\
     P3/c3 & 0.9$\,-\,$6.3 & 0.4$\,-\,$22.4 & 0.4$\,-\,$2.2 & 6.3$\,-\,$56.2 & 15.1$\,-\,$240.0 & 0.6$\,-\,$12.6 \\
     G1.6 P1/c1 & 0.9$\,-\,$1.6 & 0.4$\,-\,$0.8 & 1.0$\,-\,$2.2 & 2.5$\,-\,$5.6 & 3.2$\,-\,$15.8 & 1.6$\,-\,$6.3 \\
     P1/c2 & 0.9$\,-\,$2.2 & 0.1$\,-\,$2.5 & 0.6$\,-\,$1.6 & 5.0$\,-\,$15.8 & 5.6$\,-\,$39.8 & 2.5$\,-\,$20.0 \\
     P1/c3 & 2.0$\,-\,$4.0 & 1.6$\,-\,$6.3 & 0.6$\,-\,$1.3 & 8.9$\,-\,$20.0 & 22.4$\,-\,$63.1 & 1.6$\,-\,$5.0  \\
     P2/c1 & 0.6$\,-\,$4.0 & 0.3$\,-\,$14.1 & --$^*$ & 4.0$\,-\,$25.1 & 5.0$\,-\,$125.9 & --$^*$ \\
     P2/c2 & 2.5$\,-\,$7.9 & 1.4$\,-\,$15.8 & 0.2$\,-\,$0.5 & 6.3$\,-\,$22.4 & 31.6$\,-\,$141.3 & 1.4$\,-\,$7.9 \\
     P3/c1 & 5.0$\,-\,$39.8 & 5.6$\,-\,$56.2 & 3.2$\,-\,$31.6 & 25.1$\,-\,$100.0 & 56.2$\,-\,$562.3 & --$^*$ \\
     P3/c2 & 0.2$\,-\,$2.2 & -- & -- & -- & 0.1$\,-\,$6.3 & -- \\
     \hline\hline
    \end{tabular}
    \tablefoot{(--) Non-detections. (--$^*$) The molecule is detected, however, it has not been modelled with RADEX because either line width or peak intensity or both could not be determined properly for one or multiple of the following reasons: self-absorption, high optical depth, blending with other components,  HFS transitions.}
    \label{tab:Ncol}
\end{table*}
\begin{table*}
    \caption{Same as Table\,\ref{tab:Ncol}, but for G1.6.}
    \centering
    \begin{tabular}{rcccccc}
    \hline\hline \\[-0.3cm]
    & \multicolumn{5}{c}{$N_{\rm col} (10^{13}\,{\rm cm}^{-2})$} & $N_{\rm col} (10^{16}\,{\rm cm}^{-2})$ \\\cmidrule(lr){2-6}
    & HCO$^+$ & HNCO & SiO & SO & OCS & CH$_3$OH \\ \hline\\[-0.3cm]
     G1.3 P1/c1 & 3.5$\,-\,$22.4 & $\geq$8.9 & 1.6$\,-\,$3.2 & 6.3$\,-\,$35.5 & 10.0$\,-\,$35.5 &  0.2$\,-\,$0.6 \\
     P1/c2 & 0.6$\,-\,$5.6 & $\geq$0.8 & 0.3$\,-\,$2.2 & 1.3$\,-\,$5.6 & -- & -- \\
     P1/c3 & 1.4$\,-\,$4.0 & $\geq$2.0 & 0.4$\,-\,$1.6 & 3.2$\,-\,$10.0 & 6.3$\,-\,$15.8  & 0.2$\,-\,$0.4 \\
     P2/c1 & 1.6$\,-\,$10.0 & $\geq$6.3 & 1.3$\,-\,$5.0 & 2.5$\,-\,$22.4 & 7.9$\,-\,$25.1 & 0.1$\,-\,$0.4 \\
     P2/c2 & 1.4$\,-\,$5.0 & $\geq$2.5 & 1.3$\,-\,$3.2 & 4.0$\,-\,$10.0 & 10.0$\,-\,$25.1 & 0.4$\,-\,$0.6 \\
     P2/c3 & 1.0$\,-\,$8.9 & $\geq$2.5 & 0.8$\,-\,$2.0 & 5.0$\,-\,$20.0 & -- & --\\
     P3/c1 & 0.6$\,-\,$4.0 & $\geq$0.4 & 0.1$\,-\,$0.6 & -- & -- & -- \\
     P3/c2 & 3.4$\,-\,$15.1 & $\geq$4.0 & 1.6$\,-\,$4.0 & 10.0$\,-\,$39.8 & 12.6$\,-\,$31.6 & 0.5$\,-\,$0.8 \\
     P3/c3 & 1.9$\,-\,$12.0 & $\geq$1.3 & 1.0$\,-\,$3.5 & 4.0$\,-\,$15.8 & 5.0$\,-\,$15.8 & 0.1$\,-\,$0.6 \\
     G1.6 P1/c1 & 0.4$\,-\,$1.6 & $\geq$14.1 & 0.4$\,-\,$0.8 & 2.2$\,-\,$5.0 & 14.1$\,-\,$20.0 & 0.2$\,-\,$1.6 \\
     P1/c2 & 1.0$\,-\,$5.0 & $\geq$20.0 & 1.0$\,-\,$2.5 & 5.6$\,-\,$12.6 & 31.6$\,-\,$50.1 & $\geq$0.5 \\
     P1/c3 & 2.0$\,-\,$7.9 & $\geq$6.3 & 0.8$\,-\,$1.6 & 7.9$\,-\,$15.8 & 22.4$\,-\,$50.1 & 0.2$\,-\,$0.6 \\
     P2/c1 & 0.4$\,-\,$4.0 & $\geq$22.4 & 0.4$\,-\,$2.5 & 4.0$\,-\,$7.9 & 22.4$\,-\,$39.8 & $\geq$0.4 \\
     P2/c2 & 0.9$\,-\,$6.3 & $\geq$5.0 & 1.3$\,-\,$4.0 & 10.0$\,-\,$22.4 & 10.0$\,-\,$31.6 & 0.1$\,-\,$0.6 \\
     P3/c1 & 5.0$\,-\,$39.8 & $\geq$35.5 & 0.6$\,-\,$1.6 & 15.8$\,-\,$63.1 & 25.1$\,-\,$56.2 & 0.4$\,-\,$0.6 \\
     P3/c2 & 0.2$\,-\,$2.5 & -- & -- & -- & -- & -- \\
     \hline\hline
    \end{tabular}
    \label{tab:Ncol2}
\end{table*}

\end{appendix}

\end{document}